\documentclass[aps,prd,onecolumn,preprintnumbers,nofootinbib, amsmath,amssymb,superscriptaddress]{revtex4}

\usepackage{graphicx}
\usepackage{dcolumn}
\usepackage{bm,graphicx,hyperref,subfigure}
\usepackage{float,amsfonts,graphicx}
\usepackage{color}
\usepackage{tikz}
\usepackage{adjustbox}
\usetikzlibrary{decorations.pathreplacing}
\usetikzlibrary{arrows.meta}
\usepackage{placeins}
\usepackage{graphicx}
\usepackage{dcolumn}
\usepackage{bm}
\usepackage{amsmath}
\usepackage{cancel}
\usepackage{comment}
\usepackage{multirow}
\usepackage{xcolor}

\newcommand{\bs}{\mathbf}
\newcommand{\beq}{\begin{eqnarray}}
\newcommand{\eeq}{\end{eqnarray}}

\newcommand{\real}{{\sf I}\kern-.12em{\sf R}}
\newcommand{\comp}{{\sf I}\kern-.50em{\sf C}}
\newcommand{\unity}{{\sf I}\kern-.54em{\sf 1}}

\usepackage{hyperref}

\newcommand{\tr}{\mbox{Tr}}

\begin{document}

\title{ Charmonium radiative transitions to dileptons from lattice QCD: \\ The case of $h_c \to \eta_c \ell^+\ell^-$ and 
$\chi_{c1} \to J/\psi\,\ell^+\ell^-$ }

\author{Damir Be\v{c}irevi\'c}
\affiliation{IJCLab, P\^ole Th\'eorie (Bat.~210), CNRS/IN2P3 et Universit\'e Paris-Saclay, 91405 Orsay, France\\
CERN, Theory Department, 1211 Geneva 23, Switzerland}
\author{Roberto Di~Palma} 
\affiliation{Dipartimento di Matematica e Fisica, Universit\`a  Roma Tre and INFN, Sezione di Roma Tre, Via della Vasca Navale 84, I-00146 Rome, Italy}
\author{Roberto Frezzotti} 
\affiliation{Dipartimento di Fisica and INFN, Universit\`a di Roma ``Tor Vergata",\\ Via della Ricerca Scientifica 1, I-00133 Roma, Italy}
\author{Giuseppe Gagliardi} 
\affiliation{Istituto Nazionale di Fisica Nucleare, Sezione di Roma Tre,\\ Via della Vasca Navale 84, I-00146 Rome, Italy}

\author{Vittorio Lubicz} 
\affiliation{Dipartimento di Matematica e Fisica, Universit\`a  Roma Tre and INFN, Sezione di Roma Tre, Via della Vasca Navale 84, I-00146 Rome, Italy}

\author{Francesco Sanfilippo}
\affiliation{Istituto Nazionale di Fisica Nucleare, Sezione di Roma Tre,\\ Via della Vasca Navale 84, I-00146 Rome, Italy}

\author{Nazario Tantalo} 
\affiliation{Dipartimento di Fisica and INFN, Universit\`a di Roma ``Tor Vergata",\\ Via della Ricerca Scientifica 1, I-00133 Roma, Italy}

\begin{abstract}
\vspace{5mm}
We present a lattice QCD study of dilepton production in charmonium transitions, specifically focusing on the $1^{+-} \to 0^{-+}$ and $1^{++} \to 1^{--}$ processes: $h_c \to \eta_c \ell^+ \ell^-$ and $\chi_{c1} \to J/\psi \ell^+ \ell^-$, where $\ell = e, \mu$. The relevant hadronic matrix elements are computed using gauge field configurations generated by the Extended Twisted Mass Collaboration with $N_f = 2+1+1$ dynamical Wilson--Clover twisted-mass fermions at four lattice spacings. Simulations are performed at physical dynamical $u$, $d$, $s$, and $c$ quark masses, except for the coarsest lattice, where the lightest sea quark mass corresponds to a slightly heavier pion mass. A controlled continuum extrapolation is carried out.
In the continuum limit for the $h_c$ decays, we obtain $\Gamma(h_c \to \eta_c e^+ e^-) = 5.45(19)~\mathrm{keV}$, and $\Gamma(h_c \to \eta_c \mu^+ \mu^-) = 0.635(22)~\mathrm{keV}$.
For the $\chi_{c1}$ decays, we find:
$\Gamma(\chi_{c1} \to J/\psi e^+ e^-)= 2.869(90)~\mathrm{keV}$, and 
$\Gamma(\chi_{c1} \to J/\psi \mu^+ \mu^-) = 0.1993(72)~\mathrm{keV}$.
Our results for the $\chi_{c1}$ decays show good compatibility with experimental data. However, our prediction for the $h_c \to \eta_c e^+ e^- $ decay rate is approximately $3\sigma$ larger than the BESIII result.
We also present predictions for the differential decay widths as functions of the dilepton invariant mass, $q^2$, and for angular observables sensitive to longitudinal transition form factors, which are inaccessible in radiative decays with real photon emission.
These results constitute the first fully dynamical lattice QCD predictions for dilepton decay rates in $h_c$ and $\chi_{c1}$ charmonium transitions, including their differential distributions and angular observables. They provide benchmark predictions for future experimental studies.
\vspace{3mm}
\end{abstract}

\maketitle


\section{Introduction}\label{sec:intro}

In recent years, electromagnetic transitions in heavy quarkonia have attracted significant phenomenological and theoretical interest, as they offer a sensitive probe of the internal structure and dynamics of heavy quark--antiquark bound states. Radiative decays such as $h_{c(b)} \to \eta_{c(b)} \gamma$, $J/\psi \to \eta_c \gamma$, and $\Upsilon \to \eta_b \gamma$ are governed by electric and magnetic multipole amplitudes and they can be described using effective field theories of QCD in the nonrelativistic regime, such as nonrelativistic QCD (NRQCD) and potential nonrelativistic QCD (pNRQCD)~\cite{Bodwin:1994jh,Brambilla:1999xf,Fleming:2000ib,Brambilla:2004jw}. However, these approaches rely on an expansion in the heavy-quark velocity and, in practice, require truncation, which introduces systematic uncertainties that are often difficult to quantify. By contrast, lattice QCD provides a first-principles determination of the relevant hadronic matrix elements directly from the QCD Lagrangian, without relying on a velocity expansion or model assumptions, and with systematically improvable uncertainties.

Radiative quarkonium transitions have already been studied in lattice QCD for several channels~\cite{Dudek:2006ej,Dudek:2009kk,Becirevic:2012dc,Chen:2011kpa,Li:2022cfy,Becirevic:2025idm,Becirevic:2025ocx,Delaney:2023fsc, PhysRevD.86.094501, PhysRevD.108.014513}, yielding in some cases high-precision determinations of the corresponding multipole form factors. Building on these advances, it is natural to extend this first-principles framework to transitions mediated by an off-shell photon.

Beyond purely radiative channels, electromagnetic decays involving a virtual photon that subsequently converts into a lepton--antilepton pair provide an additional probe of quarkonium structure. In these dilepton decays, also known as Dalitz decays, the invariant mass of the lepton pair ranges from threshold ($2 m_\ell$) up to the mass difference between the initial and final quarkonium states. Because the photon is off shell, the decay amplitude includes contributions from longitudinal virtual-photon polarization, which are absent in the on-shell limit ($q^2=0$). Dilepton decays thus grant access to additional structures of the hadronic matrix elements.

Experimentally, dilepton decays of charmonium states have become accessible at modern charm factories. The branching fractions for the transitions $\chi_{c1}\to J/\psi e^+e^-$ and $\chi_{c1}\to J/\psi \mu^+\mu^-$ have been measured with good precision, enabling detailed studies of integrated decay rates and quantitative tests of lepton universality. The BESIII Collaboration reported~\cite{BESIII:2017ung, BESIII:2019yeu}~\footnote{The quantum numbers of $\chi_{c1}$ and $J/\psi$ are $J^{PC}= 1^{++}$ and $1^{--}$, respectively.}
\begin{align}
\label{eq:bes_D_1}
\frac{{\rm Br}(\chi_{c1}\to J/\psi e^{+}e^{-})}{{\rm Br}(\chi_{c1}\to J/\psi \gamma)}\bigg|_{{\rm BESIII}} &= (10.1\pm 0.3\pm 0.5)\times 10^{-3},\\[8pt]
\label{eq:bes_D_2}
\frac{{\rm Br}(\chi_{c1}\to J/\psi \mu^{+}\mu^{-})}{{\rm Br}(\chi_{c1}\to J/\psi e^{+}e^{-})}\bigg|_{{\rm BESIII}} &= (6.73 \pm 0.51 \pm 0.50)\times 10^{-2}.
\end{align}
More recently, experimental studies have been extended to dilepton transitions in other charmonium states with different spin configurations. In particular, the BESIII Collaboration reported the first observation of the electromagnetic Dalitz decay $h_c \to \eta_{c} e^+e^-$~\cite{BESIII:2024kkf}, measuring~\footnote{The respective quantum numbers of $h_{c}$ and $\eta_c$ are $J^{PC}= 1^{+-}$ and $0^{-+}$.}
\begin{align}
\frac{{\rm Br}(h_c\to \eta_{c} e^+e^-)}{{\rm Br}(h_c\to \eta_c\gamma)}\bigg|_{{\rm BESIII}} = (0.59 \pm 0.10 \pm 0.04)\times 10^{-2}~,
\end{align}
with a statistical significance exceeding five standard deviations. This result demonstrates that dilepton transitions involving the $h_c$ are experimentally accessible and provide new input for the study of electromagnetic transitions in the charmonium sector.

In addition to the standard electromagnetic contribution, dilepton decays of quarkonia may, in principle, receive contributions from non-standard interactions associated with new mediators. Precise Standard Model predictions for decay rates, differential distributions, and angular observables are therefore essential for interpreting experimental measurements and identifying possible deviations from Standard Model expectations.

Motivated by this physics potential, Dalitz decays of heavy quarkonia have recently been investigated within phenomenological frameworks that exploit heavy-quark spin symmetry to relate different transitions and infer the corresponding form factors from experimental data~\cite{Colangelo:2025yud}. While such approaches provide valuable phenomenological insights, they necessarily rely on symmetry assumptions and model input, thereby motivating first-principles lattice QCD determinations of the relevant transition form factors.

Despite this growing experimental and theoretical interest, a quantitative understanding of these decays directly from QCD remains incomplete. Radiative transitions such as $h_c\to \eta_c\gamma$ and $\chi_{c1}\to J/\psi\gamma$ have already been investigated in lattice QCD~\cite{Dudek:2006ej,Dudek:2009kk,Becirevic:2012dc,Chen:2011kpa,Li:2022cfy}, including recent high-precision determinations of the associated multipole form factors~\cite{Becirevic:2025idm,Becirevic:2025ocx}. In contrast, analogous first-principles studies of dilepton channels remain limited, as their theoretical description requires the study of the full timelike $q^{2}$-dependence of the form factors. Knowledge of this momentum-transfer dependence is crucial for reliable predictions of total decay rates and differential spectra in the dilepton invariant mass.

In this paper, we present the first unquenched lattice QCD study of the dilepton decays
\begin{align}
h_c \to \eta_c \ell^+\ell^- , \qquad
\chi_{c1} \to J/\psi \ell^+\ell^- ,
\end{align}
with $\ell = e, \mu$. The calculation uses gauge field configurations with $N_f=2+1+1$ dynamical Wilson--Clover twisted-mass fermions generated by the Extended Twisted Mass Collaboration, at four lattice spacings and with physical dynamical $u$, $d$, $s$, and $c$ quark masses, except for the coarsest ensemble where the lightest pion mass is $m_\pi \simeq 175~\mathrm{MeV}$. From our lattice determination of the relevant electromagnetic transition form factors over the full kinematic range in $q^{2}$, we compute the total decay rates for final states containing either an electron or a muon pair. We also present predictions for the differential decay widths in the dilepton invariant mass and for angular observables that provide direct sensitivity to longitudinal transition form factors, which are absent in the limit of real-photon emission.

We summarize our main results for the total decay rates as follows:
\begin{align}\label{eq:results-widths}
\Gamma\left(h_c \to \eta_c e^+e^-\right) &= 5.45(19)~\mathrm{keV},\qquad
\Gamma\left(h_c \to \eta_c \mu^+\mu^-\right) = 0.635(22)~\mathrm{keV}, \nonumber \\[8pt]
\Gamma\left(\chi_{c1} \to J/\psi e^+e^-\right) &= 2.869(90)~\mathrm{keV},\quad
\Gamma\left(\chi_{c1} \to J/\psi \mu^+\mu^-\right) = 0.1993(72)~\mathrm{keV},
\end{align}
which can be compared with the results quoted by the PDG~\cite{PDG2024} for $\chi_{c1}\to J/\psi \ell^{+}\ell^{-}$:
\begin{align}
\Gamma^{\rm PDG}\left(\chi_{c1} \to J/\psi e^+e^-\right) &= 2.91(24)~{\rm keV}, \nonumber \\[8pt]
\Gamma^{\rm PDG}\left(\chi_{c1} \to J/\psi \mu^+\mu^-\right) &= 0.196(26)~{\rm keV},
\end{align}
where we have used $\Gamma^{\rm PDG}(\chi_{c1})=0.84(4)~{\rm MeV}$ to convert the experimental branching fractions into decay widths. With this choice of the full width $\Gamma(\chi_{c1})$, we observe excellent agreement with the experimental determinations for both the electron and muon channels. For a more detailed phenomenological discussion, see Section~\ref{sec:comparison}.
In the case of $h_{c}\to \eta_{c}\ell^{+}\ell^{-}$, only the electron channel has been measured, and the PDG quotes
\begin{align}
\Gamma^{\rm PDG}(h_{c}\to \eta_{c}e^{+}e^{-}) = 2.7(1.1)~{\rm keV},
\end{align}
obtained using ${\rm Br}(h_{c}\to \eta_{c}e^{+}e^{-}) = (3.5\pm 0.7)\times 10^{-3}$ and $\Gamma^{\rm PDG}(h_{c}) = 0.78\pm 0.28~{\rm MeV}$. We thus find a difference of about $2.5\sigma$ with respect to our result~\eqref{eq:results-widths}.

The remainder of this paper is organized as follows.
In Section~\ref{sec:num_details} we introduce the form-factor decompositions and
the lattice observables used to extract them.
For clarity, the discussion is organized channel by channel:
we first present the formalism for
$\chi_{c1}\to J/\psi \ell^+\ell^-$ and then for
$h_c\to\eta_c \ell^+\ell^-$.
The corresponding numerical analyses are presented in the same order
in Section~\ref{sec:num_results}. In Section~\ref{sec:comparison} we discuss the resulting decay rates, differential distributions, and angular observables, comparing them with available experimental results where possible. Finally, in Section~\ref{sec:conclusions}, we summarize our findings and outline prospects for future improvements and extensions to other decay channels.

\section{Form Factors and Lattice Setup}
\label{sec:num_details}

We adopt the following decomposition of the transition matrix elements describing the decay $\chi_{c1} \to J/\psi  \gamma^\ast $ and $h_{c}\to \eta_{c}\gamma^{*}$:~\footnote{We take this opportunity to correct a transcription typo in Eq.~(3) of Ref.~\cite{Becirevic:2025idm}. Specifically, the factor $m_{\chi_{c1}}^{2}$ multiplying the $(E_{1}+M_{2})$ and $(E_{1}-M_{2})$ terms is missing in that equation. In addition, compared to Eq.~(3) of Ref.~\cite{Becirevic:2025idm}, we redefine $C_{1}(q^{2})/\sqrt{q^{2}} \to C_{1}(q^{2})$. 
With this definition, $C_{1}(q^{2})$ is non-zero in the limit $q^{2}\to 0$. We also remind the reader that the original decomposition of the matrix
element was first presented in Ref.~\cite{Dudek:2006ej}, where a
derivation of this covariant decomposition is provided in the Appendix. }
\begin{widetext}
\begin{align}
\label{eq:def_trans_ff}
\langle J/\psi(k,\varepsilon) | J^{\mu}_{\rm em} | \chi_{c1}(p,\eta ) \rangle &= \frac{iQ_{c}}{\sqrt{2}\lambda(m_{\chi_{c1}}^{2}, m_{J/\psi}^{2}, q^{2})} \epsilon^{\mu\nu\rho\sigma} (p
  - k)_\sigma \times  \Bigg\{ 2m_{\chi_{c1}}^{2}\left[ E_1(q^2) + M_{2}(q^{2})\right]  
  (\eta \cdot k) \, \varepsilon^{\ast}_{\nu} (p+k)_\rho + \nonumber \\
  &+ 2m_{\chi_{c1}}^{2} \frac{m_{J/\psi}}{m_{\chi_{c1}}}\left[ E_{1}(q^{2})-M_{2}(q^{2})\right]  (\varepsilon^{\ast} \cdot p)
  \eta_{\nu}\, (p+k)_\rho + C_{1}(q^2) \Big[ - \lambda(m_{\chi_{c1}}^{2}, m_{J/\psi}^{2}, q^{2}) \, \eta_{\nu} \varepsilon^{\ast}_{\rho}   \nonumber \\
&+  \big[ (m_{\chi_{c1}}^2-m_{J/\psi}^2 + q^2) ( \eta \cdot k)\; \varepsilon^{\ast}_{\nu} + (m_{\chi_{c1}}^2 - m_{J/\psi}^2 -q^2) (\varepsilon^{\ast} \cdot p) \; \eta_{\nu} \big]          \, (p+k)_\rho\Big]\Bigg\}\,, \\[8pt]
\label{eq:def_trans_ff_hc}
\langle \eta_{c}(k) | J^{\mu}_{\rm em} | h_{c}(p,\varepsilon) \rangle &=  2iQ_{c}m_{h_{c}}\left\{ F_{1}(q^{2})\left(\varepsilon^{\mu} - \frac{\varepsilon\cdot q}{q^{2}}q^{\mu}\right) 
+ \tilde{F}_{2}(q^{2})(\varepsilon\cdot q)\left[ 
\frac{q^{\mu}}{q^{2}} -\frac{(p+k)^{\mu}}{m_{h_{c}}^{2} - m_{\eta_{c}}^{2}} \right]    \right\}\,,
\end{align}
\end{widetext}
where $J^{\mu}_{\rm em}$ is the electromagnetic current,
\begin{equation}
\label{eq:em_current}
    J^{\mu}_{\text{em}}(x) = \!\!\sum_{f=u,d,s,c} \!\! J_{f}^{\mu}(x) = \!\!\sum_{f=u,d,s,c} \!\! Q_f \, \bar{q}_f(x) \gamma^{\mu} q_f(x)\, ,
\end{equation}
with the sum running over all quark flavors, $q_f(x) $ being the quark field, and $ Q_f $ its electric charge in units of $ e = \sqrt{4\pi \alpha_\mathrm{em}}$. In Eq.~\eqref{eq:def_trans_ff}, $p$ and $k$ are the momenta of $\chi_{c1} $ and $J/ \psi$ respectively, while $\eta$ and $\varepsilon$ are their respective polarization vectors. $E_1(q^2)$, $M_2(q^2)$ and $C_1(q^2)$ are the form factors parameterizing the $\chi_{c1}\to J/\psi \gamma^{*}$ matrix element, which are functions of $q^2=(p-k)^2$, while $\lambda(x,y,z)=x^2+y^2+z^2-2xy-2yz-2xz$
is the K\"all\'en function. In Eq.~\eqref{eq:def_trans_ff_hc}, instead, $p$ and $k$ are the momenta of the $h_{c}$ and $\eta_{c}$, respectively, $q=p-k$, while $\varepsilon$ is the polarization of the $h_{c}$ meson. The non-perturbative form factors relevant for the $h_{c}\to \eta_{c}\ell^{+}\ell^{-}$ decays are $F_{1}(q^{2})$ and $\tilde{F}_{2}(q^{2})$, which obey the $q^{2}=0$ constraint $F_{1}(0)=\tilde{F}_{2}(0)$. For later convenience, we also introduce the longitudinal form factor
\begin{align}
\label{eq:FL}
F_L(q^2) =
\frac{1}{2q^2}
\left[
\frac{\lambda(m_{h_c}^2,m_{\eta_c}^2,q^2)}
     {m_{h_c}^2-m_{\eta_c}^2}
\,\tilde F_2(q^2)
-
(m_{h_c}^2-m_{\eta_c}^2+q^2)
\,F_1(q^2)
\right]~.
\end{align}
In terms of the form factors $F_1(q^2)$ and $F_L(q^2)$,
the decay rate for $h_{c}\to \eta_{c}\ell^{+}\ell^{-}$ is simply the sum of two
contributions proportional to $|F_1(q^2)|^2$ and
$|F_L(q^2)|^2$, without interference terms (see Section~\ref{sec:comparison}).

In the case of real photon emission considered in Refs.~\cite{Becirevic:2025ocx,Becirevic:2025idm}, only the form factors $E_{1}(0), M_{2}(0), F_{1}(0)$ contribute. In this paper we extend these calculations, and evaluate $E_{1}(q^{2})$ and $M_{2}(q^{2})$ and $F_{1}(q^{2})$ over the full kinematic range allowed, and determine the additional form factors that contribute to virtual photon emission only, namely $C_{1}(q^{2})$ (in the case of the $\chi_{c1}$ decay) and $\tilde{F}_{2}(q^{2})$ (in the case of the $h_{c}$ decay). 
 
\begin{table}[t]
\begin{ruledtabular}
\begin{tabular}{lcccc}
\textrm{ID} & $L/a$ & $a$ \textrm{fm} & $Z_{V}$ & $am_{c}$  \\
\colrule
\textrm{A48} & $48$ & 0.0907(5) &  0.68700(15) & $0.2620$ \\
\textrm{B64} & $64$ & 0.07948(11) &  0.706354(54) & $0.23157$  \\
\textrm{C80} & $80$ & 0.06819(14)  & 0.725440(33) & $0.19840$   \\
\textrm{D96} & $96$ & 0.056850(90) & 0.744132(31)   &  $0.16490$   
\end{tabular}
\end{ruledtabular}
\caption{\small\sl$N_{f}=2+1+1$ ETMC gauge ensembles used in this work. $a$ is the lattice spacing, $L/a$ the spatial extent of the lattice, $am_{c}$ is the bare charm quark mass, and $Z_V$ the vector current renormalization constant~\cite{ExtendedTwistedMassCollaborationETMC:2024xdf}. With the exception of A48 (which corresponds to a pion mass $m_{\pi}\simeq 175~{\rm MeV}$), all the ensembles have been generated at physical values of the light, strange and charm quark masses. At each lattice spacing, the charm quark mass $m_{c}$ has been tuned to reproduce $m_{D_{s}}= 1968$~MeV. \label{tab:simudetails}}
\end{table}
For this computation we use the gauge field configurations produced by the Extended Twisted Mass Collaboration (ETMC) with $N_{f}=2+1+1$ dynamical Wilson-Clover twisted mass fermions which guarantee the automatic $\mathcal{O}(a)$ improvement of physical on-shell
observables~\cite{Frezzotti:2003ni,Frezzotti:2004wz}. Basic information regarding the four lattice ensembles used in this work is collected in Table~\ref{tab:simudetails}, and further details can be found in  Ref.~\cite{ExtendedTwistedMassCollaborationETMC:2024xdf}. To extract the form factors relevant to the $\chi_{c1}\to J/\psi \ell^{+}\ell^{-}$ and $h_{c}\to\eta_{c}\ell^{+}\ell^{-}$ decays, we work in the rest frame of the decaying meson (either the $h_{c}$ or the $\chi_{c1}$), i.e. we set $\mathbf{p}=0$.

The core non-perturbative input needed to extract the $E_{1}(q^{2})$, $M_{2}(q^{2})$ and $C_{1}(q^{2})$ form factors parameterizing the $\chi_{c1}\to J/\psi \ell^{+}\ell^{-}$ decay is given by the following Euclidean three-point Euclidean function
\begin{align}
\label{eq:three-point}
 C_{\text{3}\chi}^{ijk}(t_{\chi}; t_{J},\bs{k}) = \sum_{\mathbf{x}, \mathbf{y},\mathbf{z}} e^{i\mathbf{k}(\mathbf{x}-\mathbf{y})} \langle 0 | \mathcal{O}_{J/\psi}^{i}(\mathbf{x}, 0) J^{j}_{\rm em}(\mathbf{y}, -t_{J}) \mathcal{O}_{\chi_{c1}}^{k \dagger}(\mathbf{z}, -t_{\chi}) | 0 \rangle~,
\end{align}
where $\mathcal{O}_{J/\psi}^{i}$ and $\mathcal{O}_{\chi_{c1}}^{k}$ are the interpolating operators of the $J/\psi$ and $\chi_{c1}$ respectively, which, following previous studies, we take to be the Gaussian-smeared ones defined by
\begin{equation}
\label{eq:interpolators}
\begin{split}
\mathcal{O}_{J/\psi}^{i}(\mathbf{x},t) &= \sum_{\mathbf{y}}\bar{q}_{c}(\mathbf{x},t) G_{t}^{n}(\mathbf{x},\mathbf{y}) \gamma^{i} q_{c}(\mathbf{y},t) \,,  \\ 
\mathcal{O}_{\chi_{c1}}^{k}(\mathbf{x},t) &= \sum_{\mathbf{y}}\bar{q}_{c}(\mathbf{x},t) G_{t}^{n}(\mathbf{x},\mathbf{y}) \gamma^{k}\gamma^{5} q_{c}(\mathbf{y},t)  \,,
\end{split}
\end{equation}
where $n$ is an integer,
\begin{align}
\label{eq:G_def}
G_{t}(\mathbf{x},\mathbf{y}) = \frac{1}{1+ 6\kappa}\bigl[ \delta_{\mathbf{x},\mathbf{y}} + \kappa H_{t}(\mathbf{x},\mathbf{y})    \bigr]\,,
\end{align}
$\kappa$ is a smearing parameter (see below for more details)
and $H_{t}(\mathbf{x},\mathbf{y})$ is the Gaussian smearing operator, 
\begin{align}
\label{eq:H}
H_{t}(\mathbf{x}, \mathbf{y}) = \sum_{\mu=1}^{3}\bigl[ U^{\star}_{\mu}(\mathbf{x},t)\delta_{\mathbf{x}+\hat{\mu},\mathbf{y}} + U^{\star\dagger}_{\mu}(\mathbf{x}-\hat{\mu},t)\delta_{\mathbf{x}-\hat{\mu},\mathbf{y}}    \bigr]\,,
\end{align}
with $U^{\star}_{\mu}(x)$ being the so-called APE-smeared links, cf.~\cite{Becirevic:2012dc}. The spatial momentum $\bs{k}$ is injected along the third spatial direction $\bs{k}=(0,0,k_{z})$, and we compute the three-point function for the following values of $k_{z}$ 
\begin{align}
\label{eq:moms_for_chi}
k_{z}= 0, \pm 2.4\times 10^{-2}, 0.175, 0.275,0.337,0.3894~{\rm GeV}.
\end{align}
This allows us to cover the full kinematic range: from $q^{2}=0$ corresponding to
\begin{align}
\label{eq:momentum_JPsi}
|\mathbf{k}| = \frac{ m_{\chi_{c1}}^{2} -m_{J/\psi}^{2} }{2m_{\chi_{c1}}} \simeq 389.4~\mathrm{MeV}\,,
\end{align}
up to $q^{2}_{\rm max} = (m_{\chi_{c1}}-m_{J/\psi})^{2} \sim (0.41~{\rm GeV})^{2}$ corresponding to $|\bs{k}|=0$. Note however that, as in the case of semileptonic weak transition between pseudoscalar states, e.g. $D\to \pi\ell\nu$, the three point function at $|\bs{k}|=0$ is not sufficient by itself to extract all the form factors exactly at $q^{2}_{\rm max}$.~\footnote{As it will be clear in the next subsection only the form factor $E_{1}$ can be evaluated directly at $q^{2}_{\rm max}$ from the only knowledge of the three point function at $|\bs{k}|=0$.} For this reason we have additionally evaluated the three point function $C_{3\chi}^{ijk}$ at two very tiny values of $k_{z} \sim \pm 10^{-2}~{\rm GeV}$, which will allow us to effectively predict all three form factors at $q^{2}\sim q^{2}_{\rm max}$. This will be explained in the following subsection. 

In the case of the $h_{c}\to \eta_{c}\ell^{+}\ell^{-}$ transition, one needs to compute
\begin{align}
\label{eq:three-point_h}
 C_{\text{3}h}^{ij}(t_{h}; t_{J},\bs{k}) = \sum_{\mathbf{x}, \mathbf{y},\mathbf{z}} e^{i\mathbf{k}(\mathbf{x}-\mathbf{y})} \langle 0 | \mathcal{O}_{\eta_c}(\mathbf{x}, 0) J^{j}_{\rm em}(\mathbf{y}, -t_{J}) \mathcal{O}_{h_c}^{i\dagger}(\mathbf{z}, -t_{h}) | 0 \rangle~,
\end{align}
where the interpolators of the $\eta_{c}$ and $h_{c}$ mesons are defined as
\begin{equation}
\label{eq:interpolators_hc}
\begin{split}
\mathcal{O}_{\eta_{c}}(\mathbf{x},t) &= \sum_{\mathbf{y}}\bar{q}_{c}(\mathbf{x},t) G_{t}^{n}(\mathbf{x},\mathbf{y}) \gamma^{5} q_{c}(\mathbf{y},t) \,,  \\ 
\mathcal{O}_{h_{c}}^{i}(\mathbf{x},t) &= \sum_{\mathbf{y}}\bar{q}_{c}(\mathbf{x},t) G_{t}^{n}(\mathbf{x},\mathbf{y}) \varepsilon^{ikl}\sigma_{kl} q_{c}(\mathbf{y},t)  \,.
\end{split}
\end{equation}
Again, we inject the spatial momentum along the third direction, $\bs{k}=(0,0,k_{z})$. In the case of the $h_{c}\to \eta_{c}$ transition the spatial momenta we considered are
\begin{align}
\label{eq:moms_for_h}
k_{z} = 0, \pm 1.2 \times 10^{-6}, 0.125, 0.250, 0.375, 0.469, 0.5~{\rm GeV}~, 
\end{align}
allowing us to cover the full phase space, from $q^{2}=0$ corresponding to
\begin{align}
\label{eq:momentum_etac}
|\mathbf{k}| = \frac{ m_{h_{c}}^{2} -m_{\eta_{c}}^{2} }{2m_{h_{c}}} \simeq 0.50~{\rm GeV}~, 
\end{align}
up to $q^{2}_{\rm max}= (m_{h_{c}}-m_{\eta_{c}})^{2} \simeq (0.54~{\rm GeV})^{2}$, corresponding to $\bs{k}=0$. As in the case of the $\chi_{c1}$ decay, also here only the form factor $F_{1}(q^{2}_{\rm max})$ can be extracted directly from simulations at $\bs{k}=0$. 

In analogy with what we did for the $\chi_{c1}\to J/\psi$ decay, we have computed the three point functions for two very tiny values of $k_{z} \simeq \pm 10^{-6}~{\rm GeV}$, which then allow us to obtain $\tilde{F}_{2}(q^{2}_{\rm max})$, as it will be discussed in the next subsection. For all the meson interpolators discussed above, in analogy with our previous studies~\cite{Becirevic:2025ocx,Becirevic:2025idm},
we used the smearing parameter $\kappa=0.4$, and fixed on each ensemble the number of steps $n$ in Eq.~\eqref{eq:interpolators} to have a smearing radius $r_{0}= a\sqrt{n}/\sqrt{\kappa^{-1}+6} \simeq 0.15~{\rm fm}$.

\begin{figure*}
\centering
\includegraphics[scale=0.40]{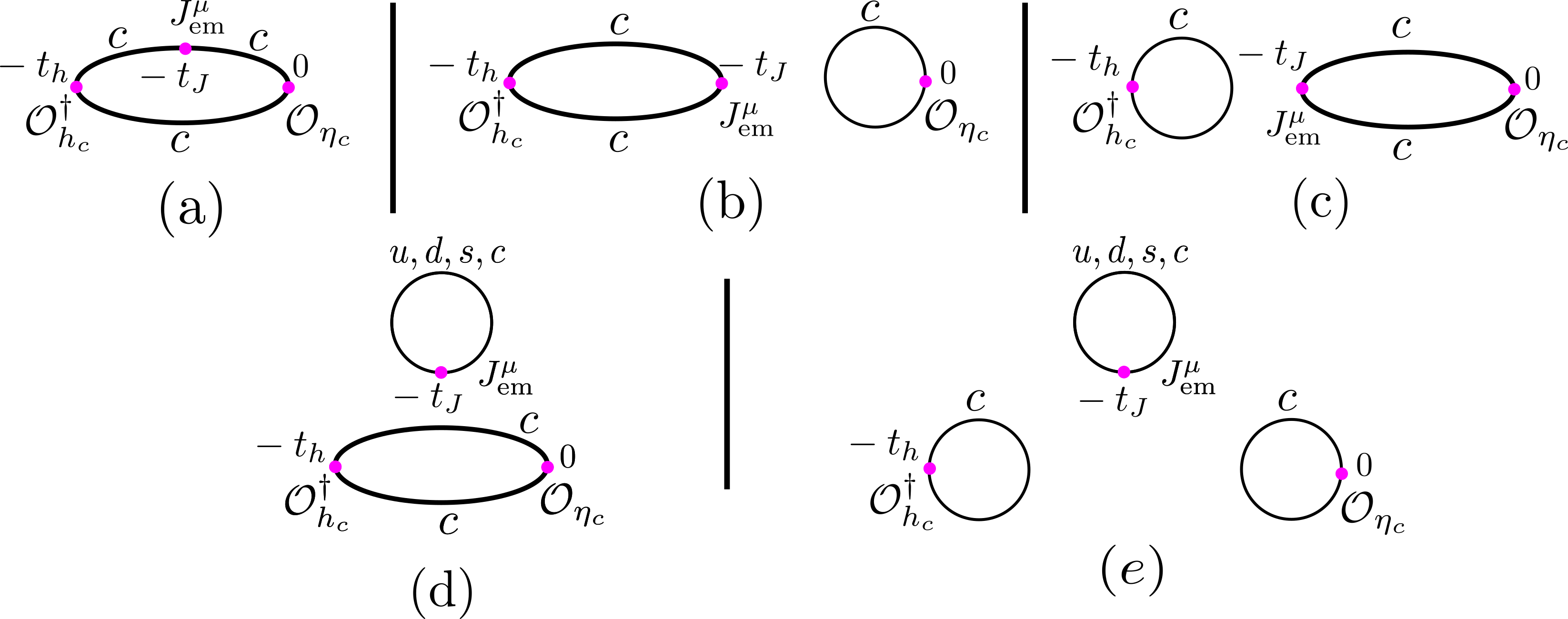}
\caption{\small\sl Wick contractions relevant to the correlation function~(\ref{eq:three-point_h}). The diagram (a) corresponds to the connected contribution, while (b), (c), (d) and (e) correspond to the disconnected ones. Diagrams (b), (c) and (e) are Zweig suppressed. For each diagram, we have indicated the flavor associated to each quark line. Light quark ($u$, $d$, $s$) contributions only appear in diagram (d) and diagram (e)  where the photon is emitted by a sea quark. These two diagrams are suppressed in the $\rm{SU}(3)$-limit $m_{u}=m_{d}=m_{s}$, while the effect of the $c$ sea quark should be negligible. \label{fig:conn-disc}}
\end{figure*}

The Wick contractions of the three point functions $C_{3\chi}^{ijk}$ and $C_{3h}^{ij}$  in Eqs.~\eqref{eq:three-point} and~\eqref{eq:three-point_h} give rise to quark-connected and quark-disconnected contributions, cf. Fig.~\ref{fig:conn-disc} where we sketch Feynman diagrams corresponding to the Wick contractions of $C_{3h}^{ij}$ (the case of $C_{3\chi}^{ijk}$ is identical). In the following, we will refer to these contributions as \textit{connected} and \textit{disconnected}, respectively.
The leftmost diagram (a) corresponds to the dominant (connected) contribution, while (b), (c), (d) and (e) correspond to the disconnected ones. The contributions corresponding to (b), (c) and (e) are expected to be very small due to the Zweig suppression. Furthermore, diagrams (d) and (e) respect the $\mathrm{SU}(3)$ suppression in which  the contributions of $u$, $d$ and $s$ cancel out in the $m_{u}=m_{d}=m_{s}$ limit. In this work we focus on the evaluation of the dominant (connected) diagram, leaving the evaluation of the disconnected contributions for future works.~\footnote{A lattice QCD study of Ref.~\cite{Hatton:2020qhk} shows that the contributions arising from disconnected diagrams indeed give a tiny contribution to the masses of charmonia.} Clearly, only the charm quark component $J_{c}^{\mu}$ of the electromagnetic current $J^{\mu}_{\rm em}$ contributes to the connected part of the correlation functions $C_{3\chi}^{ijk}$ and $C_{3h}^{ij}$.

If only the quark-connected contribution is considered, it is possible to use partially-quenched twisted boundary conditions to tune the spatial momentum $\bs{k}$ to the values given in Eqs.~\eqref{eq:moms_for_chi} and~\eqref{eq:moms_for_h}, which are in
general not integer multiples of $2\pi/L$. This is implemented by twisting the gauge links $U_{\mu}(x)$ on which one of the charm quark propagators is computed~\cite{deDivitiis:2004kq,Bedaque:2004kc,Sachrajda:2004mi}. More explicitly,\footnote{In Refs.~\cite{Becirevic:2025idm,Becirevic:2025ocx}, the phase in the corresponding formulae for the twisted gauge link should include an additional factor of $\pi$.}
\begin{align}
\label{eq:twisted_gauge}
U_\mu(x) \rightarrow U^\theta_\mu (x) = e^{i\pi a\theta_{\mu} / L} U_\mu(x), \quad  \theta_\mu = (0, \vec{\theta})\,,
\end{align}
with the twist angle $\vec{\theta}$ set to
\begin{align}
\label{eq:theta}
\vec{\theta}=( 0, 0 , \theta_{z}^{c}) , \qquad \theta_{z}^{c} = \frac{L}{\pi}k_{z}~, 
\end{align}
where $L$ is the spatial extent of the lattice.  Note that the physical timelike kinematic region relevant for the
Dalitz decays cannot be accessed in a sufficiently dense way using only
the standard Fourier lattice momenta. For this reason, we employ
partially twisted boundary conditions, which allow us to continuously
tune the hadron momenta and thereby map the full timelike $q^2$
dependence of the form factors. This is different from what has been done in previous works on the subject.

We are now in a position to discuss our specific regularization of quark bilinears, as adopted for the calculation of the quark-connected contribution to both $C_{3\chi}^{ijk}$ and $C_{3h}^{ij}$. As discussed in Refs.~\cite{Becirevic:2025idm,Becirevic:2025ocx}, when evaluating the connected diagrams using twisted-mass fermions, there are multiple options for discretizing the quark bilinears, differing in the choice of the Wilson parameter of the quark field. Specifically, the quark-connected contributions to the three-point functions can be derived from a mixed-action setup, in which—alongside the simulated twisted-mass $u$, $d$, $s$, and $c$ sea quarks—the following valence-quark action is introduced:
\begin{align}
\label{eq:tm_action}
S_{\rm val} &= \sum_{f=c,c',c''}\sum_{x} \bar{q}_{f}(x) \biggl[ \gamma_{\mu}\bar{\nabla}_{\mu}[U] - i r_{f} \gamma^{5}(W^{\rm cl}[U] + m_{\rm cr}) + m_{c} \biggr] q_{f}(x),
\end{align}
where $W^{\rm cl}[U]$ is the Wilson-Clover term~\cite{Sheikholeslami:1985ij}, $m_{\rm cr}$ is the critical mass, $m_{c}$ is the charm-quark mass, and $r_{f = c,c',c''} = \pm 1$ is the sign of the twisted-Wilson parameter of the valence quark field $q_{f}$. The primed fields are thus replica charm-quark fields differing only in the choice of the twisted-Wilson parameter, while all other bare parameters are kept identical.

In this mixed action lattice setup the quark-connected contribution to the three-point functions in Eqs.~\eqref{eq:three-point} and~\eqref{eq:three-point_h} can be obtained by representing the quark bilinear meson interpolating operators as follows:
\begin{align}
\bar{q}_{c} \Gamma_{\chi_{c1}} q_{c} &\to \bar{q}_{c'}\Gamma_{\chi_{c1}} q_{c}, \qquad \bar{q}_{c} \Gamma_{J/\psi} q_{c} \to \bar{q}_{c'}\Gamma_{J/\psi} q_{c''}, \nonumber \\
\bar{q}_{c} \Gamma_{h_{c}} q_{c} &\to \bar{q}_{c'}\Gamma_{h_{c}} q_{c}, \qquad\,\, \bar{q}_{c} \Gamma_{\eta_{c}} q_{c} \to \bar{q}_{c'}\Gamma_{\eta_{c}} q_{c''},
\end{align}
where $\bar{q}_{c}\Gamma_{P} q_{c}$ denotes the interpolating operator of the meson $P$. The charm-quark component of the electromagnetic current, $J^{\mu}_{c}$, is discretized as $J_{c}^{\mu} = Q_{c} \bar{q}_{c}\gamma^{\mu}q_{c''}$. The full Euclidean correlation function evaluated with these operators corresponds to the quark-connected contribution, up to an overall factor of two, accounting for the missing charge-conjugated diagram in which the electromagnetic current is inserted in the antiquark line.

With twisted-mass fermions, the parameters $r_{c}$, $r_{c'}$, and $r_{c''}$ can be freely chosen as $\pm 1$. We restrict ourselves to $r_{c} = r_{c''} = 1$. This still leaves open the choice between $r_{c} = r_{c'}$ and $r_{c} = -r_{c'}$. The first choice corresponds to the Osterwalder-Seiler (OS) regularization of the meson interpolating fields, while the second corresponds to the twisted-mass (TM) regularization of the same fields. In Refs.~\cite{Becirevic:2025ocx,Becirevic:2025idm}, we demonstrated that for the $\chi_{c1}$ decay (resp. $h_{c}$ decay), the OS (resp. TM) regularization is preferable, because, due to $H(3)$, charge conjugation and (only for the TM case) parity times $G$-parity symmetries of the valence quark action, it avoids unwanted mixing of the interpolated meson states with lighter states of different $J^{PC}$ quantum numbers. For the present calculation, we adopt the same choices as in Refs.~\cite{Becirevic:2025ocx,Becirevic:2025idm}, using OS regularization for the $\chi_{c1} \to J/\psi \ell^+ \ell^-$ decay and TM regularization for the $h_{c} \to \eta_{c} \ell^+ \ell^-$ decay. Since the electromagnetic current is discretized in both cases as $\bar{q}_{c}\gamma^{i}q_{c''}$ with $r_{c} = r_{c''} = 1$, the current renormalizes with the vector renormalization constant $Z_{V}(\beta)$, whose values are collected in Table~\ref{tab:simudetails}.

In our mixed-action setup, after denoting by $S_{c}^{\pm}(x,y)$ and $S_{c}^{\pm,\theta}(x,y)$ the charm-quark propagators evaluated with $r = \pm 1$ on the background field configurations corresponding to $U_{\mu}(x)$ and $U_{\mu}^{\theta}(x)$, respectively, the quark-connected contribution to $C_{\rm 3\chi}^{ijk}(t_{\chi}; t_{J},\bs{k})$ and $C_{\rm 3h}^{ij}(t_{h}; t_{J},\bs{k})$ is evaluated as~\footnote{For simplicity, we consider local interpolating operators, i.e., $\kappa = 0$. In practice, we implement Gaussian smearing as discussed in the text.}
\begin{align}
\label{eq:3pt_explicit}
C_{\rm 3\chi}^{ijk}(t_{\chi}; t_{J},\bs{k}) &= 2Q_{c} \sum_{\mathbf{x},\mathbf{y},\mathbf{w}} \langle \tr \left[ S_{c}^{+}(x, w) \gamma^{k}\gamma^{5} S_{c}^{+}(w,y) \gamma^{j} S_{c}^{+\theta}(y, x) \gamma^{i} \right] \rangle, \nonumber \\
C_{\rm 3h}^{ij}(t_{h}; t_{J},\bs{k}) &= 2 Q_{c}\sum_{\mathbf{x},\mathbf{y},\mathbf{z}} \langle \tr \left[ S_{c}^{-}(x, z) \varepsilon^{ikl}\sigma_{kl} S_{c}^{+}(z,y) \gamma^{j} S_{c}^{+\theta}(y, x) \gamma^{5} \right] \rangle,
\end{align}
where $x = (\mathbf{x},0)$, $y = (\mathbf{y}, -t_{J})$, $w = (\mathbf{w}, -t_{\chi})$, and $z = (\mathbf{z}, -t_{h})$. The trace $\tr\left[ \dots \right]$ is taken over color and Dirac indices, and $\langle \dots \rangle$ denotes the average over the $\rm{SU}(3)$ gauge field configurations $U_{\mu}(x)$. The factor of two in Eq.~\eqref{eq:3pt_explicit} accounts for the contribution of the charge-conjugated diagram. The sum over $\mathbf{x}$ is evaluated stochastically by inverting the Dirac charm-quark operator on $N_{s}$ spatial stochastic sources placed at time $t = 0$. The number of gauge configurations and stochastic sources used for the evaluation of the three-point functions differs slightly between the $h_{c}$ and $\chi_{c1}$ cases and will be detailed in the corresponding numerical sections. We now proceed to the extraction of the relevant form factors from the three-point functions, beginning with the $\chi_{c1}$ case.

\subsection{Extraction of the $\chi_{c1} \to J/\psi \ell^+ \ell^-$ form factors $E_{1}$, $M_{2}$, and $C_{1}$}
\label{sec:extr_form_factor_charm_chi}

We first consider the two-point functions of the interpolating fields $\mathcal{O}_{J/\psi}^{i}$ and $\mathcal{O}_{\chi_{c1}}^{k}$ to extract their masses/energies and couplings:
\begin{equation}
\label{eq:2pt_def_chi}
\begin{split}
C_{2,J/\psi}^{i}(t,\bs{k}) &= \sum_{\mathbf{x}} e^{i\mathbf{k}\mathbf{x}} \langle 0 | \mathcal{O}_{J/\psi}^{i}(\mathbf{x}, t) \mathcal{O}_{J/\psi}^{i \dagger}(\mathbf{0}, 0) | 0 \rangle, \\
C_{2,\chi_{c1}}^{k}(t) &= \sum_{\mathbf{x}} \langle 0 | \mathcal{O}_{\chi_{c1}}^{k}(\mathbf{x}, t) \mathcal{O}_{\chi_{c1}}^{k \dagger}(\mathbf{0}, 0) | 0 \rangle,
\end{split}
\end{equation}
neglecting the Zweig-suppressed disconnected contributions. In the limit of large Euclidean time $t$, the two-point correlation functions behave as:
\begin{equation}
\label{eq:2pt_asympt_chi}
\begin{split}
C_{2,J/\psi}^{i}(t,\bs{k}) &= \frac{|Z_{J/\psi}^{i}(\bs{k})|^2}{2 E_{J/\psi}(\bs{k})} \left( e^{-E_{J/\psi}(\bs{k}) t} + e^{-E_{J/\psi}(\bs{k}) (T-t)} \right) + \ldots, \\
C_{2,\chi_{c1}}^{k}(t) &= \frac{|Z_{\chi_{c1}}^{k}|^2}{2 m_{\chi_{c1}}} \left( e^{-m_{\chi_{c1}} t} + e^{-m_{\chi_{c1}} (T-t)} \right) + \ldots,
\end{split}
\end{equation}
with $m_{\chi_{c1}}$ and $E_{J/\psi}(\bs{k})$ being the mass and energy of the corresponding ground states, and
\begin{equation}
\begin{split}
Z_{J/\psi}^{i}(\bs{k}) = \langle 0 | \mathcal{O}_{J/\psi}^{i} | J/\psi(\bs{k},\varepsilon_{i}) \rangle, \qquad
Z_{\chi_{c1}}^{k} = \langle \chi_{c1}(\bs{0},\eta_{k}) | \mathcal{O}_{\chi_{c1}}^{k \dagger} | 0 \rangle,
\end{split}
\end{equation}
where the dependence of $Z_{J/\psi}^{i}$ on the momentum $\bs{k}$ arises from the smearing. For $\chi_{c1}$, the interpolator $\mathcal{O}_{\chi_{c1}}^{k \dagger}$ ($k = 1,2,3$) creates a $\chi_{c1}$ state with polarization $\eta_{k}^{\mu} \propto \delta^{\mu}_{k}$. For $J/\psi$, the interpolator $\mathcal{O}_{J/\psi}^{i}$ ($i = 1,2$) annihilates a $J/\psi$ state with polarization $\varepsilon_{i}^{\mu} \propto \delta^{\mu}_{i}$, while for $i = 3$ it annihilates a longitudinally polarized $J/\psi$ with $\varepsilon_{3}^{\mu} = (\alpha, 0, 0, \sqrt{1 + \alpha^{2}})$ and $\alpha = |\bs{k}|/m_{J/\psi}$. The ellipses in Eq.~\eqref{eq:2pt_asympt_chi} denote contributions that vanish in the large-time limit, $0 \ll t \ll T$, and are further suppressed by the smearing (cf. Eq.~\eqref{eq:G_def}). With our choice of kinematics, we have $Z_{\chi_{c1}}^{1} = Z_{\chi_{c1}}^{2} = Z_{\chi_{c1}}^{3}$ and $Z_{J/\psi}^{1}(\bs{k}) = Z_{J/\psi}^{2}(\bs{k})$.

Using Eq.~\eqref{eq:def_trans_ff}, we combine the above quantities to construct the following estimators:
\begin{align}
\label{eq:FF_estimators}
\overline{E}_{1}(t_{\chi}; t_{J},\bs{k}) &= \frac{i Z_{V}(\beta)}{\sqrt{2} Q_{c}} 4 E_{J/\psi}(\bs{k}) e^{E_{J/\psi}(\bs{k}) t_{J}} e^{m_{\chi_{c1}}(t_{\chi} - t_{J})} \times \left[ \frac{C_{\rm 3\chi}^{213}(t_{\chi}; t_{J},\bs{k})}{Z_{J/\psi}^{2}(\bs{k}) Z_{\chi_{c1}}^{3}} - \frac{C_{\rm 3\chi}^{312}(t_{\chi}; t_{J},\bs{k})}{Z_{J/\psi}^{3}(\bs{k}) Z_{\chi_{c1}}^{2}} \right], \nonumber \\
\overline{M}_{2}(t_{\chi}; t_{J},\bs{k}) &= \frac{i Z_{V}(\beta)}{\sqrt{2} Q_{c}} 4 E_{J/\psi}(\bs{k}) e^{E_{J/\psi}(\bs{k}) t_{J}} e^{m_{\chi_{c1}}(t_{\chi} - t_{J})} \times \left[ \frac{C_{\rm 3\chi}^{213}(t_{\chi}; t_{J},\bs{k})}{Z_{J/\psi}^{2}(\bs{k}) Z_{\chi_{c1}}^{3}} + \frac{C_{\rm 3\chi}^{312}(t_{\chi}; t_{J},\bs{k})}{Z_{J/\psi}^{3}(\bs{k}) Z_{\chi_{c1}}^{2}} \right], \nonumber \\
\overline{C}_{1}(t_{\chi}; t_{J},\bs{k}) &= \frac{i Z_{V}(\beta)}{\sqrt{2} Q_{c}} 4 E_{J/\psi}(\bs{k}) \frac{m_{\chi_{c1}}}{|\bs{k}|} e^{E_{J/\psi}(\bs{k}) t_{J}} e^{m_{\chi_{c1}}(t_{\chi} - t_{J})} \times \left[ \frac{C_{\rm 3\chi}^{102}(t_{\chi}; t_{J},\bs{k})}{Z_{J/\psi}^{1}(\bs{k}) Z_{\chi_{c1}}^{2}} - \frac{C_{\rm 3\chi}^{201}(t_{\chi}; t_{J},\bs{k})}{Z_{J/\psi}^{2}(\bs{k}) Z_{\chi_{c1}}^{1}} \right],
\end{align}
which, for large Euclidean time separations, yield the desired form factors $E_{1}$, $M_{2}$, and $C_{1}$:
\begin{equation}
\begin{split}
\overline{E}_{1}(t_{\chi}; t_{J},\bs{k}) \xrightarrow{\begin{array}{c} t_{J} \to \infty \\ t_{\chi} - t_{J} \to \infty \end{array}} E_{1}(q^{2}), \quad
\overline{M}_{2}(t_{\chi}; t_{J},\bs{k}) \xrightarrow{\begin{array}{c} t_{J} \to \infty \\ t_{\chi} - t_{J} \to \infty \end{array}} M_{2}(q^{2}), \quad
\overline{C}_{1}(t_{\chi}; t_{J},\bs{k}) \xrightarrow{\begin{array}{c} t_{J} \to \infty \\ t_{\chi} - t_{J} \to \infty \end{array}} C_{1}(q^{2}).
\end{split}
\end{equation}
The relation between $|\bs{k}|$ and $q^{2}$ is given by
\begin{align}
q^{2} = m_{\chi_{c1}}^{2} + m_{J/\psi}^{2} - 2 m_{\chi_{c1}} \sqrt{m_{J/\psi}^{2} + |\bs{k}|^{2}}.
\end{align}

In $C_{\rm 3\chi}^{ijk}(t_{\chi}; t_{J},\bs{k})$, the interpolating field of $J/\psi$ is placed at $t = 0$, the charm-quark component of the electromagnetic current is inserted at $-t_{J} < 0$, and the time $-t_{\chi} \ll -t_{J}$, at which the $\chi_{c1}$ is created, corresponds to the \textit{sink} of the correlation function. This setup allows us to monitor the dominance of the $\chi_{c1}$ state. Since $t_{J}$ is fixed in our computational setup, it must be chosen large enough to isolate the $J/\psi$ state. Placing the $J/\psi$ interpolator at the source of the correlation function is advantageous, as $t_{J}$ can be chosen reasonably large: the signal-to-noise ratio (SNR) of $C_{\rm 3\chi}^{ijk}(t_{\chi}; t_{J})$ for fixed $|t_{\chi} - t_{J}|$ decreases only as $e^{-(E_{J/\psi} - m_{\eta_{c}}) t} \simeq e^{-\Delta_{c} t}$, where $\Delta_{c} \simeq 113~{\rm MeV}$ is the (experimental) hyperfine splitting. We considered two values of $t_{J} \simeq 1.6, 2.4~\mathrm{fm}$ and observed only minor differences in our results, which we include in the final systematic error.

We introduce two improvements to the estimators for $M_{2}(q^{2})$ and $C_{1}(q^{2})$ that significantly reduce their statistical uncertainties:

\begin{enumerate}
\item {\bf Case of $C_{1}(q^{2})$:}
The three-point correlation functions $C_{3\chi}^{102}$ and $C_{3\chi}^{201}$ vanish in the limit $\bs{k} \to 0$ (see Eq.~\eqref{eq:three-point}). However, on a finite statistical sample, they vanish only within statistical uncertainties. For small but non-zero $\bs{k}$, the noise associated with the correlation function is expected to be similar to that at $\bs{k} = 0$. We therefore use
\begin{align}
C_{\rm 3\chi}^{201}(t_{\chi}; t_{J},\bs{k}) &\to C_{\rm 3\chi}^{201}(t_{\chi}; t_{J},\bs{k}) - C_{\rm 3\chi}^{201}(t_{\chi}; t_{J},\bs{0}), \nonumber \\
C_{\rm 3\chi}^{102}(t_{\chi}; t_{J},\bs{k}) &\to C_{\rm 3\chi}^{102}(t_{\chi}; t_{J},\bs{k}) - C_{\rm 3\chi}^{102}(t_{\chi}; t_{J},\bs{0}),
\end{align}
since part of the statistical noise cancels in the correlated differences. This subtraction is applied to all considered momenta $\bs{k}$ and is particularly effective at very small momenta.

At $\bs{k} = 0$, the form factor $C_{1}$ cannot be evaluated directly. However, by performing a correlated subtraction between the three-point function at very small momentum $|\bs{k}| \simeq 10^{-2}~{\rm GeV}$ and that at vanishing momentum, we obtain an estimate of $C_{1}(q^{2} \simeq q^{2}_{\rm max})$.

\item {\bf Case of $M_{2}(q^{2})$:}
For vanishing momentum $\bs{k} = 0$, the relation $C_{3\chi}^{213}(t_{\chi}; t_{J},\bs{0}) = -C_{3\chi}^{312}(t_{\chi}; t_{J},\bs{0})$ holds, implying $M_{2}(q^{2}_{\rm max}) = 0$. Since both $C_{3\chi}^{213}$ and $C_{3\chi}^{312}$ are even functions of the spatial momentum, it follows that for small $|\bs{k}|$, $M_{2}(q^{2} \sim q^{2}_{\rm max}) \propto |\bs{k}|^{2}$. We perform a zero-momentum subtraction to reduce the statistical uncertainty:
\begin{align}
\label{eq:zero_mom_M2}
\overline{M}_{2}(t_{\chi}; t_{J},\bs{k}) \to \overline{M}_{2}(t_{\chi}; t_{J},\bs{k}) - \overline{M}_{2}(t_{\chi}; t_{J},\bs{0}).
\end{align}

This procedure is applied to all momenta $k_z$ except for the smallest momentum $k_{z} = 2.4 \times 10^{-2}~{\rm GeV}$, for which a further improved estimator is constructed by exploiting the behavior $M_{2}(q^{2} \sim q^{2}_{\rm max}) \propto |\bs{k}|^{2}$. In this case, we use
\begin{align}
\label{eq:second_der}
\overline{M}_{2}(t_{\chi}; t_{J},\bs{k}) \to \frac{1}{2} \left[ \overline{M}_{2}(t_{\chi}; t_{J},\bs{k}) + \overline{M}_{2}(t_{\chi}; t_{J},-\bs{k}) \right].
\end{align}

This subtraction, involving $\pm \bs{k}$ together with $\bs{k} = 0$, is more effective than the zero-momentum subtraction of Eq.~\eqref{eq:zero_mom_M2}, as it cancels noise components that are either momentum-independent or linear in $\bs{k}$. At very small momentum, this improvement is crucial, as the signal scales as $|\bs{k}|^{2}$ and would otherwise be overwhelmed by statistical noise.

In the limit $\bs{k} \to 0$, the improved estimator divided by $|\bs{k}|^{2}$ directly provides an estimate of the second derivative of the form factor with respect to $|\bs{k}|$, allowing a clean extraction of the slope
\begin{align}
\alpha = \lim_{|\bs{k}| \to 0} \frac{M_{2}(q^{2}(\bs{k}))}{|\bs{k}|^{2}}~.
\end{align}
We emphasize that these improvements are effective only if the zero-momentum subtraction and the averaging in Eq.~\eqref{eq:second_der} are performed in a fully correlated manner, using identical gauge configurations and stochastic sources for all terms.
\end{enumerate}

Having described our strategy for determining the form factors $E_{1}(q^{2})$, $M_{2}(q^{2})$, and $C_{1}(q^{2})$, we now turn to the $h_{c} \to \eta_{c} \ell^+ \ell^-$ case.

\subsection{Extraction of the $h_{c}\to \eta_{c}\ell^{+}\ell^{-}$ form factors $F_{1}(q^{2})$ and $\tilde{F}_{2}(q^{2})$}
\label{sec:extr_form_factor_charm_hc}
We now discuss the extraction of the two-point functions for the interpolating fields $\mathcal{O}_{\eta_{c}}$ and $\mathcal{O}_{h_{c}}^{i}$, which are necessary to determine their masses/energies and couplings. We define these two-point functions as:
\begin{equation}
\label{eq:2pt_def_hc}
\begin{split}
C_{2, \eta_c}(t,\bs{k}) &= \sum_{\mathbf{x}} e^{i\mathbf{k}\mathbf{x}} \langle 0 | \mathcal{O}_{\eta_c}(\mathbf{x}, t) \mathcal{O}_{\eta_c}^\dagger(\mathbf{0}, 0) | 0 \rangle, \\
C_{2,h_c}^{i}(t) &= \sum_{\mathbf{x}} \langle 0 | \mathcal{O}_{h_c}^{i}(\mathbf{x}, t) \mathcal{O}_{h_c}^{i\dagger}(\mathbf{0}, 0) | 0 \rangle,
\end{split}
\end{equation}
neglecting the Zweig-suppressed disconnected contributions to $C_{2,\eta_{c}}(t,\bs{k})$ and $C_{2,h_c}^{i}(t)$.

In the limit of large Euclidean time $t$, the two-point correlation functions behave as:
\begin{equation}
\label{eq:2pt_asympt}
\begin{split}
C_{2,\eta_c}(t,\bs{k}) &= \frac{|Z_{\eta_c}(\bs{k})|^2}{2 E_{\eta_c}(\bs{k})} \left( e^{-E_{\eta_c}(\bs{k}) t} + e^{-E_{\eta_c}(\bs{k}) (T-t)} \right) + \ldots, \\
C_{2,h_c}^{i}(t) &= \frac{|Z_{h_c}^{i}|^2}{2 m_{h_c}} \left( e^{-m_{h_c} t} + e^{-m_{h_c} (T-t)} \right) + \ldots,
\end{split}
\end{equation}
where $m_{h_c}$ and $E_{\eta_c}$ are the mass and energy of the corresponding mesons, and
\begin{align}
Z_{\eta_c}(\bs{k}) = \langle 0 | \mathcal{O}_{\eta_c} | \eta_c(\bs{k}) \rangle, \qquad
Z_{h_c}^{i} = \langle h_c(\bs{0},\epsilon_i) | \mathcal{O}_{h_c}^{i\dagger} | 0 \rangle\,.
\end{align}
The dependence of $Z_{\eta_c}$ on $\bs{k}$ is due to the smearing of the interpolator. With our choice of kinematics $Z_{h_c}^{1}=Z_{h_c}^{2}=Z_{h_c}^{3}=Z_{h_{c}}$.

The ellipses in Eq.~\eqref{eq:2pt_asympt} denote contributions that vanish in the large time limit, $0 \ll t \ll T$. The interpolating field $\mathcal{O}_{h_c}^{i\dagger}$ in Eq.~\eqref{eq:interpolators_hc} creates an $h_c$ meson with polarization $\varepsilon_i^\mu \propto \delta_i^\mu$. Given our choice of kinematics, $\bs{p} = 0$ and $\bs{k} = (0,0,k_z)$, we use Eq.~\eqref{eq:def_trans_ff_hc} to define the following estimators:
\begin{align}
\label{eq:FF_estimator_hc}
\overline{F}_1(t_h; t_J,\bs{k}) &= \frac{i Z_V(\beta)}{2Q_c} \frac{4 E_{\eta_c}(\bs{k})}{Z_{\eta_c}(\bs{k}) Z_{h_c}} e^{E_{\eta_c}(\bs{k}) t_J} e^{m_{h_c}(t_h - t_J)} C_{\rm 3h}^{11}(t_h; t_J, \bs{k}), \nonumber \\
\overline{R}_1(t_h; t_J,\bs{k}) &= \frac{i Z_V(\beta)}{2Q_c} \frac{4 E_{\eta_c}(\bs{k})}{Z_{\eta_c}(\bs{k}) Z_{h_c}} e^{E_{\eta_c}(\bs{k}) t_J} e^{m_{h_c}(t_h - t_J)} C_{\rm 3h}^{33}(t_h; t_J, \bs{k}),
\end{align}
which, for large Euclidean time separations, allow us to extract the form factors $F_1(q^2)$ and $\tilde{F}_2(q^2)$:
\begin{align}
\label{eq:aux}
\overline{F}_1(t_h; t_J,\bs{k}) &\xrightarrow{\begin{array}{c} t_J \to \infty \\ t_h - t_J \to \infty \end{array}} F_1(q^2), \nonumber \\
\overline{R}_1(t_h; t_J,\bs{k}) &\xrightarrow{\begin{array}{c} t_J \to \infty \\ t_h - t_J \to \infty \end{array}} \left(1 + \frac{|\bs{k}|^2}{q^2}\right) F_1(q^2) -\left[ \frac{1}{(m_{h_c}^2 - m_{\eta_c}^2)} + \frac{1}{q^2} \right] |\bs{k}|^2 \tilde{F}_2(q^2).
\end{align}

For convenience, we define an estimator that converges directly to $\tilde{F}_2(q^2)$:
\begin{align}
\label{eq:F2_estimator_hc}
\overline{F}_2(t_h; t_J, \bs{k}) \equiv \frac{\overline{R}_1^{\rm sub}(t_h; t_J, \bs{k})}{\delta(q^2)} \equiv \frac{\overline{R}_1(t_h; t_J, \bs{k}) - \beta(q^2) \overline{F}_1(t_h; t_J, \bs{k})}{\delta(q^2)},
\end{align}
with
\begin{align}
\beta(q^2) = 1 + \frac{|\bs{k}|^2}{q^2}, \qquad \delta(q^2) = -|\bs{k}|^2 \left[ \frac{1}{(m_{h_c}^2 - m_{\eta_c}^2)} + \frac{1}{q^2} \right].
\end{align}

We remind the reader that, in our setup, the relation between $|\bs{k}|$ and $q^2$ is given by:
\begin{align}
q^2 = m_{h_c}^2 + m_{\eta_c}^2 - 2 m_{h_c} \sqrt{m_{\eta_c}^2 + |\bs{k}|^2}~.
\end{align}

In analogy with the $\chi_{c1} \to J/\psi \ell^+ \ell^-$ transition discussed in Section~\ref{sec:extr_form_factor_charm_chi}, we compute $C_{\rm 3h}^{ij}(t_h; t_J, \bs{k})$ by placing the interpolating field of $\eta_c$ at $t = 0$, inserting the charm-quark component of the electromagnetic current at a fixed time $-t_J < 0$, while the time $-t_h \ll -t_J$, at which the $h_c$ state is created, corresponds to the \textit{sink} of the correlation function. This setup allows us to monitor the onset of the $h_c$ ground-state dominance.

The insertion time $t_J$ must be chosen large enough to isolate the $\eta_c$ ground state. The advantage of our setup is that $t_J$ can be taken reasonably large without significant loss of signal, since the SNR of $C_{\rm 3h}^{ij}(t_h; t_J, \bs{k})$ for fixed time separation $|t_h - t_J|$ scales as
\begin{align}
e^{-(E_{\eta_c}(\bs{k}) - m_{\eta_c}) t_J} \simeq e^{-|\bs{k}|^2 t_J / (2 m_{\eta_c})}~,
\end{align}
with $|\bs{k}|^2 / (2 m_{\eta_c}) \leq 40~{\rm MeV}$. The SNR therefore decreases only slowly as $t_J$ increases for fixed $|t_{h}-t_{J}|$. As in the previous case, we consider two values, $t_J \simeq 1.6$ and $2.4~{\rm fm}$, and verify that the results for $\overline{F}_{1(2)}(t_h; t_J, \bs{k})$ are consistent within statistical uncertainties.

To further reduce statistical errors, we employ additional numerical improvements, particularly for $\tilde{F}_2(q^2)$. This procedure closely parallels the strategy used for the $M_2(q^2)$ form factor. In the limit $\bs{k} \to 0$, we see from Eq.~\eqref{eq:aux} that $\overline{R}_1 = \overline{F}_1$, and with $\beta(q^2) \to 1$, the estimator $\overline{R}_1^{\rm sub}(t_h; t_J, \bs{0})$ vanishes. However, with finite statistics, this cancellation is not exact, and the noise of $\overline{R}_1(t_h; t_J, \bs{0})$ is strongly correlated with that of $\overline{R}_1(t_h; t_J, \bs{k})$, at least for moderately small momenta $\bs{k}$. For this reason, the zero-momentum subtraction:
\begin{align}
\label{eq:impro_hc_1}
\overline{F}_2(t_h; t_J, \bs{k}) \to \frac{\overline{R}_1^{\rm sub}(t_h; t_J, \bs{k}) - \overline{R}_1^{\rm sub}(t_h; t_J, \bs{0})}{\delta(q^2)}~,
\end{align}
leads to a significantly reduced statistical noise.

The zero-momentum subtraction is applied to all values of $\bs{k}$ in Eq.~\eqref{eq:moms_for_h}, except for the very small momentum $k_z \simeq 10^{-6}~{\rm GeV}$. In this case, we introduce a further improvement by subtracting noise components that are momentum-independent or linear in $\bs{k}$:
\begin{align}
\label{eq:impro_hc_2}
\overline{F}_2(t_h; t_J, \bs{k}) \to \frac{1}{2} \left[ \overline{F}_2(t_h; t_J, \bs{k}) + \overline{F}_2(t_h; t_J, -\bs{k}) \right],
\end{align}
exploiting the fact that $\overline{F}_2$ is an even function of $\bs{k}$. This correlated subtraction is essential for determining $\tilde{F}_2(q^2_{\rm max})$, which would otherwise suffer from very large statistical uncertainties.

\section{Numerical Results}
\label{sec:num_results}

\subsection{Lattice Results for the $\chi_{c1} \to J/\psi \ell^+ \ell^-$ Form Factors}

\begin{table}[t]
\begin{ruledtabular}
\begin{tabular}{lccc}
\textrm{ID} & $N_g$ & $N_s^0$ & $N_s$ \\
\colrule
\textrm{A48} & $800$ & $140$ & $140$ \\
\textrm{B64} & $400$ & $32$ & $32$ \\
\textrm{C80} & $640$ & $24$ & $8$ \\
\textrm{D96} & $300$ & $35$ & $8$
\end{tabular}
\end{ruledtabular}
\caption{\small\sl The number, $N_g$, of gauge configurations considered for the calculation of the $\chi_{c1} \to J/\psi \ell^+ \ell^-$ form factors, for each of the $N_f = 2+1+1$ ETMC gauge ensembles in Table~\ref{tab:simudetails}. $N_s^0$ is the number of (spin and color diluted) stochastic sources adopted in Ref.~\cite{Becirevic:2025idm} for the calculation of the $E_1$ and $M_2$ form factors at $q^2 = 0$, while $N_s$ is the number of stochastic sources considered for the present calculation in which the photon is generally off-shell. \label{tab:simu_chi}}
\end{table}
Having described in Section~\ref{sec:extr_form_factor_charm_chi} the extraction strategy for the
$\chi_{c1}\to J/\psi\ell^+\ell^-$ form factors,
we now turn to the corresponding numerical analysis. In Ref.~\cite{Becirevic:2025idm}, we determined the form factors $E_1(0)$ and $M_2(0)$ through a high-statistics analysis of the relevant two- and three-point correlation functions, employing a large number of gauge field configurations and stochastic sources. The statistics used in that work are summarized for each ensemble in Table~\ref{tab:simu_chi}, where $N_s^0$ denotes the number of stochastic sources employed in that calculation.

To extend the analysis to the case of a virtual photon, we performed additional simulations for all momentum values listed in Eq.~\ref{eq:moms_for_chi}, using the same gauge configurations as in Ref.~\cite{Becirevic:2025idm} but with newly generated stochastic sources, whose number is denoted by $N_s$. For the computationally more demanding C80 and D96 ensembles, a reduced number of stochastic sources was employed; the corresponding values are reported in Table~\ref{tab:simu_chi}. To benefit from the high-statistics computation of Ref.~\cite{Becirevic:2025idm} within the present analysis, we implemented the following improvements.

First, the mass and couplings of the $\chi_{c1}$ are always determined from the high-statistics dataset, achieving higher precision. Moreover, for the form factors $E_1(q^2)$ and $M_2(q^2)$, we introduce the following preconditioning of the estimators $\overline{E}_1(t_\chi; t_J, \bs{k})$ and $\overline{M}_2(t_\chi; t_J, \bs{k})$:
\begin{align}
\label{eq:preconditioning}
\overline{E}_1(t_\chi; t_J, \bs{k}) &= \overline{E}_1^{N_s^0}(t_\chi; t_J, \bs{k}_{\rm max}) \frac{\overline{E}_1^{N_s}(t_\chi; t_J, \bs{k})}{\overline{E}_1^{N_s}(t_\chi; t_J, \bs{k}_{\rm max})}, \nonumber \\
\overline{M}_2(t_\chi; t_J, \bs{k}) &= \overline{M}_2^{N_s^0}(t_\chi; t_J, \bs{k}_{\rm max}) \frac{\overline{M}_2^{N_s}(t_\chi; t_J, \bs{k})}{\overline{M}_2^{N_s}(t_\chi; t_J, \bs{k}_{\rm max})},
\end{align}
where the superscripts $N_s^0$ and $N_s$ indicate determinations obtained with $N_s^0$ and $N_s$ stochastic sources, respectively.

The rationale behind this strategy is that results obtained for a generic momentum $|\bs{k}|$ and at the reference momentum $|\bs{k}_{\rm max}| \simeq 389.4~{\rm MeV}$ are partially correlated, with the correlation increasing as $\bs{k}$ approaches $\bs{k}_{\rm max}$. This allows the ratio between form factors at $q^2$ and at $q^2 = 0$ to be determined using a reduced set of stochastic sources without significant loss of precision. The high-statistics determinations of $E_1(0)$ and $M_2(0)$ obtained with $N_s^0$ stochastic sources can then be used to reconstruct $E_1(q^2)$ and $M_2(q^2)$ over the full kinematic range.

This preconditioning procedure proves highly effective in reducing statistical uncertainties and, more generally, can be advantageous for reducing computational costs whenever form factors must be evaluated across the allowed kinematic region. In particular, when accessing different kinematic points requires additional inversions of the Dirac operator, as in calculations employing twisted boundary conditions, it is convenient to determine the form factors with high precision at a reference kinematic point (here $q^2 = 0$) and then compute ratios with respect to this reference using a smaller number of stochastic sources.

\begin{figure}
\centering
\includegraphics[width=0.47\linewidth]{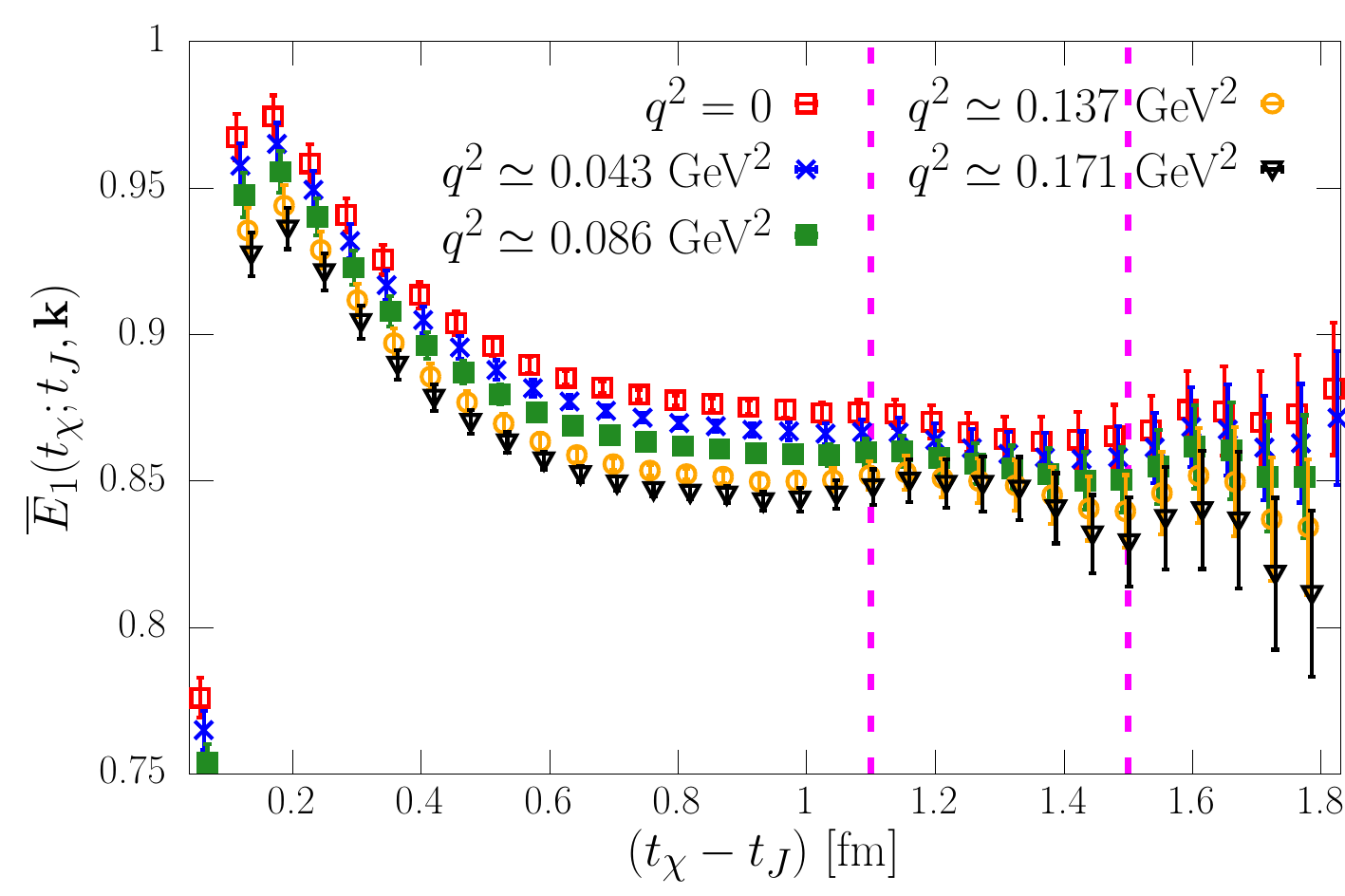}
\includegraphics[width=0.47\linewidth]{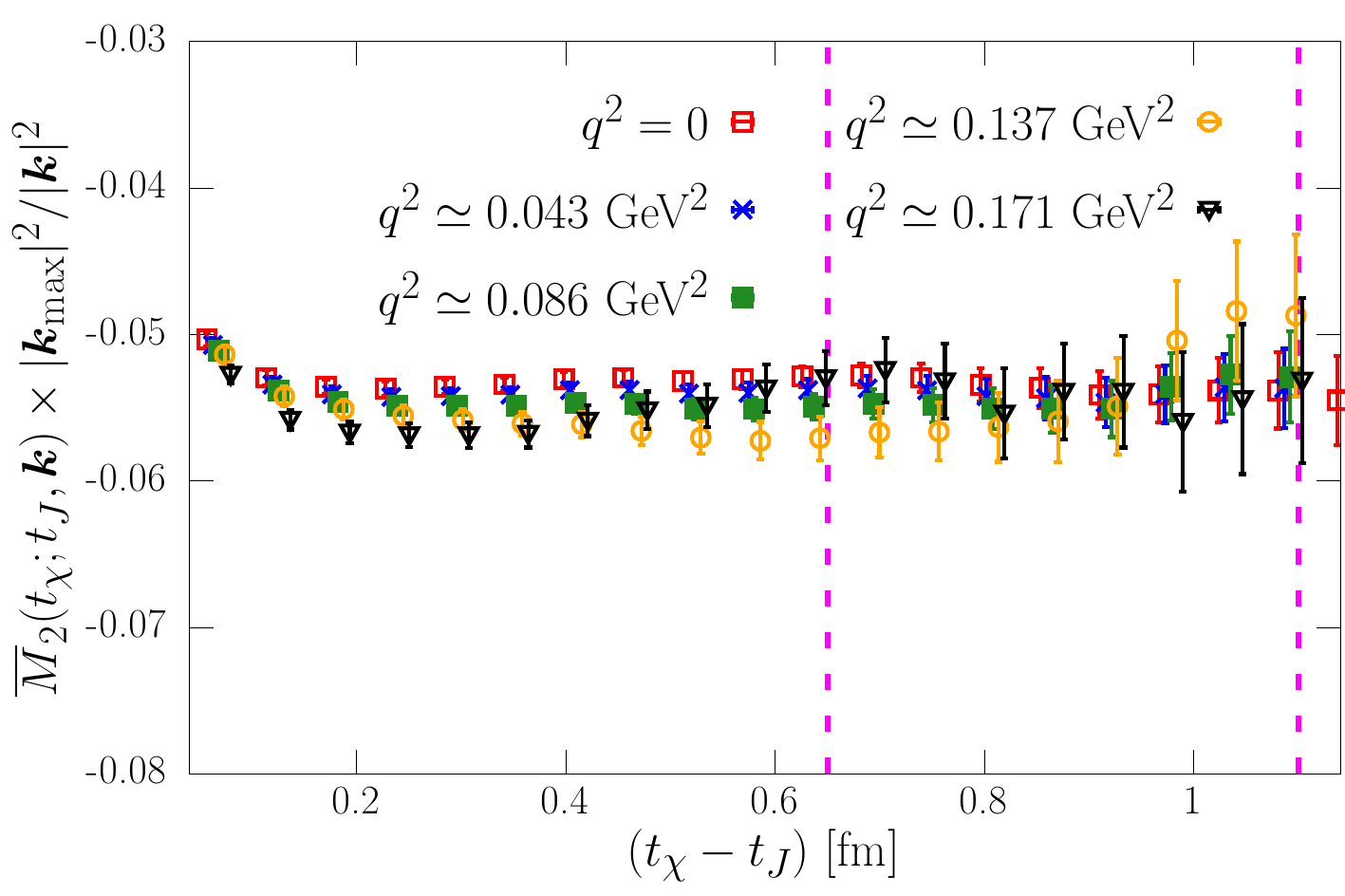} \\
\includegraphics[width=0.47\linewidth]{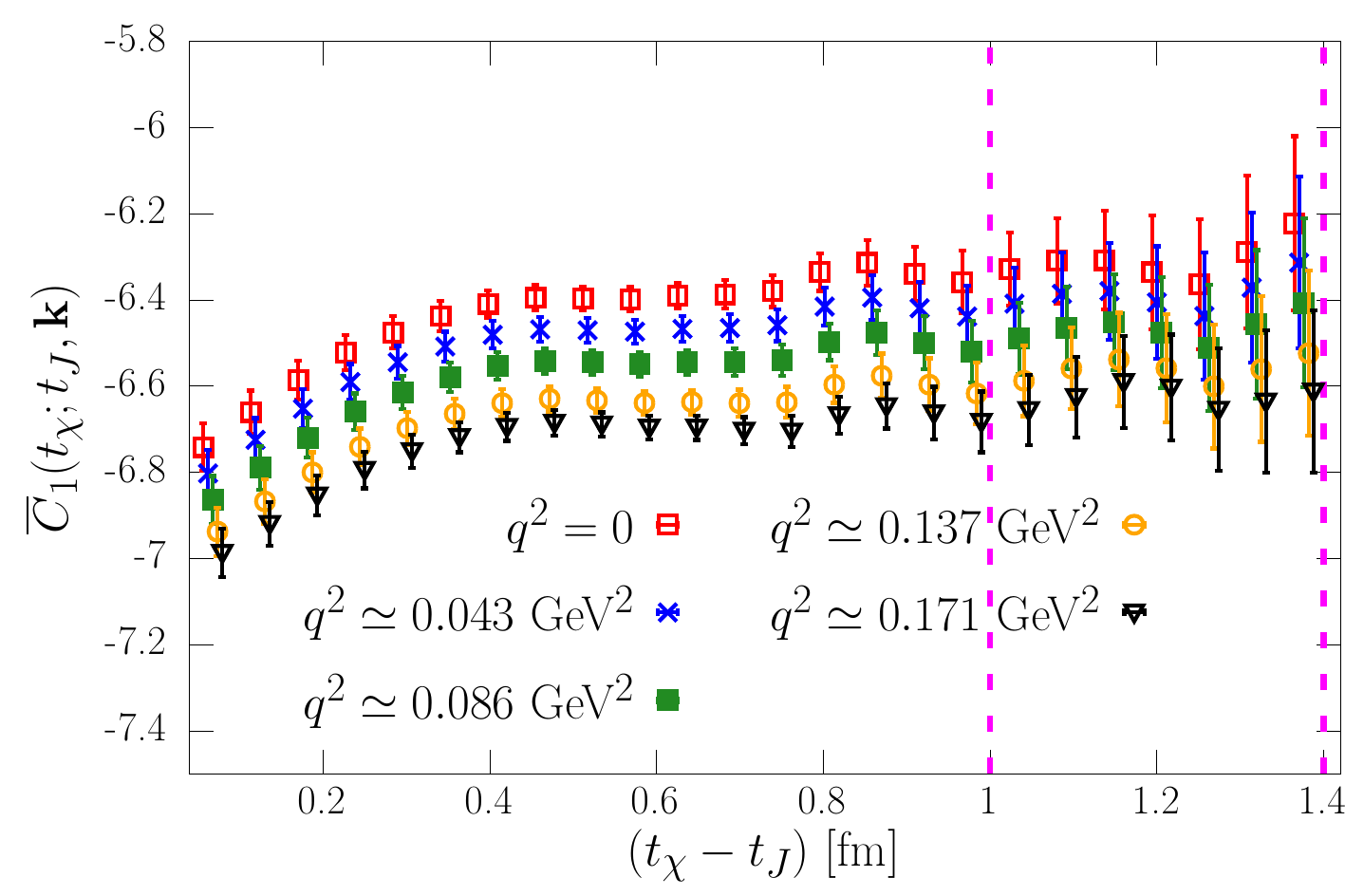}
\caption{\small The estimators $\overline{E}_1(t_\chi; t_J, \bs{k})$, $\overline{M}_2(t_\chi; t_J, \bs{k})/|\bs{k}|^2$, and $\overline{C}_1(t_\chi; t_J, \bs{k})$ for a fixed value of $t_J \simeq 1.6~{\rm fm}$ on the finest lattice spacing ensemble adopted for this study (ensemble D96). In each plot, the different curves correspond to the different simulated values of $q^2$. The data points at the different $q^2$ have been slightly shifted horizontally for better visualization. The magenta vertical lines (approximately) delimit the time interval in which we perform a constant fit to extract the form factors.}
\label{fig:chi}
\end{figure}

In Fig.~\ref{fig:chi}, we show, as an illustrative example, the estimators of the form factors $E_1(q^2)$, $M_2(q^2)/|\bs{k}|^2$, and $C_1(q^2)$ on the finest lattice-spacing ensemble (D96) for a fixed value of $t_J \simeq 1.6~{\rm fm}$.~\footnote{Note that for small values of the $J/\psi$ three-momentum, $\bs{k}$, in the $\chi_{c1}$ rest frame, $M_2(q^2) \propto |\bs{k}|^2$. We factorize the leading $|\bs{k}|^2$ dependence and present results for the dimensionless combination $M_2(q^2) |\bs{k}_{\rm max}|^2 / |\bs{k}|^2$, where $|\bs{k}_{\rm max}| \simeq 389.4~{\rm MeV}$, which exhibits a much weaker dependence on $q^2$.} We observe that the $q^2$ dependence of $E_1$, $M_2/|\bs{k}|^2$, and $C_1$ is smooth, and the quality of the plateaus remains comparable when moving away from $q^2 = 0$. As $q^2_{\rm max}$ is approached, an increase in statistical noise is observed for $E_1$ and $M_2$. This can be traced back to the reduced efficiency of the preconditioning in Eq.~\eqref{eq:preconditioning}, since the correlation between $q^2 = 0$ and $q^2 \simeq q^2_{\rm max}$ becomes weaker. Note also that the longitudinal form factor, $C_1(q^2)$, is new. It was not considered in our study of $\chi_{c1} \to J/\psi \gamma$~\cite{Becirevic:2025idm} since it does not contribute in the real-photon limit. The estimators for $C_1(q^2)$ are also shown in Fig.~\ref{fig:chi}.

The form factors are extracted from correlated constant fits to the
effective estimators in the plateau region. For illustration,
the vertical dashed lines in Fig.~\ref{fig:chi} approximately delimit the fit
intervals adopted for the D96 ensemble, namely
$t\in[1.1,1.5]$~fm for $E_1$,
$t\in[1.0,1.4]$~fm for $C_1$,
and $t\in[0.65,1.1]$~fm for $M_2$.
Similar intervals are used for the other ensembles and other kinematic
points, with variations of at most about $0.2$~fm.
In the case of $C_1(q^2)$, the constant fit yields a statistical precision
of approximately $2\%$.

The dependence of the results on the insertion time of the electromagnetic current, $t_J$, is found to be very small, in agreement with our previous study~\cite{Becirevic:2025idm}. This investigation was performed on a single gauge ensemble (the B64 ensemble), for which all three-point correlation functions were computed using two values of $t_J$, namely $t_J \simeq 1.6~{\rm fm}$ and $t_J \simeq 2.4~{\rm fm}$.

\begin{figure}
\centering
\includegraphics[width=0.47\linewidth]{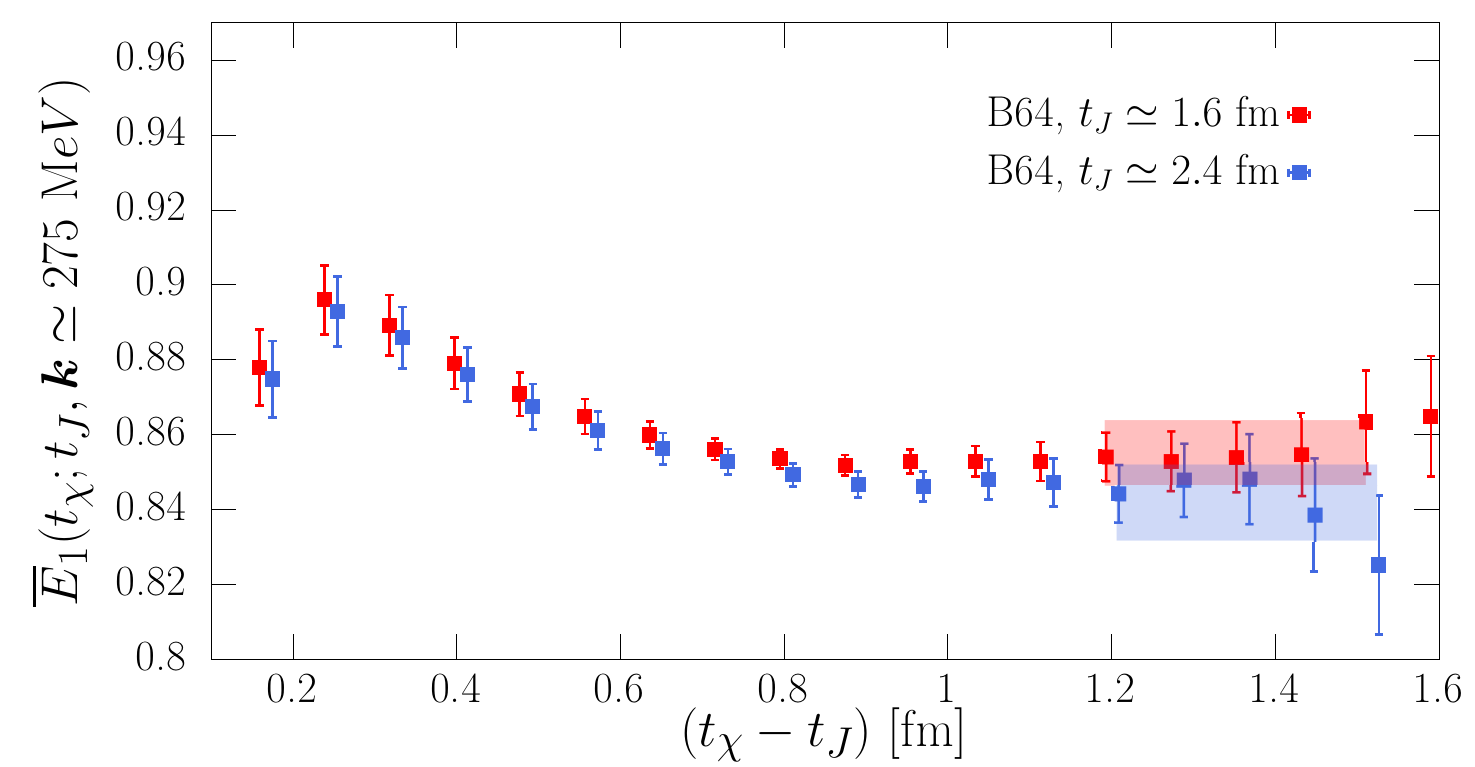}
\includegraphics[width=0.47\linewidth]{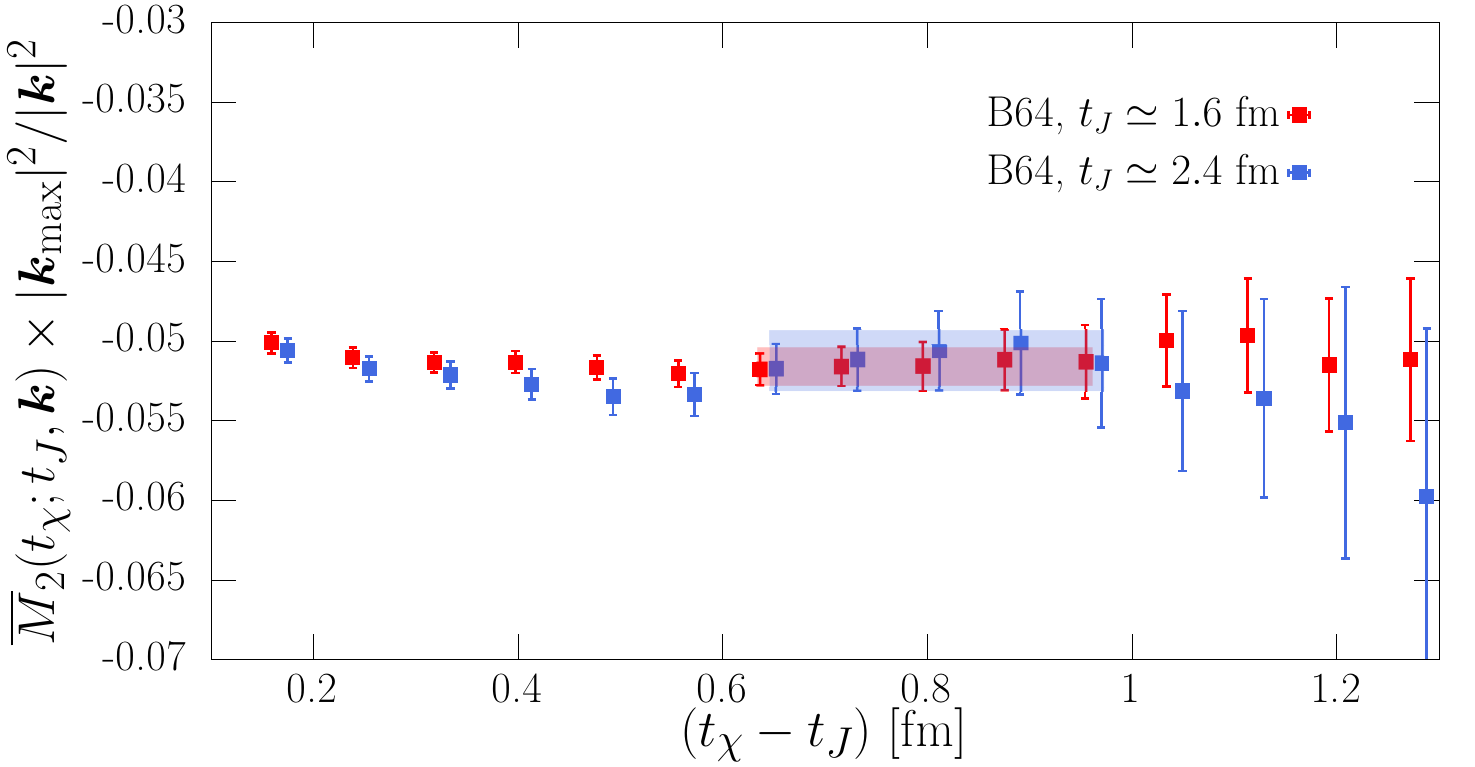} \\
\includegraphics[width=0.47\linewidth]{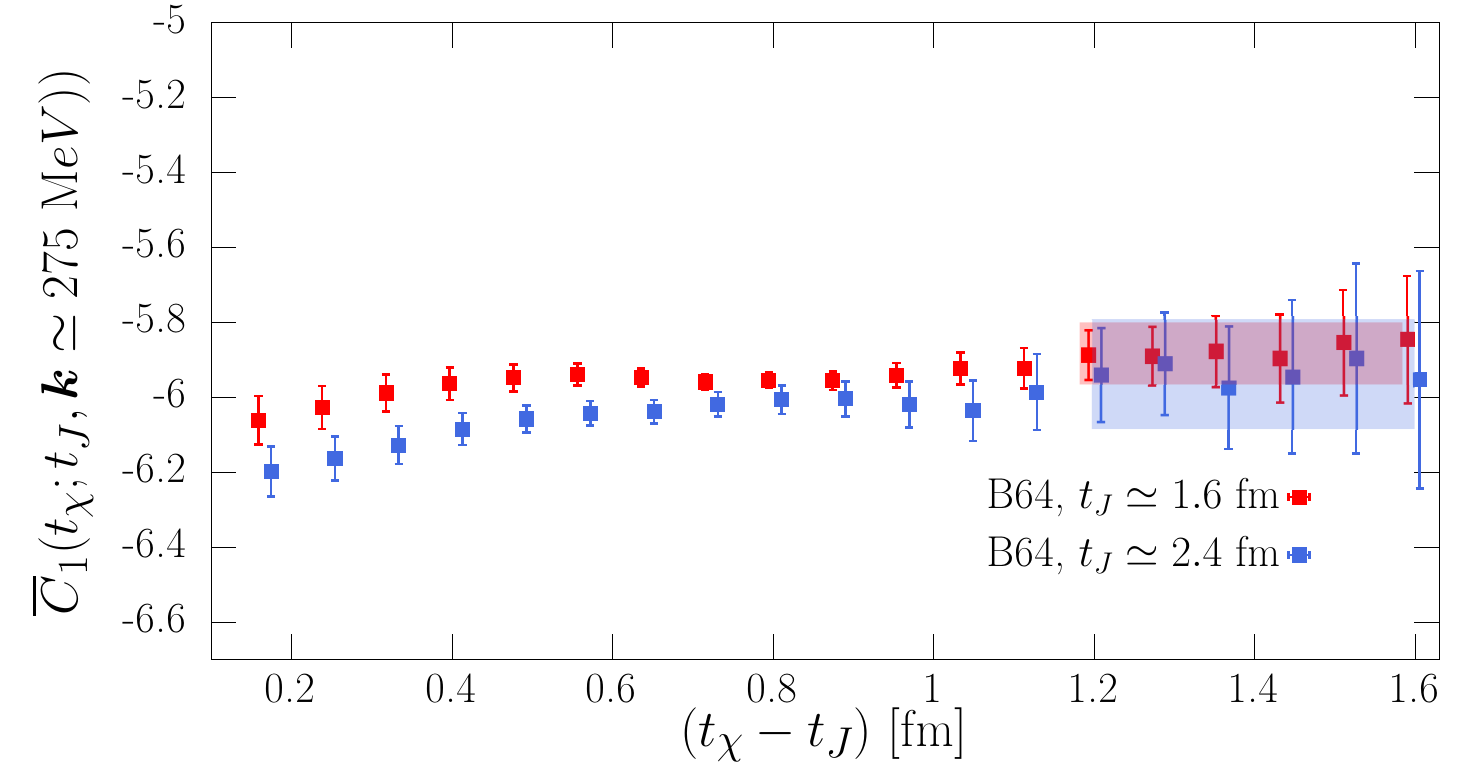}
\caption{\small The estimators $\overline{E}_1(t_\chi; t_J, \bs{k})$, $\overline{M}_2(t_\chi; t_J, \bs{k}) |\bs{k}_{\rm max}|^2 / |\bs{k}|^2$, and $\overline{C}_1(t_\chi; t_J, \bs{k})$ are shown for two values of $t_J \simeq 1.6, 2.4~{\rm fm}$. The results correspond to our determination on the B64 ensemble for $k_z = 275~{\rm MeV}$, which corresponds to $q^2 \simeq 0.086~{\rm GeV}^2$. The coloured bands indicate the results of constant fits to the estimators over the selected time intervals, from which the form factors are determined.}
\label{fig:tj_dep_chi}
\end{figure}

A comparison between the estimators of the form factors obtained with these two choices of $t_J$ is shown in Fig.~\ref{fig:tj_dep_chi} for an intermediate value $q^2 \simeq (0.293~{\rm GeV})^2$. The behavior observed in all other cases is qualitatively the same. Within statistical uncertainties, no significant dependence on $t_J$ is observed in the plateau regions, indicating that our smallest choice, $t_J \simeq 1.6~{\rm fm}$, is already sufficient to render finite-$t_J$ effects negligible at the present level of precision.

To estimate any residual $t_J$ dependence, we associate a systematic uncertainty to the results obtained at $t_J \simeq 1.6~{\rm fm}$, defined as:
\begin{align}
\Sigma_F &= \left| F(t_J \simeq 2.4~{\rm fm}) - F(t_J \simeq 1.6~{\rm fm}) \right|, \nonumber \\
\Delta F &= \Sigma_F \, {\rm erf}\!\left(\frac{\Sigma_F}{\sqrt{2} \sigma_{\Sigma_F}}\right), \qquad F = E_1(q^2), M_2(q^2), C_1(q^2),
\end{align}
where $\sigma_{\Sigma_F}$ denotes the statistical uncertainty on $\Sigma_F$ and ${\rm erf}(x)$ is the error function. The quantities $\Delta E_1, \Delta M_{2}$ and $\Delta C_{1}$ therefore represent the observed spreads weighted by the probability that they do not originate from statistical fluctuations.

\begin{figure}
\centering
\includegraphics[width=0.47\linewidth]{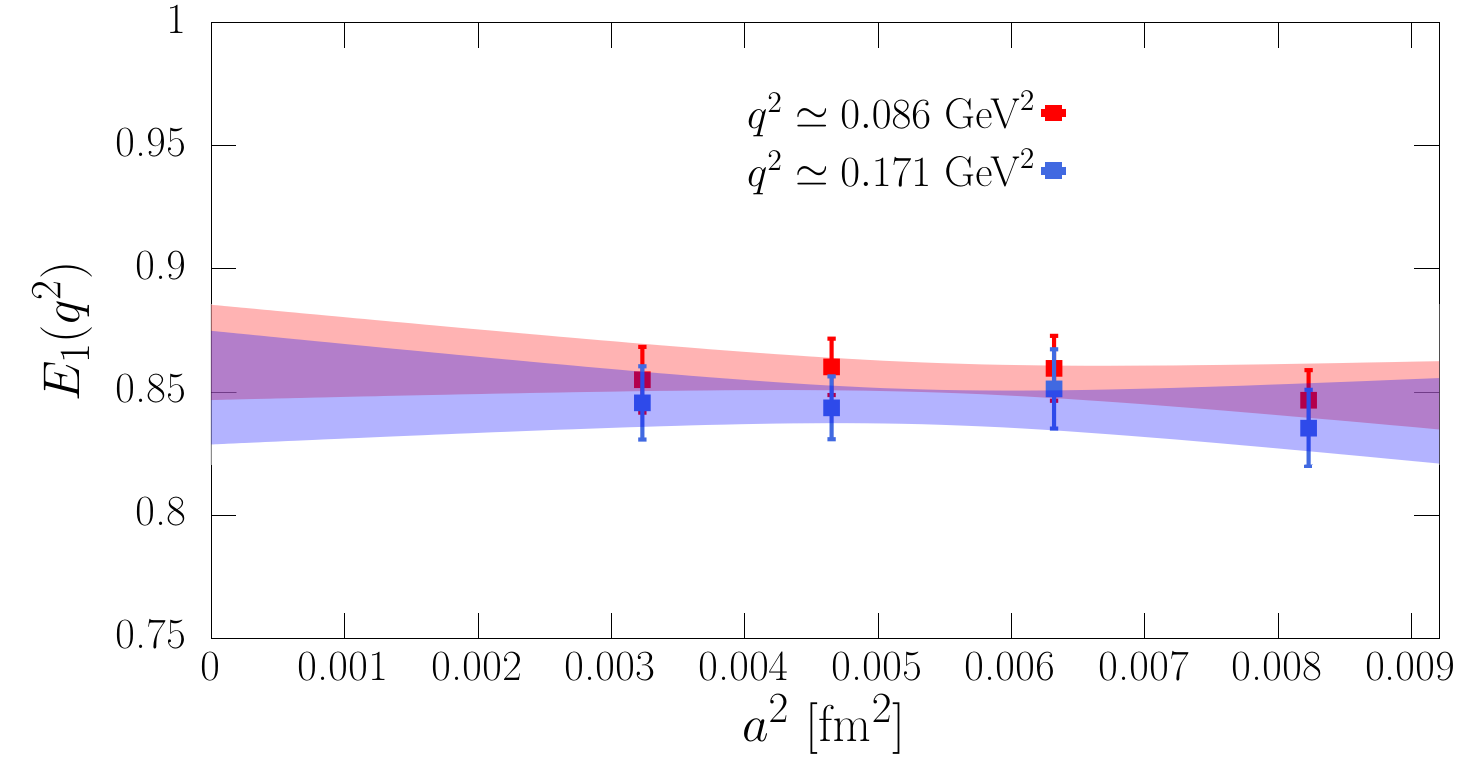}
\includegraphics[width=0.47\linewidth]{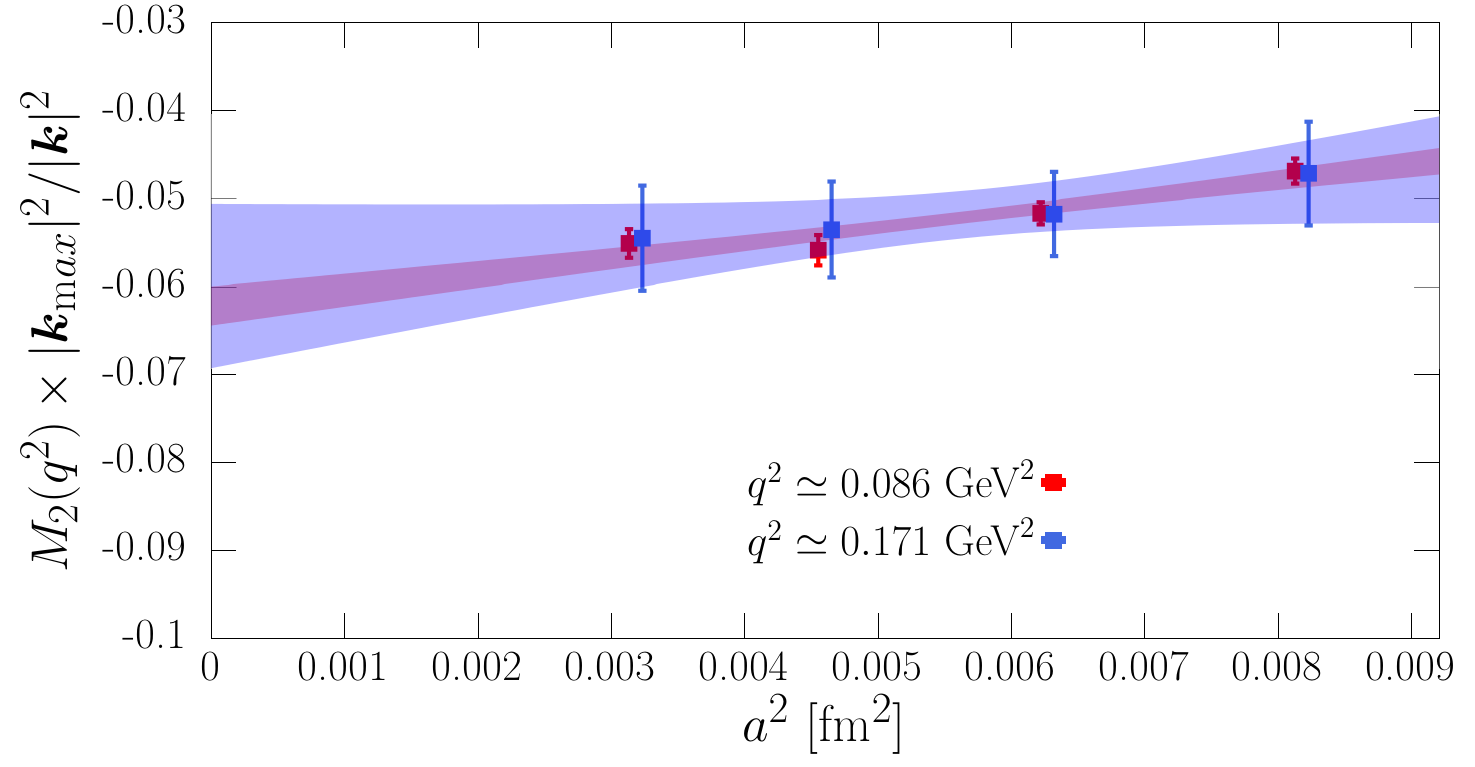} \\
\includegraphics[width=0.47\linewidth]{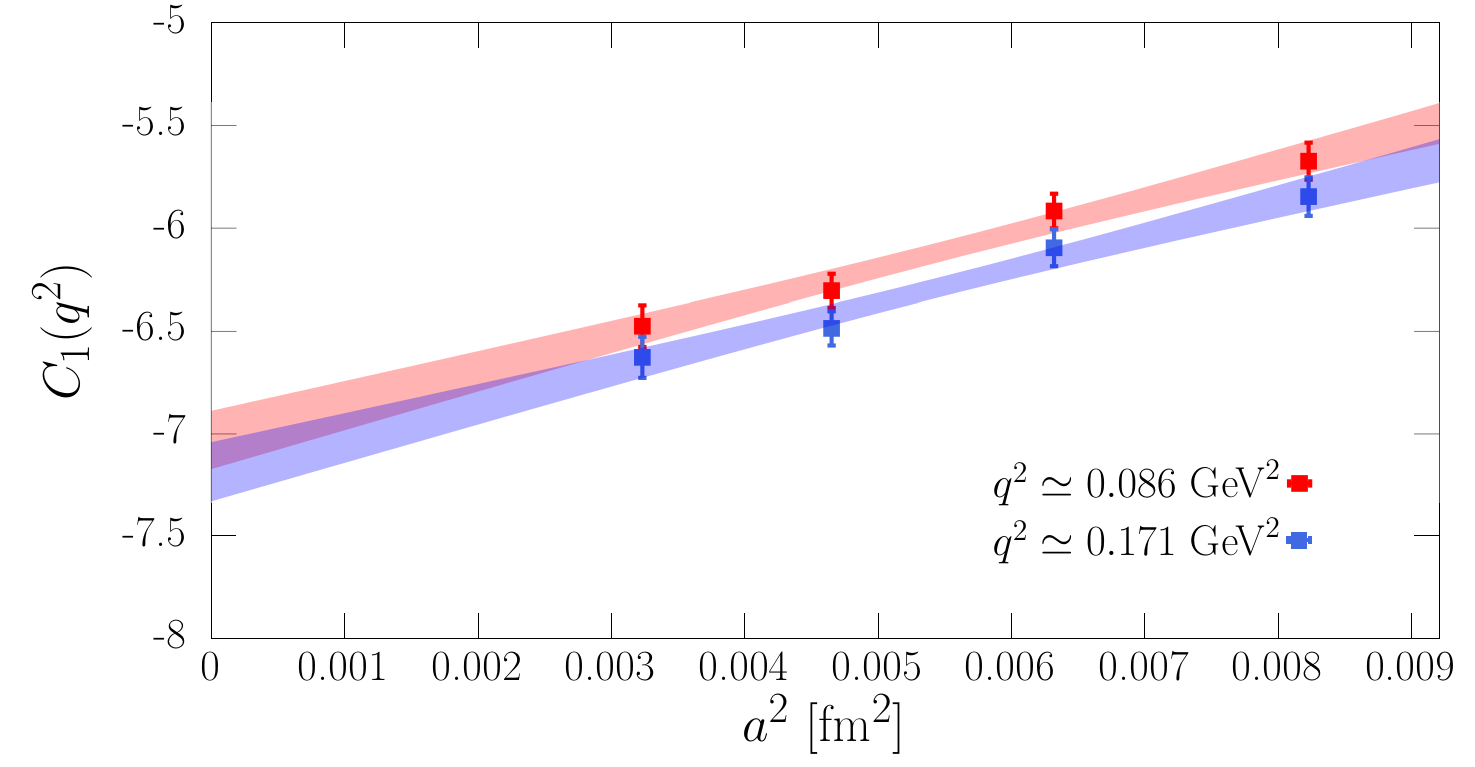}
\caption{\small Lattice spacing dependence of the form factors $E_1(q^2)$, $M_2(q^2)$, and $C_1(q^2)$ for two values of the three-momentum transfer $\bs{k}$ corresponding to $q^2 \simeq 0.086~{\rm GeV}^2$ and $q^2 \simeq 0.171~{\rm GeV}^2 \simeq q^2_{\rm max}$.}
\label{fig:cont_chi}
\end{figure}

As for the lattice-spacing dependence of the results, we illustrate in Fig.~\ref{fig:cont_chi} the form factors as functions of $a^2$ for an intermediate $q^2 \simeq (0.293~{\rm GeV})^2$ and for a value close to $q^2_{\rm max}$. The lattice-spacing dependence of the $E_1(q^2)$ and $M_2(q^2)$ form factors closely resembles that observed at $q^2 = 0$, cf. Ref.~\cite{Becirevic:2025idm}. While $E_1(q^2)$ displays a rather mild dependence on $a^2$, we again observe a more pronounced cutoff dependence for $M_2(q^2)$. As $q^2 \to q^2_{\rm max}$, the statistical uncertainties increase, particularly for the form factor $M_2(q^2)$.

The longitudinal form factor $C_1(q^2)$, instead, exhibits sizable cutoff effects of order $\mathcal{O}(10\text{--}15\%)$, which are nevertheless well described by an $a^2$ scaling. To extrapolate our results to the continuum limit, we adopt the same strategy as in Ref.~\cite{Becirevic:2025idm}. Specifically, for each form factor and each value of $q^2$, we perform two linear fits in $a^2$: the first using the full dataset and the second excluding the coarsest lattice spacing. The fits obtained from the full dataset are shown in Fig.~\ref{fig:cont_chi} as colored bands. In all cases, the fit quality is satisfactory, with reduced $\chi^2$ values of order $\mathcal{O}(1)$.

\begin{figure}
\centering
\includegraphics[width=0.9\linewidth]{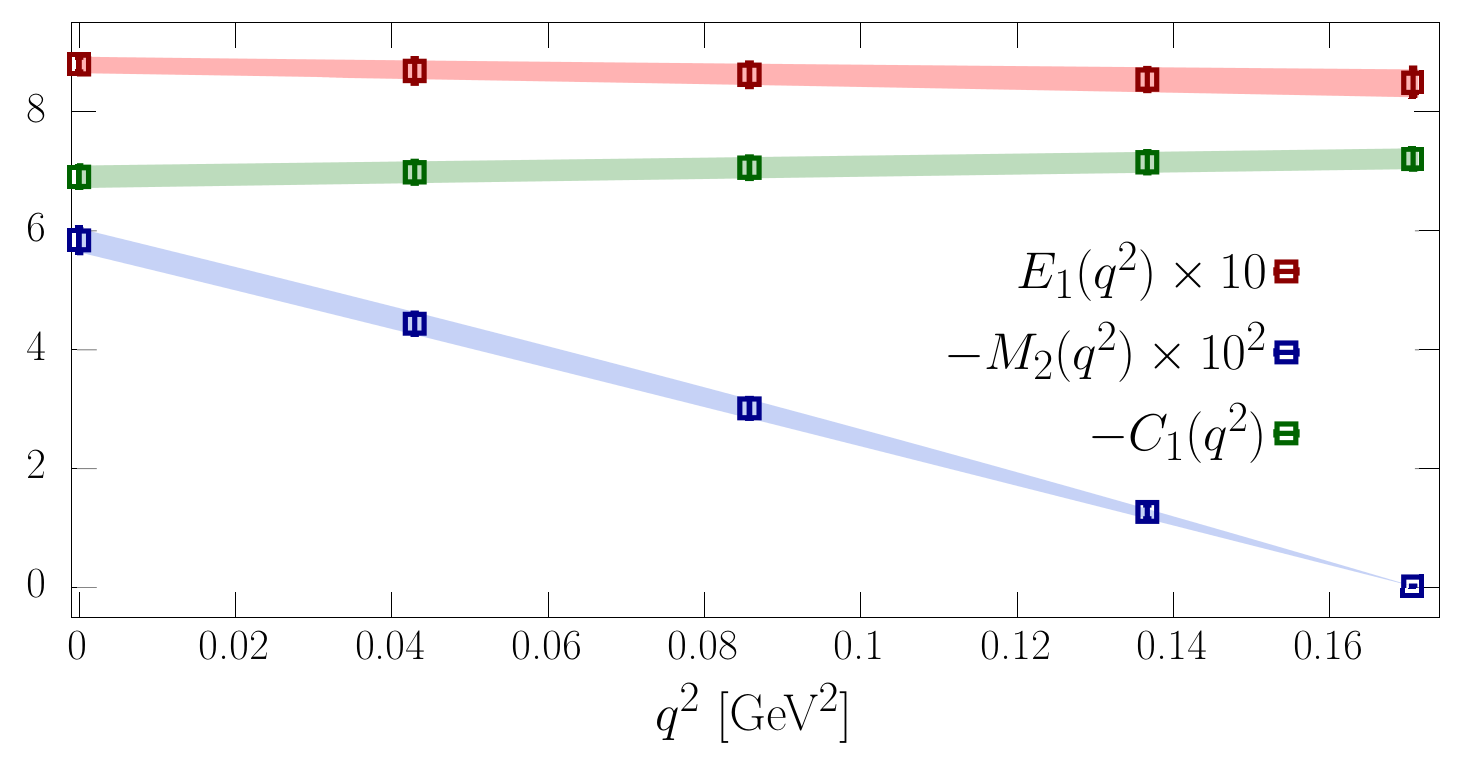}
\caption{\small The three form factors $E_1(q^2)$, $M_2(q^2)$, and $C_1(q^2)$ describing the $\chi_{c1} \to J/\psi \ell^+ \ell^-$ decay as a function of $q^2$ in the range $q^2 \in [0, q^2_{\rm max}]$. For illustrative purposes, we have rescaled the form factors or reversed their sign to fit them in a single plot. Note that this time we show directly $M_2(q^2)$ and not $M_2(q^2) \times |\bs{k}_{\rm max}|^2 / |\bs{k}|^2$. The colored bands correspond to the results of the $q^2$ fits performed according to Eq.~\eqref{eq:q2_fit_ansatz_chi}.}
\label{fig:final_FF_chi}
\end{figure}

Our final results are obtained by averaging the two fits using the Bayesian Akaike Information Criterion (BAIC).~\footnote{See Eqs.~(21,22) of Ref.~\cite{Becirevic:2025ocx} for details.} The resulting continuum-extrapolated form factors $E_1(q^2)$, $M_2(q^2)$, and $C_1(q^2)$ are displayed in Fig.~\ref{fig:final_FF_chi}, and the values are given in Tab.~\ref{tab:chi_c1_FF}.

\begin{table}[t]
\begin{ruledtabular}
\begin{tabular}{lccc}
$q^2\,[\mathrm{GeV}^2]$ & $E_1(q^2)$ & $M_2(q^2)\times |\bs{k}_{\rm max}|^{2}/|\bs{k}|^{2}$ & $C_1(q^2)$ \\
\colrule
0        & 0.879(12) & -0.0584(21) & -6.90(18) \\
0.043 & 0.869(20) & -0.0593(24) & -6.98(18) \\
0.086 & 0.863(19) & -0.0605(34) & -7.05(18) \\
0.137 & 0.855(18) & -0.0631(57) & -7.15(17) \\
0.171 & 0.849(24) & -0.0591(91)  & -7.20(17) \\
\end{tabular}
\end{ruledtabular}
\caption{Continuum extrapolated values of the form factors $E_{1}(q^{2}), M_{2}(q^{2})\times |\bs{k}_{\rm max}|^{2}/|\bs{k}|^{2}$, and $C_{1}(q^{2})$ for all values of $q^{2}$ considered.}
\label{tab:chi_c1_FF}
\end{table}

We observe that the $q^2$-dependence of $E_1(q^2)$, $M_2(q^2) \times |\bs{k}_{\rm max}|^2 / |\bs{k}|^2$, and $C_1(q^2)$ is smooth, and we fit them to:
\begin{align}
\label{eq:q2_fit_ansatz_chi}
F(q^2) = C_F [1 + D_F q^2], \qquad F \in \left\{ E_1, M_2 \frac{|\bs{k}_{\rm max}|^2}{|\bs{k}|^2}, C_1 \right\},
\end{align}
where $C_F$ and $D_F$ are the fit parameters. The resulting fit curves are shown in Fig.~\ref{fig:final_FF_chi} as colored bands, while the values of $F(0)$ and $D_F$, together with their correlations, can be found in Table~\ref{tab:coeff_chi} for each form factor. We note that the $q^2$ dependence is very mild, with $|D_Fq^2|\ll1$ over the full physical range, which is densely covered by our lattice data. Alternative smooth parametrizations, such as pole or dipole forms, would therefore lead to numerically very similar interpolations.

Note that only in the case of the longitudinal form factor $C_1(q^2)$ do we observe a non-zero slope in $q^2$. The form factor $M_2(q^2)$ is, to a very good approximation, proportional to $|\bs{k}|^2$, and no statistically significant slope is observed when fitting the ratio $M_2(q^2)/|\bs{k}|^2$. In Section~\ref{sec:comparison}, we use the linear $q^2$ parametrization of the form factors to determine the decay rates and differential observables for $\chi_{c1} \to J/\psi e^+ e^-$ and $\chi_{c1} \to J/\psi \mu^+ \mu^-$.

We now turn to the numerical results for the form factors $F_{1}(q^{2})$ and $\tilde{F}_{2}(q^{2})$ relevant for the $h_{c}\to \eta_{c}\ell^{+}\ell^{-}$ decay.

\begin{table}[t]
\begin{ruledtabular}
\begin{tabular}{lccc}
parameter & $E_{1}$ & $M_{2}\times|\bs{k}_{\rm max}|^{2}/|\bs{k}|^{2}$  & $C_{1}$     \\
\colrule
$C_{F}$  & $0.879(14)$ & $ -0.0584(20)$ & $-6.90(19)$  \\
$D_{F}~[{\rm GeV^{-2}}]$  & $-0.20(10)$ & $0.39(35)$ & $0.258(61)$  \\
${\rm Corr}\,(F(0), D_{F})$ & $0.31$ & $-0.42$ & $-0.49$
\end{tabular}
\end{ruledtabular}
\caption{\small\sl Values of the fit parameters $C_{F}$ and $D_{F}$ for the three form factors describing the $\chi_{c1}\to J/\psi \ell^{+}\ell^{-}$ decay. \label{tab:coeff_chi}}
\end{table}

\subsection{Lattice results for the  $h_{c}\to \eta_{c}\ell^{+}\ell^{-}$ form factors}
We now present the numerical analysis of the
$h_c\to\eta_c\ell^+\ell^-$ transition,
following the extraction strategy described in
Section~\ref{sec:extr_form_factor_charm_hc}.
For the calculation of the two- and three-point correlation functions required to determine the form factors $F_1(q^2)$ and $\tilde{F}_2(q^2)$ of the $h_c \to \eta_c \ell^+ \ell^-$ decay, we employed the same number of gauge configurations and stochastic sources for each of the $N_f = 2+1+1$ gauge ensembles listed in Table~\ref{tab:simudetails} as in our previous study~\cite{Becirevic:2025ocx}. This information is summarized in Table~\ref{tab:simu_hc}. Note that the number of stochastic sources used for this calculation is significantly smaller than that used for the $\chi_{c1} \to J/\psi \ell^+ \ell^-$ transition. Given the relatively lower computational cost, we did not implement the preconditioning described in the previous subsection (Eq.~\eqref{eq:preconditioning}), but instead computed the two- and three-point correlation functions using the same number of gauge configurations and sources for all kinematic points, corresponding to both $q^2 = 0$ and $q^2 \neq 0$.

\begin{table}[t]
\begin{ruledtabular}
\begin{tabular}{lcc}
\textrm{ID} & $ N_g $ & $ N_s $   \\
\colrule
\textrm{A48} & 300 & 32   \\
\textrm{B64} & 203 & 8   \\
\textrm{C80} & 609 & 4     \\
\textrm{D96} & 150 & 8
\end{tabular}
\end{ruledtabular}
\caption{\small The number of gauge configurations ($N_g$) and stochastic sources ($N_s$) used for the calculation of the $h_c \to \eta_c \ell^+ \ell^-$ form factors, for each of the $N_f = 2+1+1$ ETMC gauge ensembles in Table~\ref{tab:simudetails}. \label{tab:simu_hc}}
\end{table}

Unlike in Ref.~\cite{Becirevic:2025ocx}, where the form factor $F_1(0)$ relevant to the $h_c \to \eta_c \gamma$ decay was computed including the E112 ensemble ($a \simeq 0.049~\mathrm{fm}$), we did not perform simulations on this ensemble in the present calculation, as cutoff effects in this channel were found to be very small. Nevertheless, some of the results obtained in Ref.~\cite{Becirevic:2025ocx} at $q^2 = 0$ can still be reused in the present analysis, as discussed below.

\begin{figure}
\centering
\includegraphics[width=0.49\linewidth]{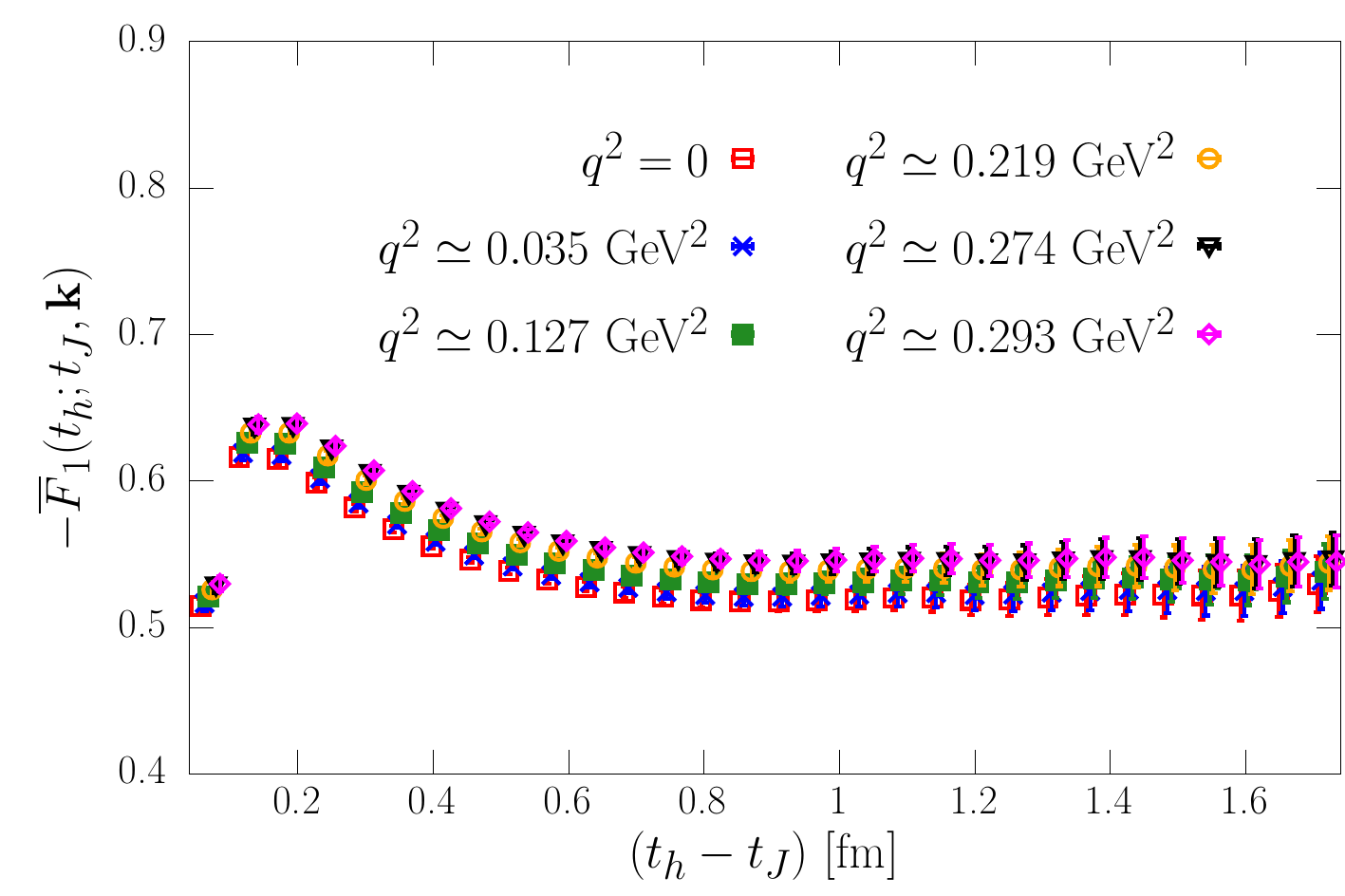}
\includegraphics[width=0.49\linewidth]{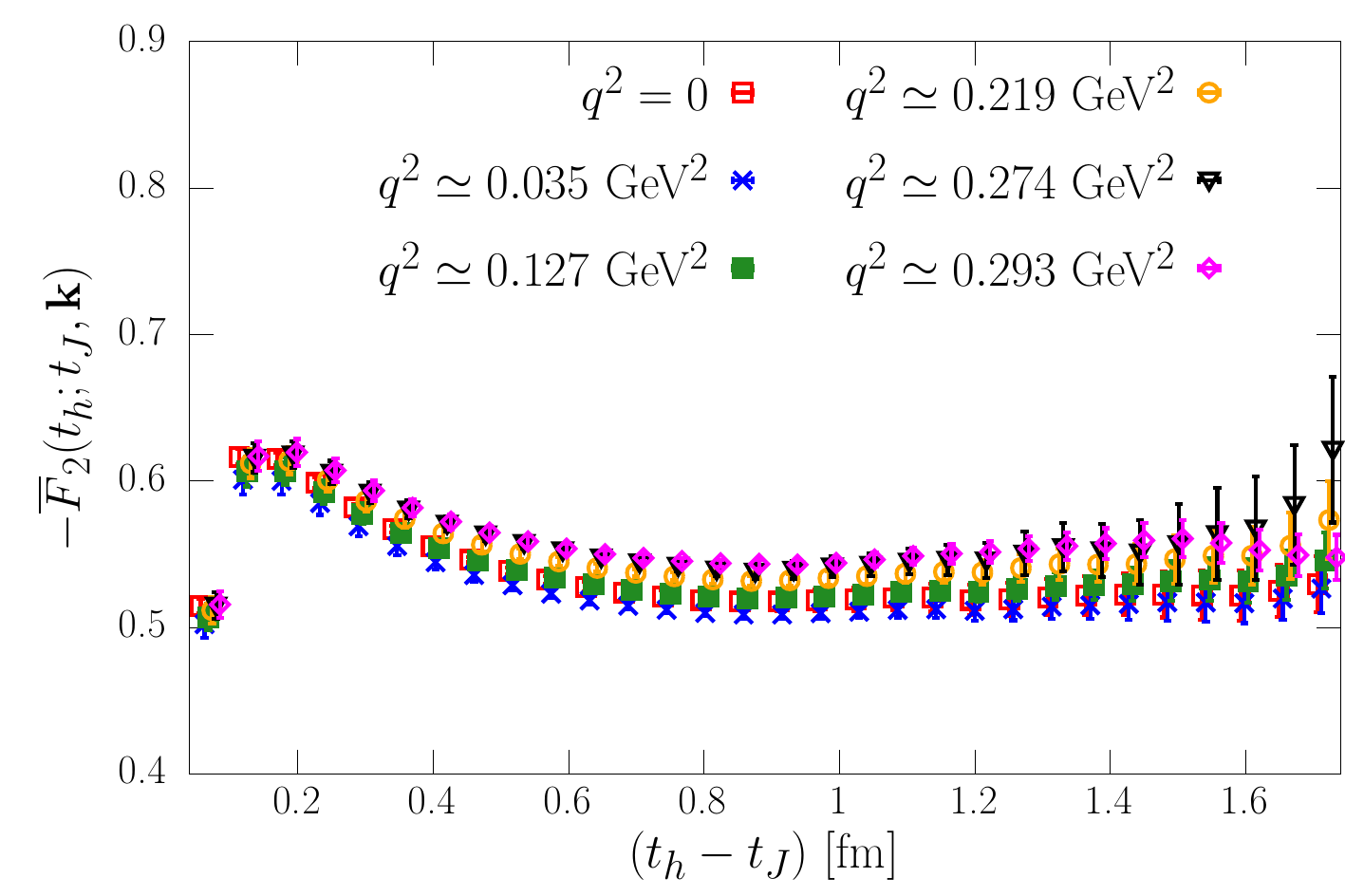}
\caption{\small The estimators $\overline{F}_1(t_h; t_J, \bs{k})$ and $ \overline{F}_2(t_h; t_J, \bs{k})$ for a fixed value of $t_J \simeq 1.6~\mathrm{fm}$ on the finest lattice spacing ensemble (D96). In each plot, the different curves correspond to the different simulated values of $q^2$. The data points corresponding to the different $q^2$ values have been slightly shifted horizontally for better visualization.}
\label{fig:hc}
\end{figure}

We recall that the two form factors $F_1(q^2)$ and $\tilde{F}_2(q^2)$ are extracted from the estimators defined in Eqs.~\eqref{eq:FF_estimator_hc}--\eqref{eq:F2_estimator_hc}, after applying the improvements described in Eqs.~\eqref{eq:impro_hc_1}--\eqref{eq:impro_hc_2}. In Fig.~\ref{fig:hc}, we show the improved estimators of the form factors $F_1(q^2)$ and $ \tilde{F}_2(q^2)$ on the finest lattice-spacing ensemble used in this study (the D96 ensemble), for a fixed value of the electromagnetic-current insertion time $t_J \simeq 1.6~\mathrm{fm}$. The figure shows clear and stable plateaus for both form factors at all kinematic points.

\begin{figure}
\centering
\includegraphics[width=0.49\linewidth]{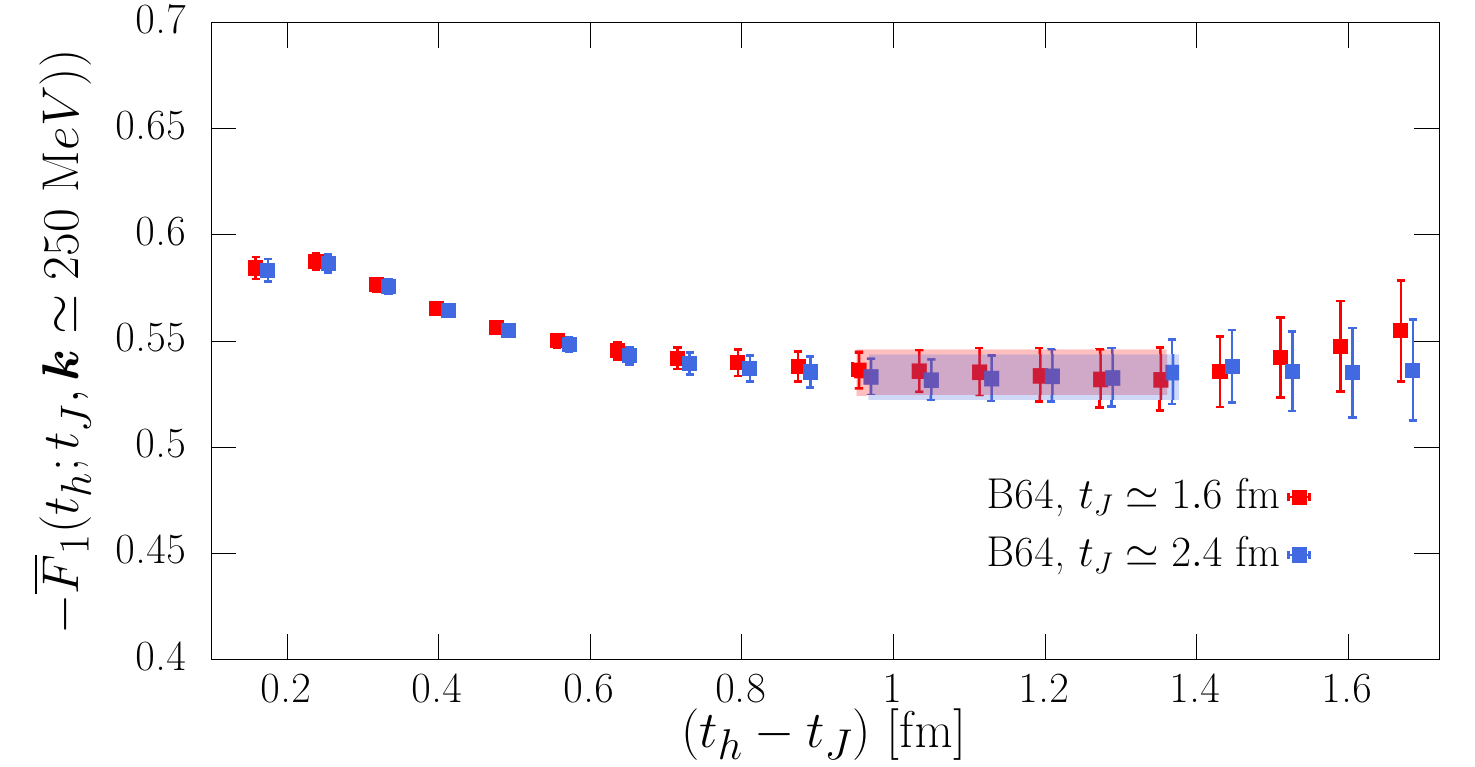}
\includegraphics[width=0.49\linewidth]{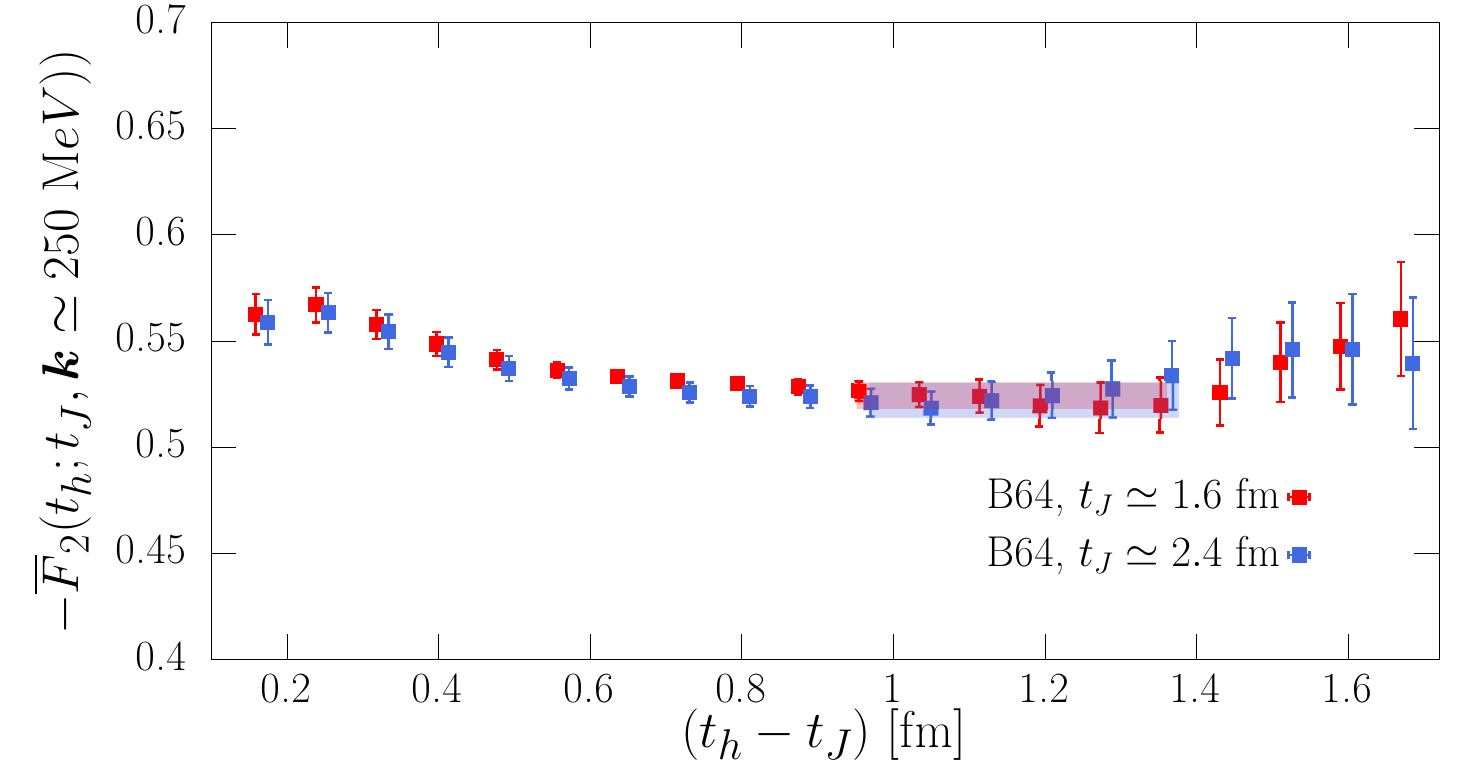}
\caption{\small The estimators $\overline{F}_1(t_h; t_J, \bs{k})$ and $ \overline{F}_2(t_h; t_J, \bs{k})$ are shown for two values of $t_J \simeq 1.6, 2.4~\mathrm{fm}$. The results correspond to our determination on the B64 ensemble for $k = 250~\mathrm{MeV}$, which corresponds to $q^2 \simeq 0.219~\mathrm{GeV}^2$. The coloured bands indicate the results of constant fits to the estimators over the selected time intervals, from which the form factors are determined.}
\label{fig:tj_dep_h}
\end{figure}

The $q^2$ dependence of the form factors $F_1(q^2)$ and $\tilde{F}_{2}(q^{2})$ is very mild, varying by about $5\%$ between $q^2 = 0$ and $q^2 = q^2_{\mathrm{max}} \simeq 0.293~\mathrm{GeV}^2$. We also investigated finite-$t_J$ effects by performing simulations at two values, $t_J \simeq 1.6~\mathrm{fm}$ and $t_J \simeq 2.4~\mathrm{fm}$, on a single ensemble (B64). The comparison, shown in Fig.~\ref{fig:tj_dep_h} for an intermediate $q^2 \simeq 0.219~\mathrm{GeV}^2$, reveals excellent agreement between the two choices. This indicates that $t_J \simeq 1.6~\mathrm{fm}$ is already sufficient to suppress finite-$t_J$ effects for both the $F_1$ and $\tilde{F}_2$ form factors at the current level of precision.

For all ensembles other than B64, the three-point correlation functions are computed at a single value $t_J \simeq 1.6~\mathrm{fm}$, as done for the $ \chi_{c1} \to J/\psi \ell^+ \ell^-$ analysis. The estimators $\overline{F}_1$ and $\overline{F}_2$ are fitted to a constant in the interval $t_h - t_J \in [1, 1.4]~\mathrm{fm}$, where a clear plateau is observed, in order to extract the form factors $F_1(q^2)$ and $ \tilde{F}_2(q^2)$.

For the extrapolation to the continuum limit, we proceed slightly differently compared to the previous subsection. To exploit the $q^2 = 0$ simulations performed in Ref.~\cite{Becirevic:2025ocx}, which included an ultrafine lattice-spacing ensemble with $a \simeq 0.049~\mathrm{fm}$ (the E112 ensemble), we first extrapolate to the continuum limit the ratios:
\begin{align}\label{eq:R12}
R_1(q^2) \equiv \frac{F_1(q^2)}{F_1(0)}, \qquad
R_2(q^2) \equiv \frac{\tilde{F}_2(q^2)}{F_1(0)},
\end{align}
which are, by definition, equal to unity at $q^2 = 0$ [recall that in the on-shell photon limit one has $F_1(0) = \tilde{F}_2(0)$]. The continuum extrapolation of these ratios is performed using the four gauge ensembles A48, B64, C80, and D96. We then combine the extrapolated ratios with our determination $F_1(0) = -0.522(10)$ from Ref.~\cite{Becirevic:2025ocx}, obtained from a continuum extrapolation including five lattice spacings (i.e., including the E112 ensemble), in order to reconstruct both  $F_1(q^2) $ and $\tilde{F}_2(q^2)$.

\begin{figure}
\centering
\includegraphics[width=0.47\linewidth]{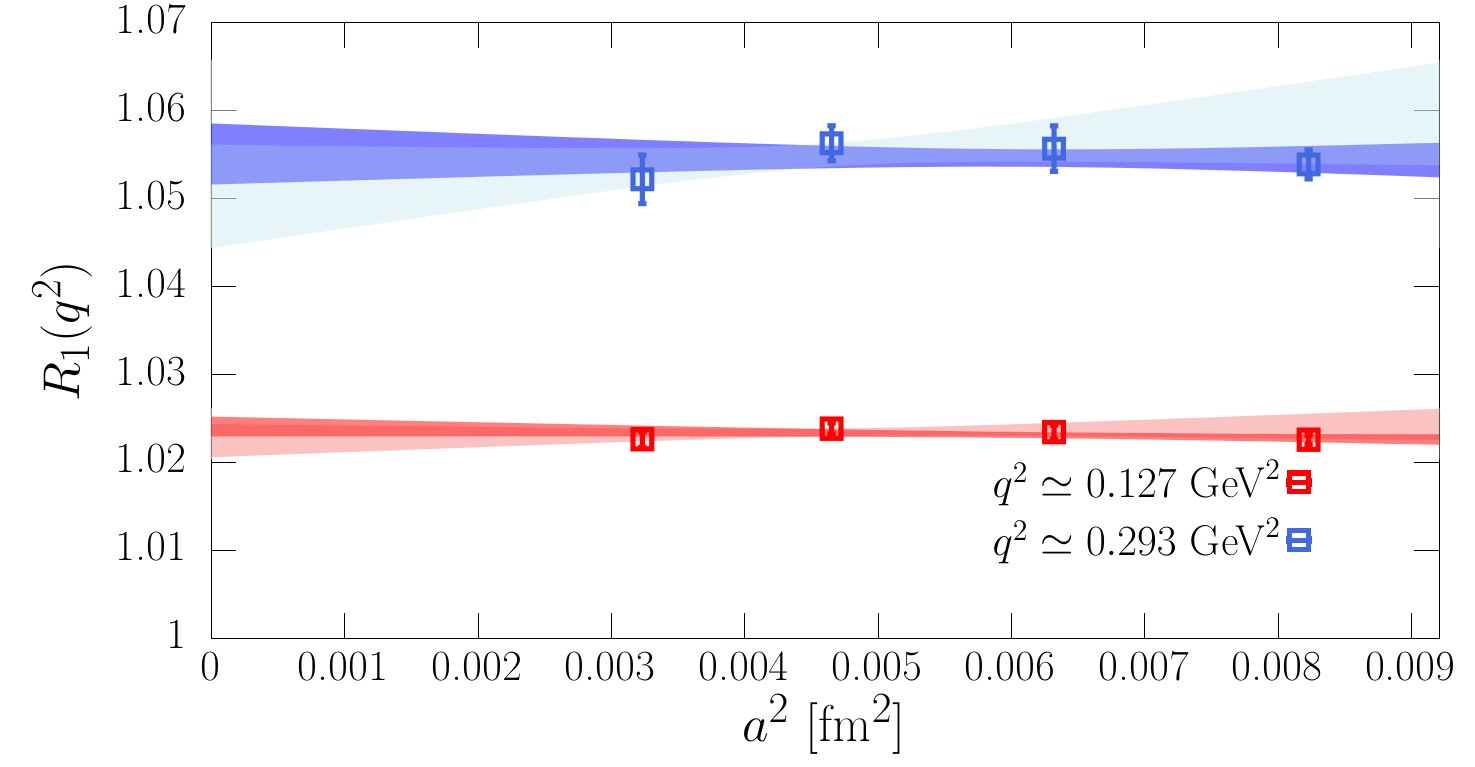}
\includegraphics[width=0.47\linewidth]{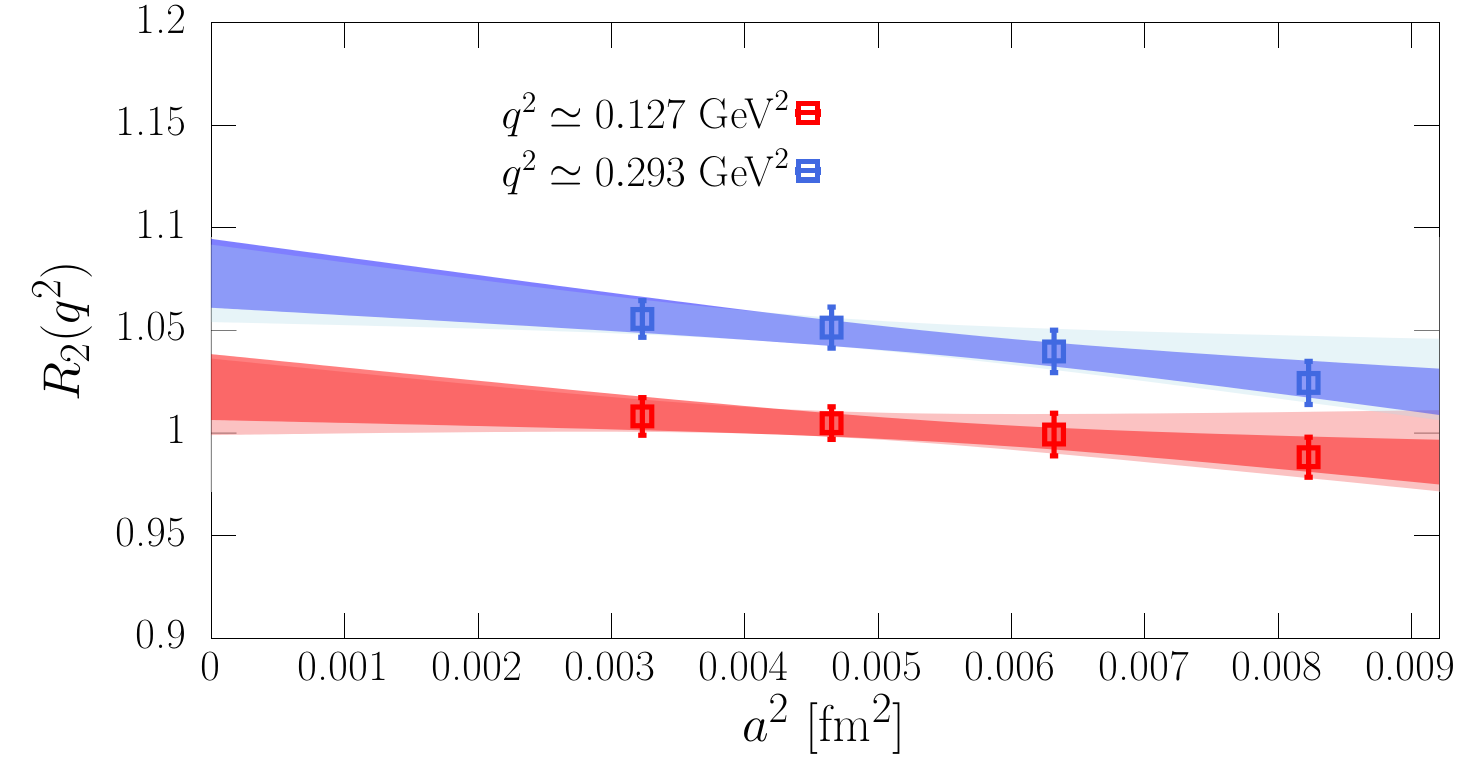}
\caption{\small Lattice spacing dependence of the ratios $R_1(q^2)$ and $ R_2(q^2)$ for two values of the momentum transfer corresponding to $q^2 \simeq 0.127~\mathrm{GeV}^2$ and $q^2 \simeq q^2_{\mathrm{max}} \simeq 0.293~\mathrm{GeV}^2$.}
\label{fig:cont_hc}
\end{figure}

In Fig.~\ref{fig:cont_hc}, we illustrate the ratios $R_1(q^2)$ and $R_2(q^2) $ as functions of the squared lattice spacing $a^2$ for an intermediate $q^2 = 0.127~\mathrm{GeV}^2$ and for a value close to $q^2_{\mathrm{max}} = 0.293~\mathrm{GeV}^2$. The lattice spacing dependence is very mild, and for the extrapolation of $R_1(q^2)$ and $R_2(q^2)$ to the continuum limit, we follow the same procedure as in the $\chi_{c1} \to J/\psi \ell^+ \ell^-$ case, i.e., we perform two linear fits in $a^2$: one using the full dataset and a second excluding the coarsest lattice spacing. The results of both fits are shown in Fig.~\ref{fig:cont_hc}, where the lighter colored bands correspond to fits excluding the coarsest lattice spacing.

\begin{figure}
\centering
\includegraphics[width=0.8\linewidth]{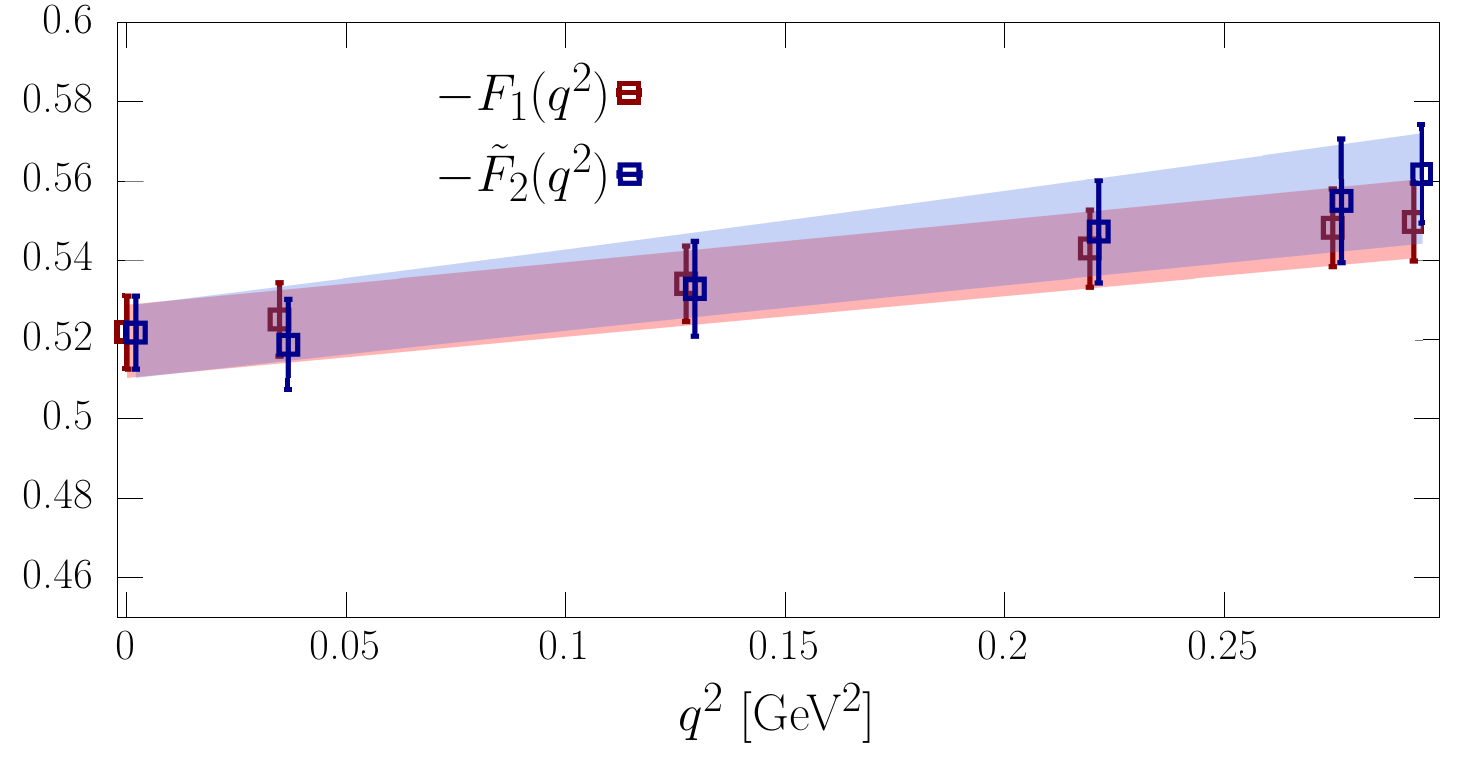}
\caption{\small The two form factors $F_1(q^2)$ and $\tilde{F}_2(q^2)$ describing the $h_c \to \eta_c \ell^+ \ell^-$ decay as a function of $q^2$ in the range $q^2 \in [0, q^2_{\mathrm{max}}]$, where $q^2_{\mathrm{max}} \simeq 0.293~\mathrm{GeV}^2$. The colored bands correspond to the results of the $q^2$ fits performed according to Eq.~\eqref{eq:q2_fit_ansatz_hc}.}
\label{fig:final_FF_hc}
\end{figure}

For all fits, the reduced $\chi^2$ values do not exceed $\chi^2/\mathrm{dof} = 1.1$. Our results for $R_1(q^2)$ and $R_2(q^2)$ are obtained by averaging the two fits using the Bayesian Akaike Information Criterion (BAIC). Our final results for $F_1(q^2)$ and $\tilde{F}_2(q^2)$ are shown in Fig.~\ref{fig:final_FF_hc}, where, as mentioned above, we use $F_1(0) = \tilde{F}_2(0) = -0.522(10)$ from Ref.~\cite{Becirevic:2025ocx} to determine $F_1(q^2)$ and $\tilde{F}_2(q^2)$ from the knowledge of the ratios $R_1(q^2) $ and $R_2(q^2)$. The continuum-extrapolated values of the ratios $R_{1}(q^{2})$ and $R_{2}(q^{2})$ are collected in Tab.~\ref{tab:hc_ratios}.

\begin{table}[t]
\begin{ruledtabular}
\begin{tabular}{lcc}
$q^2\,[\mathrm{GeV}^2]$ & $R_1(q^2)$ & $R_2(q^2)$ \\
\colrule
0.035 & 1.00635(41) & 0.994(15) \\
0.127 & 1.0235(15)  & 1.021(16) \\
0.219 & 1.0404(27)  & 1.049(18) \\
0.274 & 1.0504(38)  & 1.063(24) \\
0.293 & 1.0532(45)  & 1.076(16) 
\end{tabular}
\end{ruledtabular}
\caption{Continuum extrapolated values of the ratios $R_{1}(q^{2})$ and $R_{2}(q^{2})$, defined in Eq.~\eqref{eq:R12}, for all values of $q^{2}$ considered in this work.}
\label{tab:hc_ratios}
\end{table}

As in the case of the $\chi_{c1} \to J/\psi \ell^+ \ell^-$ form factors, having determined $F_1(q^2)$ and $\tilde{F}_2(q^2)$ at discrete $q^2$ values over the full kinematic range, we fit them using a simple polynomial in $q^2 $. We again use a linear ansatz in $q^2$ to describe the form factor $F_1(q^2)$ and $\tilde{F}_{2}(q^{2})$. We perform a joint fit given by:
\begin{align}
\label{eq:q2_fit_ansatz_hc}
F_1(q^2) &= C \times \left[ 1 + D_{F_1} q^2 \right], \nonumber\\
\tilde{F}_2(q^2) &= C \times \left[ 1 + D_{F_2} q^2  \right],
\end{align}
where $C$, $D_{F_1}$, and $D_{F_2}$ are free fit parameters. The result of the global fit is given by the colored bands in Fig.~\ref{fig:final_FF_hc}, while in Table~\ref{tab:coeff_hc} we provide the values of the fit parameters, including their correlations, which can be easily implemented in phenomenological analyses. As in the case of the $\chi_{c1}\to J/\psi \ell^{+}\ell^{-}$ form factors, the $q^2$ dependence is very mild, with $|D_{F_{1(2)}}q^2|\ll1$ over the full physical range. 

\begin{table}[t]
\centering
\begin{ruledtabular}
\begin{tabular}{c c|ccc}
\multirow{2}{*}{Parameter} & \multirow{2}{*}{Value} & \multicolumn{3}{c}{Correlation matrix} \\
\cline{3-5}
& & $C$ & $D_{F_1}$ & $D_{F_2}$  \\
\hline
$-C$ & $0.520 \pm 0.010$ & 1 & -0.13 & 0.10 \\
$D_{F_1} \, [\mathrm{GeV}^{-2}]$ & $0.202 \pm 0.022$ & -0.13 & 1 & -0.63  \\
$D_{F_2} \, [\mathrm{GeV}^{-2}]$ & $0.252 \pm 0.058$ & 0.10 & -0.63 & 1  \\
\end{tabular}
\end{ruledtabular}
\caption{Values of the fit parameters entering Eq.~\eqref{eq:q2_fit_ansatz_hc} and associated correlation matrix.}
\label{tab:coeff_hc}
\end{table}

\begin{figure}
\centering
\includegraphics[width=0.7\linewidth]{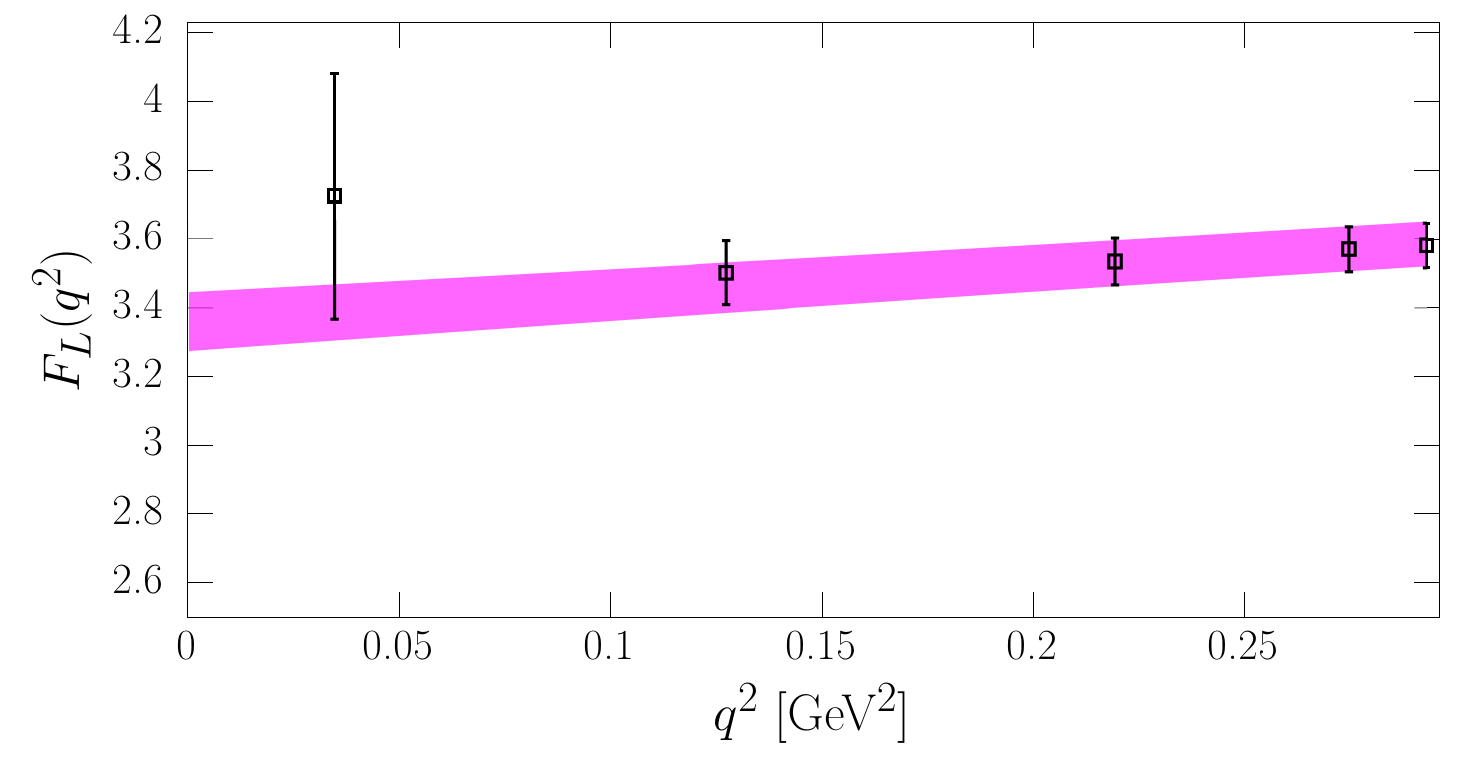}
\caption{\small The form factor $F_L(q^2)$, obtained from our data for $ F_1(q^2)$ and $\tilde{F}_2(q^2)$ through Eq.~\eqref{eq:FL}.}
\label{fig:FL}
\end{figure}

The results for the longitudinal form factor $F_{L}(q^{2})$,
introduced in Eq.~(\ref{eq:FL}), are shown in Fig.~\ref{fig:FL}.

\section{Comparison with Experimental Data}
\label{sec:comparison}
In this section, we compare our predictions for decay rates and differential distributions with the available experimental data and discuss observables with varying sensitivity to the underlying form factors. The existing measurements include extensive studies by the BESIII Collaboration of the $\chi_{c1} \to J/\psi e^{+} e^{-}$ and $\chi_{c1} \to J/\psi \mu^{+} \mu^{-}$ decays~\cite{BESIII:2017ung,BESIII:2019yeu}, as well as the recent observation of the $h_c \to \eta_c e^{+} e^{-}$ transition~\cite{BESIII:2024kkf}.

\subsection{$\chi_{c1} \to J/\psi \ell^{+} \ell^{-}$ Decay}

The decay rate for $\chi_{c1} \to J/\psi \ell^{+} \ell^{-}$, after summing over the $J/\psi$ polarizations and averaging over the $\chi_{c1}$ polarizations, can be expressed as:
\begin{align}
\Gamma(\chi_{c1} \to J/\psi \ell^{+} \ell^{-}) = \int_{4m_{\ell}^{2}}^{q_{\mathrm{max}}^{2}} dq^{2} \int_{-1}^{1} d\cos\theta \ \Gamma^{''\ell}_{\chi_{c1}}(q^{2}, \cos\theta),
\end{align}
where $\theta$ is the angle between $\ell^{+}$ and $\chi_{c1}$ in the dilepton rest frame, $q_{\mathrm{max}}^{2} = (m_{\chi_{c1}} - m_{J/\psi})^{2}$, and the doubly-differential decay rate is given by:
\begin{align}
\label{eq:double_diff}
\Gamma^{''\ell}_{\chi_{c1}}(q^{2}, \cos\theta) &= \frac{\alpha_{\mathrm{em}}^{2} Q_{c}^{2}}{48\pi} \frac{\beta_{\ell} \sqrt{\lambda(m_{\chi_{c1}}^{2}, m_{J/\psi}^{2}, q^{2})}}{q^{2} m_{\chi_{c1}}^{3}} \left[ A(q^{2}) + B(q^{2}) \cos^{2}\theta \right] \nonumber \\
&= \Gamma^{''\ell}_{\chi_{c1}, A}(q^{2}) + \Gamma^{''\ell}_{\chi_{c1}, B}(q^{2}) \cos^{2}\theta,
\end{align}
where $\lambda(x, y, z)$ is the Källén function, $\beta_{\ell} = \sqrt{1 - 4m_{\ell}^{2}/q^{2}}$, $m_{\ell}$ is the lepton mass, and the functions $A(q^{2})$ and $B(q^{2})$ are defined as:
\begin{align}
\label{eq:A_and_B}
A(q^{2}) &= q^{2} |C_{1}(q^{2})|^{2} + m_{\chi_{c1}}^{2} \left(1 + \frac{4m_{\ell}^{2}}{q^{2}}\right) \left(|E_{1}(q^{2})|^{2} + |M_{2}(q^{2})|^{2}\right), \nonumber \\
B(q^{2}) &= -\beta_{\ell}^{2} \left[ q^{2} |C_{1}(q^{2})|^{2} - m_{\chi_{c1}}^{2} \left(|E_{1}(q^{2})|^{2} + |M_{2}(q^{2})|^{2}\right) \right].
\end{align}

The $q^{2}$-differential decay rate is then given by:
\begin{align}
\label{eq:single_diff}
\Gamma'^{\ell}_{\chi_{c1}}(q^{2}) \equiv \int_{-1}^{1} d\cos\theta \ \Gamma^{''\ell}_{\chi_{c1}}(q^{2}, \cos\theta) = \frac{\alpha_{\mathrm{em}}^{2} Q_{c}^{2}}{48\pi} \frac{\beta_{\ell} \sqrt{\lambda(m_{\chi_{c1}}^{2}, m_{J/\psi}^{2}, q^{2})}}{q^{2} m_{\chi_{c1}}^{3}} \left[ 2A(q^{2}) + \frac{2}{3} B(q^{2}) \right].
\end{align}

\begin{figure}
\centering
\includegraphics[width=0.7\linewidth]{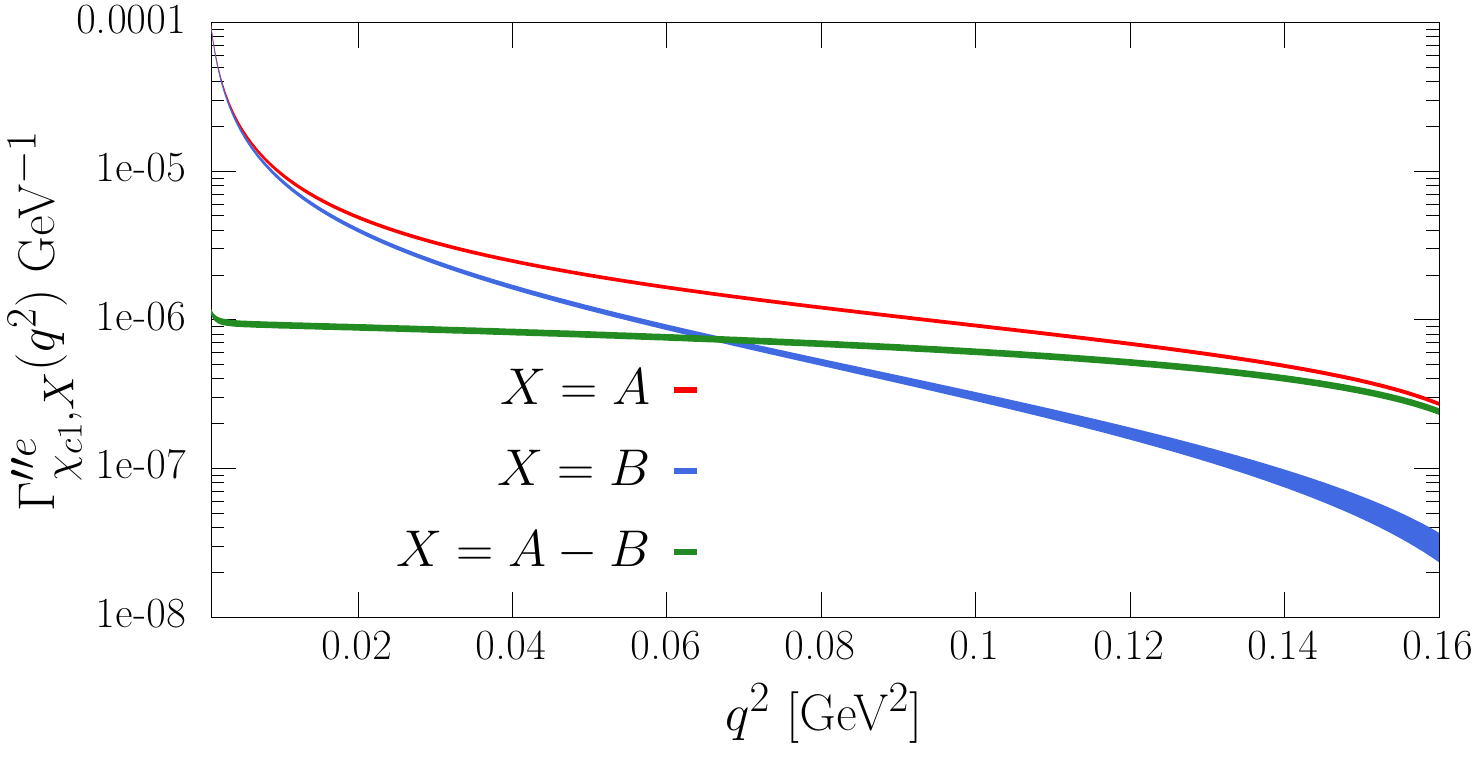}
\caption{Our results for the $A(q^{2})$ and $B(q^{2})$ contributions to $\Gamma^{''\ell}_{\chi_{c1}}(q^{2}, \cos\theta)$ of the $\chi_{c1} \to J/\psi e^{+} e^{-}$ decay, as defined in Eq.~\eqref{eq:double_diff}. We also show the behavior of $\Gamma^{''\ell}_{\chi_{c1}, A-B}(q^{2})$, defined in Eq.~\eqref{eq:angular_obs}, which is primarily sensitive to the longitudinal form factor $C_{1}(q^{2})$.}
\label{fig:double_diff}
\end{figure}

In Fig.~\ref{fig:double_diff}, we show our determination of the $A(q^{2})$ and $B(q^{2})$ contributions to $\Gamma^{''\ell}_{\chi_{c1}}(q^{2}, \cos\theta)$ for the electron ($\ell = e$) mode, as defined in Eq.~\eqref{eq:double_diff}.

In Fig.~\ref{fig:double_diff}, we also show the result for the following angular observable:
\begin{align}
\label{eq:angular_obs}
\Gamma^{''\ell}_{\chi_{c1}, A-B}(q^{2}) \equiv \Gamma^{''\ell}_{\chi_{c1}, A}(q^{2}) - \Gamma^{''\ell}_{\chi_{c1}, B}(q^{2}),
\end{align}
which, in the massless limit, is only sensitive to the longitudinal form factor $C_{1}(q^{2})$, since:
\begin{align}
A(q^{2}) - B(q^{2}) = 2q^{2} |C_{1}(q^{2})|^{2} + \mathcal{O}(m_{\ell}^{2}).
\end{align}

Since both $A(q^{2})$ and $B(q^{2})$ can, in principle, be extracted experimentally from the angular analysis of the decay products, measuring the quantity in Eq.~\eqref{eq:angular_obs} for final-state electrons allows for a direct measurement of the longitudinal form factor $C_{1}(q^{2})$, which does not contribute in the case of real photon emission.

\begin{figure}
\centering
\includegraphics[width=0.7\linewidth]{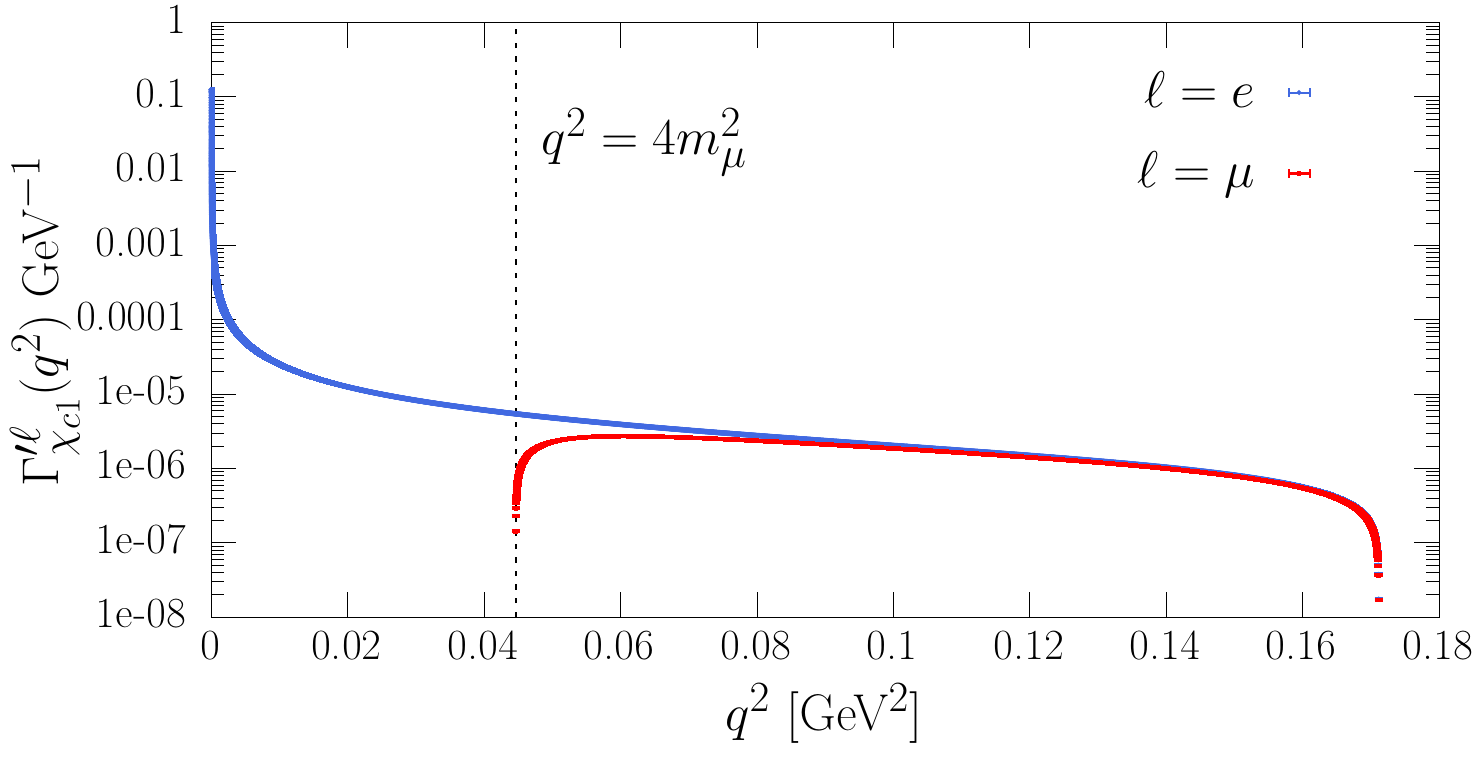}
\caption{Our results for the $q^{2}$-differential decay rates $\Gamma'^{\ell}_{\chi_{c1}}(q^{2})$ for both the electron mode (blue band) and muon mode (red band).}
\label{fig:single_diff}
\end{figure}

In Fig.~\ref{fig:single_diff}, we show our determination of the $q^{2}$-differential decay rate in Eq.~\eqref{eq:single_diff} for both the electron ($\ell = e$) and muon ($\ell = \mu$) modes. All the differential rates, as expected, peak at $q^{2} \gtrsim 4m_{\ell}^{2}$ and decrease rapidly as $q^{2}$ approaches $q_{\mathrm{max}}^{2}$. After integrating the differential rates over $q^{2}$, we obtain the following values for the total decay rates:
\begin{align}
\label{eq:final_res_width_chi}
\Gamma(\chi_{c1} \to J/\psi e^{+} e^{-}) &= 2.869(90)~{\rm keV}, \nonumber \\
\Gamma(\chi_{c1} \to J/\psi \mu^{+} \mu^{-}) &= 0.1993(72)~{\rm keV},
\end{align}
achieving a precision of about $3.1\%$ and $3.6\%$ for the electron and muon modes, respectively.

To compare with experiment, we convert the PDG results for the branching fractions~\cite{PDG2024}:
\begin{align}
\label{eq:PDG_branchings}
\mathrm{Br}^{\mathrm{PDG}}(\chi_{c1} \to J/\psi e^{+} e^{-}) &= (3.46 \pm 0.24) \times 10^{-3},\nonumber\\
\mathrm{Br}^{\mathrm{PDG}}(\chi_{c1} \to J/\psi \mu^{+} \mu^{-}) &= (2.33 \pm 0.29) \times 10^{-4},
\end{align}
to decay widths using $\Gamma(\chi_{c1}) = 0.84(4)~{\rm MeV}$~\cite{PDG2024}.

\begin{figure}
\centering
\includegraphics[width=0.35\linewidth]{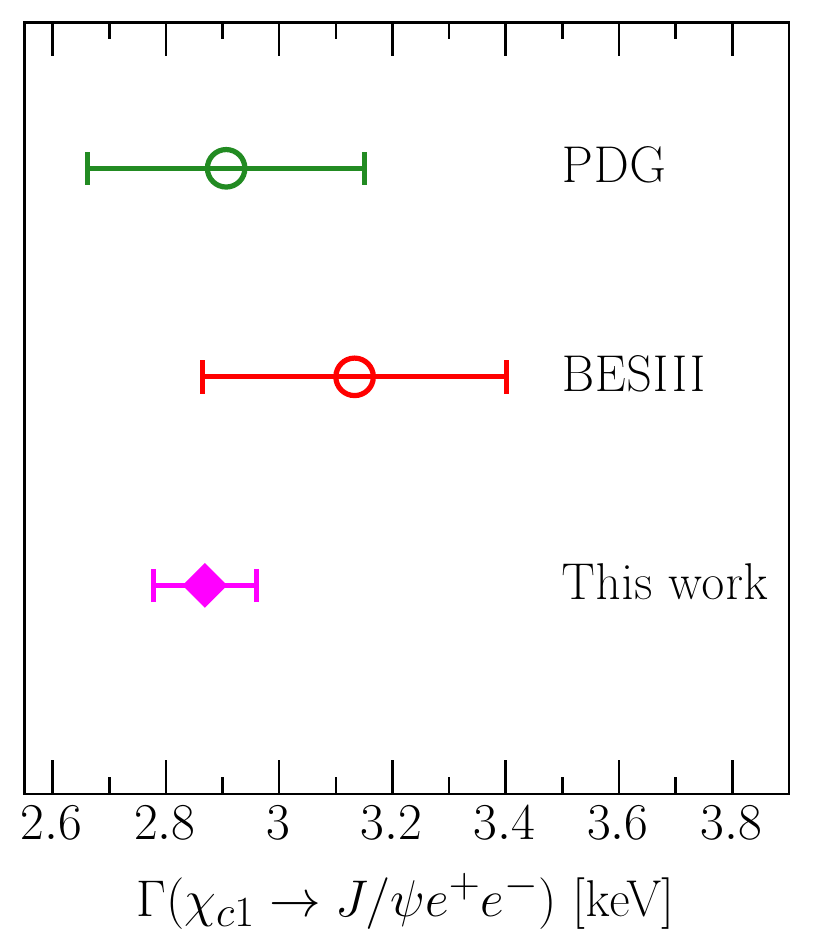}\hspace{2cm}
\includegraphics[width=0.35\linewidth]{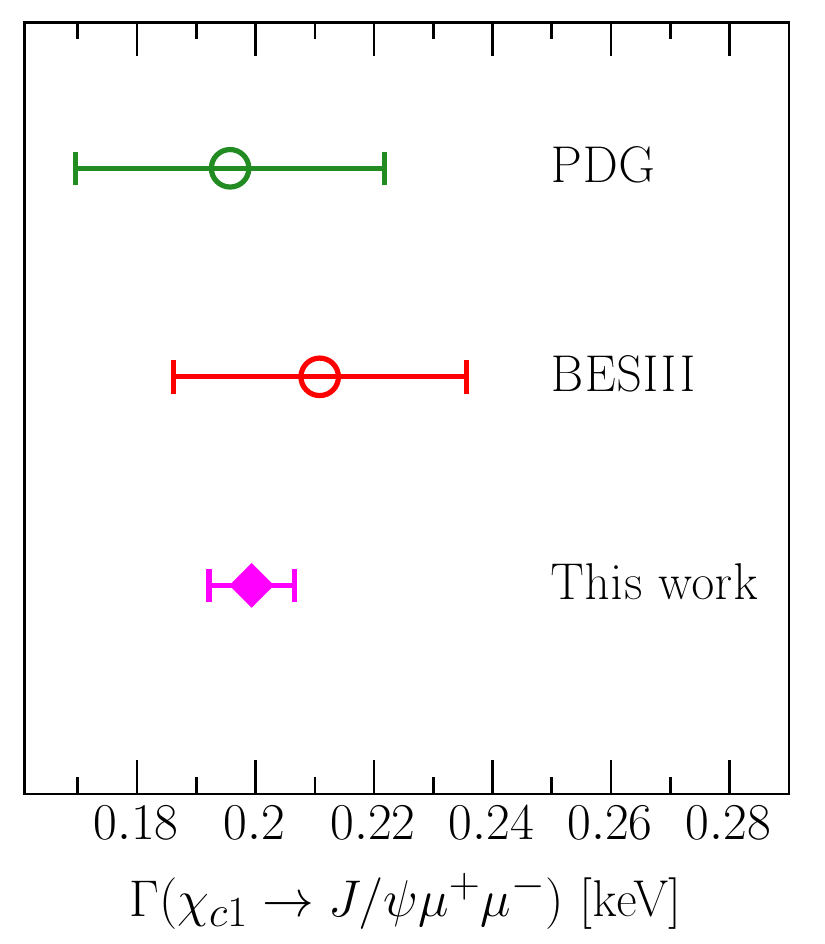}
\caption{Comparison between our lattice QCD predictions for the decay rates $\Gamma(\chi_{c1} \to J/\psi e^{+} e^{-})$ (left) and $\Gamma(\chi_{c1} \to J/\psi \mu^{+} \mu^{-})$ (right) and the results quoted by the PDG~\cite{PDG2024} and by BESIII~\cite{BESIII:2017ung,BESIII:2019yeu}.}
\label{fig:comparison_PDG}
\end{figure}

Our results are in excellent agreement with, and significantly more precise than, the experimental results in both channels, as shown in Fig.~\ref{fig:comparison_PDG}. A similar comparison with the direct experimental results quoted by BESIII~\cite{BESIII:2017ung,BESIII:2019yeu},
\begin{align}
\label{eq:bes_branchings}
\mathrm{Br}^{\mathrm{BESIII}}(\chi_{c1} \to J/\psi e^{+} e^{-}) &= (3.73 \pm 0.09 \pm 0.25) \times 10^{-3}, \nonumber \\
\mathrm{Br}^{\mathrm{BESIII}}(\chi_{c1} \to J/\psi \mu^{+} \mu^{-}) &= (2.51 \pm 0.18 \pm 0.20) \times 10^{-4}~,
\end{align}
leads to results that  are still compatible with our Standard Model predictions within one standard deviation (see Fig.~\ref{fig:comparison_PDG}).

Alternatively, after dividing our lattice predictions for the partial decay
widths in Eq.~\eqref{eq:final_res_width_chi} by the experimentally
measured branching fractions, we extract the total decay
width of the $\chi_{c1}$. We obtain:
\begin{align}
\label{eq:62}
\Gamma(\chi_{c1})_{e^{+} e^{-}} &= 0.77(6)~{\rm MeV} \ [0.83(6)~{\rm MeV}], \nonumber \\
\Gamma(\chi_{c1})_{\mu^{+} \mu^{-}} &= 0.79(9)~{\rm MeV} \ [0.86(11)~{\rm MeV}],
\end{align}
where the values in square brackets are obtained using the PDG branching fractions (Eq.~\ref{eq:PDG_branchings}), while the others are obtained using the BESIII results (Eq.~\ref{eq:bes_branchings}).

A more direct comparison with the BESIII measurements, which avoids relying on the PDG value of the total decay width of $\chi_{c1}$, can be made by comparing our results for:
\begin{align}
\label{eq:lqcd_D_2}
\frac{\mathrm{Br}(\chi_{c1} \to J/\psi \mu^{+} \mu^{-})}{\mathrm{Br}(\chi_{c1} \to J/\psi e^{+} e^{-})} &= 6.95(10) \times 10^{-2}, \\
\label{eq:lqcd_D_1}
\frac{\mathrm{Br}(\chi_{c1} \to J/\psi e^{+} e^{-})}{\mathrm{Br}(\chi_{c1} \to J/\psi \gamma)} &= 8.78(2) \times 10^{-3}.
\end{align}
While the lepton-flavor-universality ratio is in excellent agreement with the experimental result given in Eq.~\eqref{eq:bes_D_2}, our result in Eq.~\eqref{eq:lqcd_D_1} is smaller by about $2.3\sigma$ than the BESIII determination in Eq.~\eqref{eq:bes_D_1}.

A more detailed comparison between our predictions and the experimental data can be made for the $q^{2}$-differential decay rate. BESIII provided such results for the electron channel in Ref.~\cite{BESIII:2017ung}, where the number of events was measured from threshold up to $\sqrt{q^{2}} \simeq 0.3675~{\rm GeV}$ using short bins of width $5~{\rm MeV}$.

\begin{figure}
\centering
\includegraphics[width=0.75\linewidth]{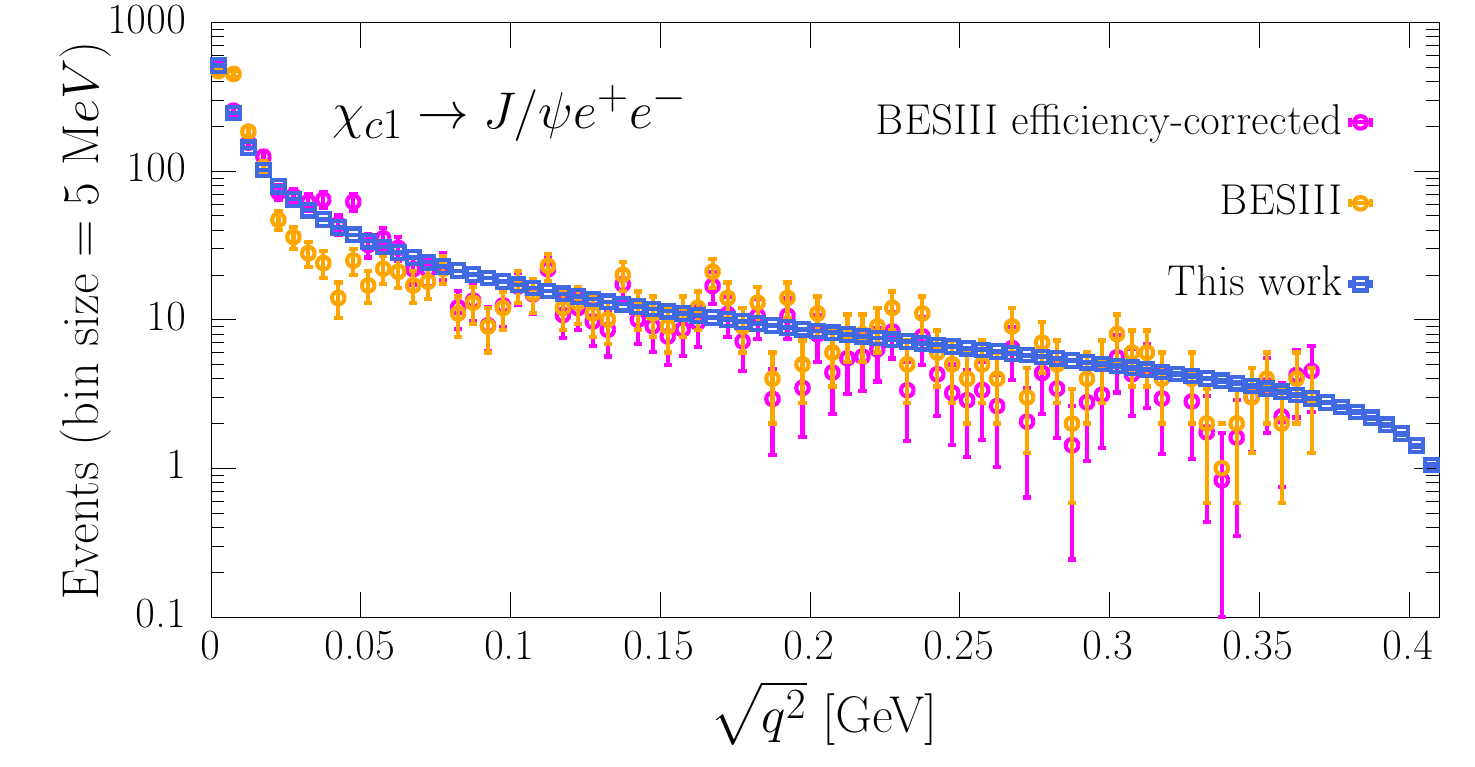}
\caption{Number of bin-per-bin events measured by BESIII in Ref.~\cite{BESIII:2017ung} before (orange data points) and after (magenta data points) efficiency corrections. Our corresponding theoretical predictions are shown by the blue data points. Each bin has a size of $5~{\rm MeV}$.}
\label{fig:BESIII_comp_diff}
\end{figure}

We first integrate our differential decay rate over the same bins, divide our results by the total decay width in Eq.~\eqref{eq:final_res_width_chi}, and then multiply the resulting values by the total number of events recorded by BESIII ($N_{\mathrm{events}} = 1980$). The comparison is shown in Fig.~\ref{fig:BESIII_comp_diff}, where the orange data points correspond to the BESIII measurements and the blue ones to our theoretical predictions.

An important aspect of this comparison is that the bin-by-bin event yields reported in Ref.~\cite{BESIII:2017ung} are obtained after applying selection criteria. Therefore, a meaningful comparison requires efficiency corrections. The authors of Ref.~\cite{BESIII:2017ung} kindly provided us with the corresponding correction factors for each bin. The efficiency-corrected event yields are shown in Fig.~\ref{fig:BESIII_comp_diff} as magenta data points.

As illustrated in the figure, the efficiency correction is essential to achieve agreement in the small-$q^{2}$ region, where a large discrepancy between lattice QCD predictions and experimental data would otherwise be observed. After applying the efficiency corrections, the BESIII data are found to be in reasonably good agreement with our lattice QCD results, although some differences remain in individual bins.

\begin{figure}
\centering
\includegraphics[width=0.75\linewidth]{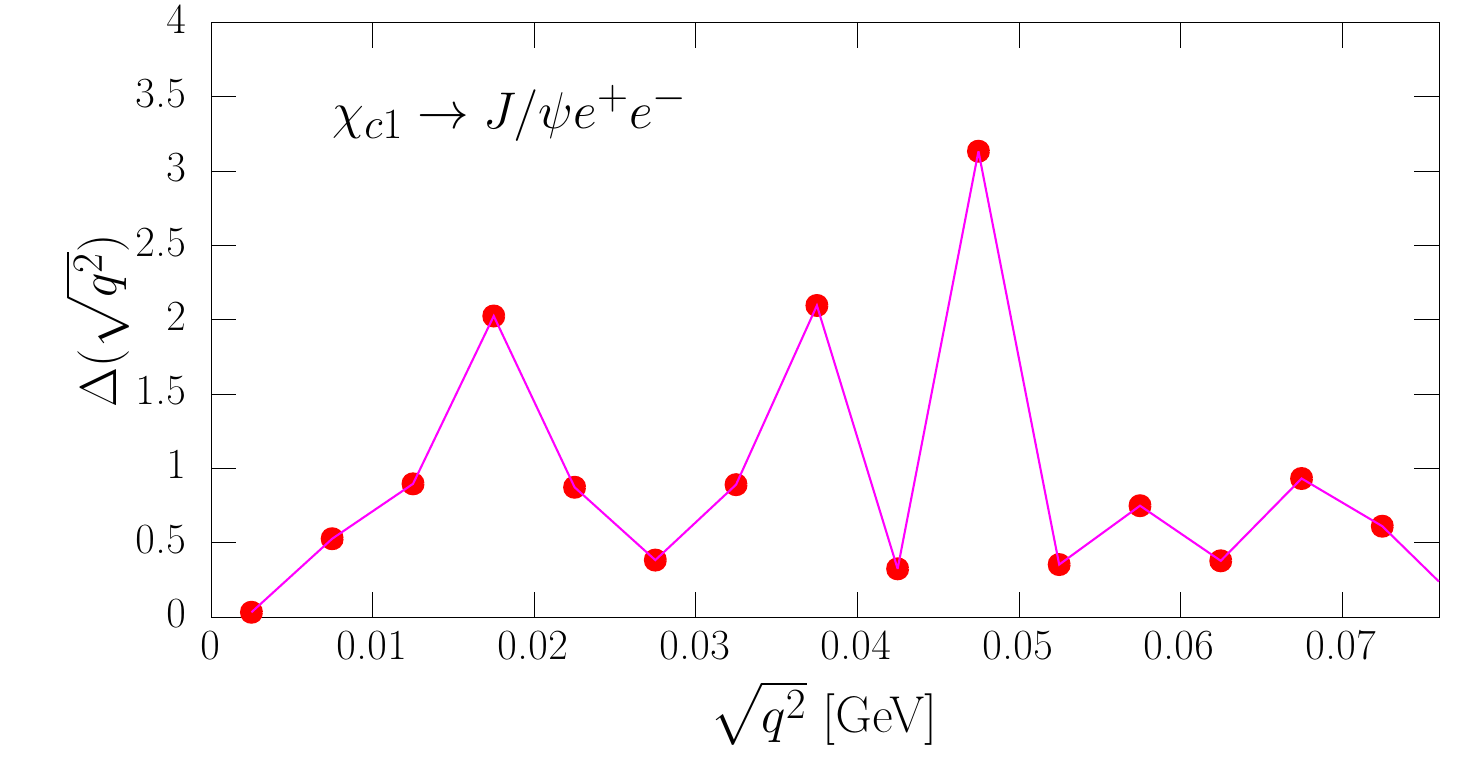}
\caption{Relative difference $\Delta(\sqrt{q^{2}})$ in Eq.~\eqref{eq:delta} in the low-$q^{2}$ region up to $\sqrt{q^{2}} \simeq 0.075~{\rm GeV}$.}
\label{fig:BESIII_comp_spread}
\end{figure}

In Fig.~\ref{fig:BESIII_comp_spread}, we display the bin-by-bin relative difference between the BESIII measurements and our lattice QCD predictions in the small-$q^{2}$ region, where the event statistics are higher and the uncertainties are more reliable. In practice, we restrict the comparison to bins containing more than 20 events. The pull $\Delta(\sqrt{q^{2}})$ is defined as:
\begin{align}
\label{eq:delta}
\Delta(\sqrt{q^{2}}) = \frac{\left| N_{\mathrm{ev}}^{\mathrm{BESIII}}(\sqrt{q^{2}}) - N_{\mathrm{ev}}^{\mathrm{LQCD}}(\sqrt{q^{2}}) \right|}{\sqrt{\left(\sigma^{\mathrm{BESIII}}(\sqrt{q^{2}})\right)^{2} + \left(\sigma^{\mathrm{LQCD}}(\sqrt{q^{2}})\right)^{2}}},
\end{align}
where $N_{\mathrm{ev}}^{\mathrm{BESIII}}(\sqrt{q^{2}})$ denotes the efficiency-corrected number of events measured by BESIII in a given $\sqrt{q^{2}}$ bin, and $N_{\mathrm{ev}}^{\mathrm{LQCD}}(\sqrt{q^{2}})$ the corresponding lattice QCD prediction. The quantities $\sigma^{\mathrm{BESIII}}(\sqrt{q^{2}})$ and $\sigma^{\mathrm{LQCD}}(\sqrt{q^{2}})$ represent the associated uncertainties.

As the figure shows, there is overall agreement between our lattice QCD predictions and the experimental results, with the exception of the bins $\sqrt{q^{2}} \simeq 17~{\rm MeV}$, $37~{\rm MeV}$, and $47~{\rm MeV}$. In these bins, we observe a tension of about $2.2\sigma$ at $17$ and $37~{\rm MeV}$, and approximately $3.2\sigma$ at $47~{\rm MeV}$. Although no statistically significant deviation from the Standard Model is observed, the excess in the $17~\mathrm{MeV}$ bin is qualitatively compatible with previously reported hints from the Atomki~\cite{Krasznahorkay:2015iga} and Padme~\cite{PADME:2025dla} experiments, which should be further experimentally scrutinized. A similar comparison cannot be performed using the more recent BESIII data reported in Ref.~\cite{BESIII:2025otp}, which provide a precise measurement of the $\chi_{c1} \to J/\psi e^{+} e^{-}$ decay width in the small-$q^{2}$ region. The published event yields correspond to data after the selection criteria have been applied, but the corresponding efficiency corrections were not available to us.

\subsection{$h_c \to \eta_c \ell^+ \ell^-$ Decay}

Similarly to the case of the $ \chi_{c1} \to J/\psi \ell^+ \ell^- $ decay discussed in the previous section, the decay rate for $h_c \to \eta_c \ell^+ \ell^-$, after averaging over the $h_c$ polarizations, can be written as:
\begin{align}
\Gamma(h_c \to \eta_c \ell^+ \ell^-) = \int_{4m_{\ell}^{2}}^{q_{\mathrm{max}}^{2}} dq^{2} \int_{-1}^{1} d\cos\theta \ \Gamma^{''\ell}_{h_c}(q^{2}, \cos\theta),
\end{align}
where $q_{\mathrm{max}}^{2} = (m_{h_c} - m_{\eta_c})^{2}$, and $ \theta$ is the angle between $\ell^+$ and the decaying $h_c$ in the dilepton rest frame. The doubly-differential decay rate is given by:
\begin{align}
\label{eq:double_diff_hc}
\Gamma^{''\ell}_{h_c}(q^{2}, \cos\theta) &= \frac{\alpha_{\mathrm{em}}^{2} Q_{c}^{2}}{12\pi} \frac{\beta_{\ell} \sqrt{\lambda(m_{h_c}^{2}, m_{\eta_c}^{2}, q^{2})}}{q^{2} m_{h_c}^{3}} \left[ A(q^{2}) + B(q^{2}) \cos^{2}\theta \right] \nonumber \\
&= \Gamma^{''\ell}_{h_c, A}(q^{2}) + \Gamma^{''\ell}_{h_c, B}(q^{2}) \cos^{2}\theta,
\end{align}
where $A(q^{2})$ and $ B(q^{2})$ are defined similarly to Eq.~\eqref{eq:A_and_B}, but with the replacements:
\begin{align}
|C_{1}(q^{2})|^{2} \to |F_{L}(q^{2})|^{2}, \qquad |E_{1}(q^{2})|^{2} + |M_{2}(q^{2})|^{2} \to |F_{1}(q^{2})|^{2}.
\end{align}

The $q^{2}$-differential decay rate is then given by:
\begin{align}
\label{eq:single_diff_hc}
\Gamma'^{\ell}_{h_c}(q^{2}) \equiv \int_{-1}^{1} d\cos\theta \ \Gamma^{''\ell}_{h_c}(q^{2}, \cos\theta) = \frac{\alpha_{\mathrm{em}}^{2} Q_{c}^{2}}{12\pi} \frac{\beta_{\ell} \sqrt{\lambda(m_{h_c}^{2}, m_{\eta_c}^{2}, q^{2})}}{q^{2} m_{h_c}^{3}} \left[ 2A(q^{2}) + \frac{2}{3} B(q^{2}) \right].
\end{align}

We define the following angular observable, analogous to the $ \chi_{c1} \to J/\psi \ell^+ \ell^-$ case:
\begin{align}
\label{eq:angular_obs_hc}
\Gamma^{''\ell}_{h_c, A-B}(q^{2}) \equiv \Gamma^{''\ell}_{h_c, A}(q^{2}) - \Gamma^{''\ell}_{h_c, B}(q^{2}),
\end{align}
which, in the massless limit, is only sensitive to the longitudinal form factor $F_L$, since:
\begin{align}
A(q^{2}) - B(q^{2}) = 2q^{2} |F_L(q^{2})|^{2} + \mathcal{O}(m_{\ell}^{2}).
\end{align}

\begin{figure}
\centering
\includegraphics[width=0.7\linewidth]{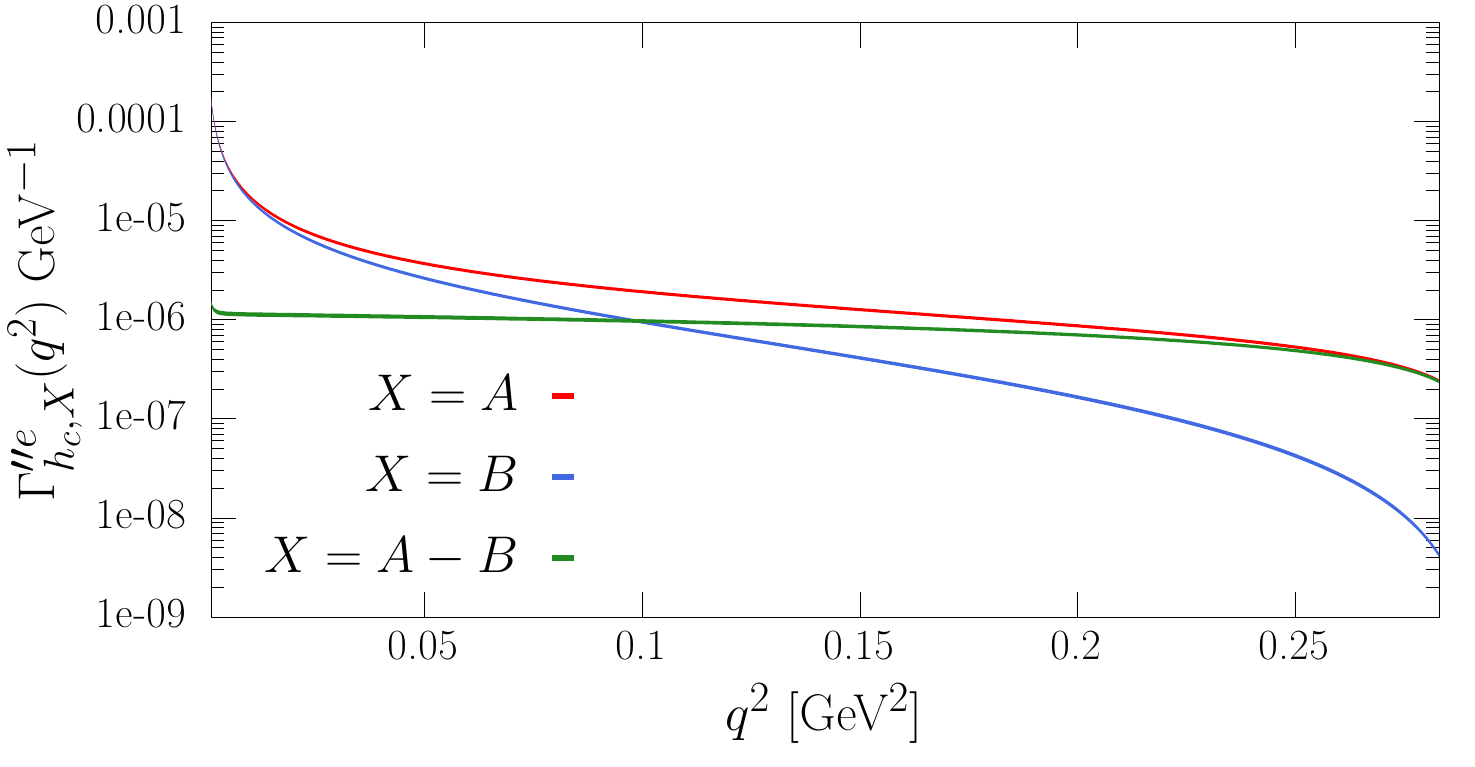}
\caption{Our results for the $A(q^{2})$ and $B(q^{2})$ contributions to $\Gamma^{''\ell}_{h_c}(q^{2}, \cos\theta)$ in Eq.~\eqref{eq:double_diff_hc}, as well as the results for the angular observable $\Gamma^{''\ell}_{h_c, A-B}(q^{2})$ in Eq.~\eqref{eq:angular_obs_hc}, which is primarily sensitive to the longitudinal form factor $F_L$.}
\label{fig:double_diff_hc}
\end{figure}

\begin{figure}
\centering
\includegraphics[width=0.7\linewidth]{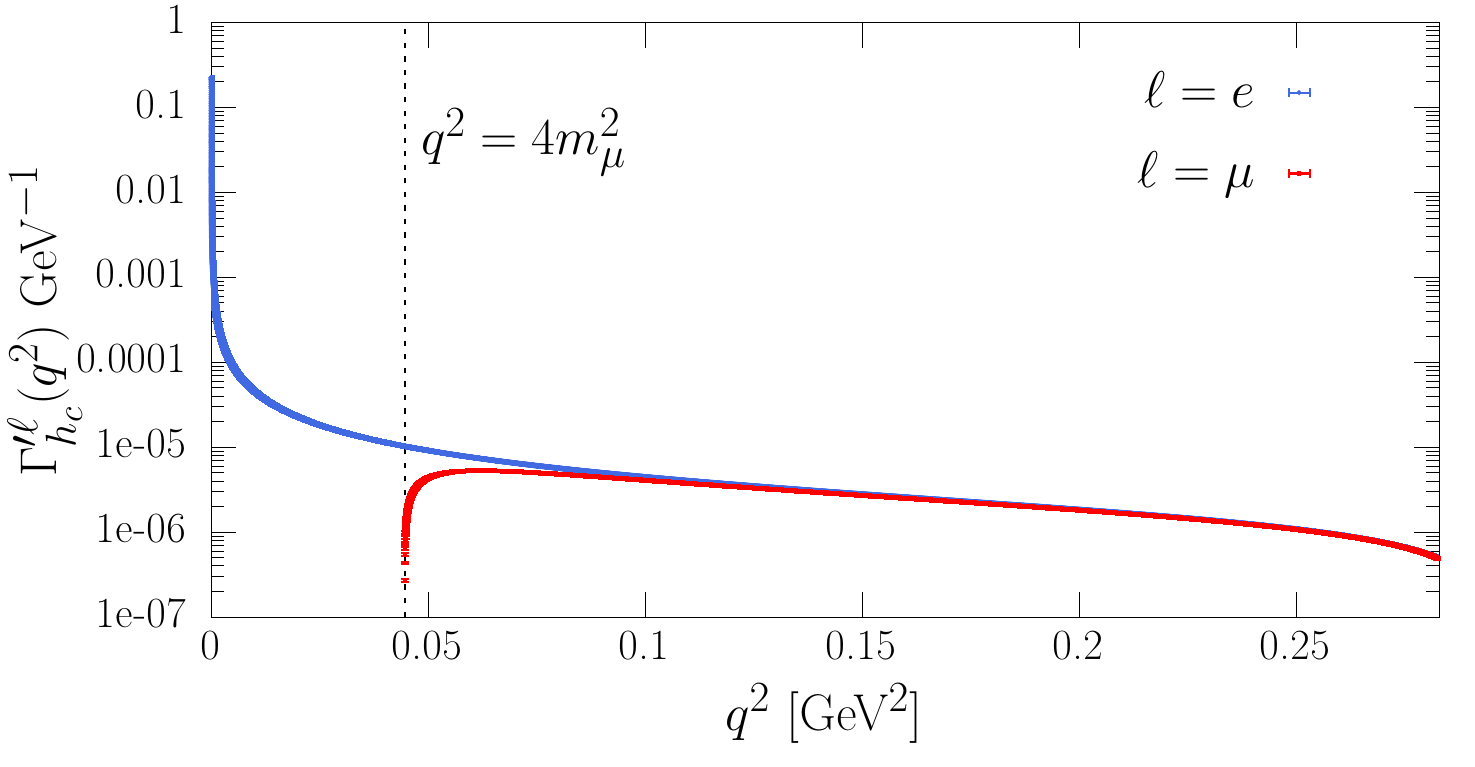}
\caption{Our results for the $q^{2}$-differential decay rates $ \Gamma'^{\ell}_{h_c}(q^{2})$ for both the electron (red band) and muon modes (blue band).}
\label{fig:single_diff_hc}
\end{figure}

In Fig.~\ref{fig:double_diff_hc}, we show our determination of $ \Gamma'^{\ell}_{h_c, A}(q^{2})$, $ \Gamma'^{\ell}_{h_c, B}(q^{2})$, and $ \Gamma'^{\ell}_{h_c, A-B}(q^{2})$ for the electron mode. In Fig.~\ref{fig:single_diff_hc}, we show the $q^{2}$-differential decay rate $\Gamma'^{\ell}_{h_c}(q^{2})$ for both the electron and muon channels ($\ell = e, \mu $).

After integrating $\Gamma'^{\ell}_{h_c}(q^{2})$, we obtain:
\begin{align}
\label{eq:final_res_width_hc}
\Gamma(h_c \to \eta_c e^+ e^-) &= 5.45(19)~{\rm keV}, \nonumber \\
\Gamma(h_c \to \eta_c \mu^+ \mu^-) &= 0.635(22)~{\rm keV},
\end{align}
achieving a precision similar to that obtained in the case of $ \chi_{c1} \to J/\psi \ell^+ \ell^-$, namely about $3.1\%$ for the electron mode and about $3.5\%$ for the muon mode.

To compare our result for the electron channel with the experimentally measured branching fraction~\cite{PDG2024}:
\begin{align}
\mathrm{Br}^{\mathrm{PDG}}(h_c \to \eta_c e^+ e^-) = (3.5 \pm 0.7) \times 10^{-3},
\end{align}
we use the total width of the $h_c$ quoted by PDG, $\Gamma(h_c) = 0.78(28)~{\rm MeV}$~\cite{PDG2024}, and obtain:
\begin{align}
\Gamma^{\mathrm{PDG}}(h_c \to \eta_c e^+ e^-) = 2.7(1.1)~{\rm keV},
\end{align}
which is in tension with our result in Eq.~\eqref{eq:final_res_width_hc} by about $2.5$ standard deviations.

We arrive at a similar tension by directly comparing our result with the BESIII result for the ratio~\cite{BESIII:2024kkf}:~\footnote{Note that the experimental result quoted in Ref.~\cite{BESIII:2024kkf}, also given in Eq.~\eqref{eq:rel_ph_hc}, is a weighted average of two different modes to produce $h_c$. In the first it comes from $\Psi(2S)\to h_c\pi^0$ and the ratio is measured $0.46(12)(5)\%$, while in the second $h_c$ is produced by $e^+e^-\to h_c \pi^+\pi^-$, and the ratio is $0.89(18)(9)\times 10^{-2}$.}
\begin{align}
\label{eq:rel_ph_hc}
\frac{\mathrm{Br}(h_c \to \eta_c e^+ e^-)}{\mathrm{Br}(h_c \to \eta_c \gamma)} \bigg|_{\mathrm{BESIII}} = (0.59 \pm 0.10 \pm 0.04)\times 10^{-2}~.
\end{align}
Indeed, using:
\begin{align}
\Gamma(h_c \to \eta_c \gamma) = \frac{2 Q_{c}^{2}}{3} \alpha_{\mathrm{em}} \frac{m_{h_c}^{2} - m_{\eta_c}^{2}}{m_{h_c}} |F_1(0)|^{2},
\end{align}
we obtain:
\begin{align}
\frac{\mathrm{Br}(h_c \to \eta_c e^+ e^-)}{\mathrm{Br}(h_c \to \eta_c \gamma)} = 0.9341(14)\%,
\end{align}
which is higher than the BESIII result by $3.2$ standard deviations. In the future, it will be important to improve the experimental determination to monitor whether this difference persists or disappears. Unfortunately, in the case of $h_c \to \eta_c \mu^+ \mu^-$, no experimental information is currently available.

\section{Conclusions}
\label{sec:conclusions}

In this work, we have presented the first fully dynamical lattice QCD determination of the electromagnetic transition form factors governing the decays $h_c \to \eta_c \ell^+ \ell^-$ and $ \chi_{c1} \to J/\psi \ell^+ \ell^-$, where $\ell = e, \mu$, thereby providing new information on the non-perturbative dynamics of charmonium systems. The calculation was performed using $N_f = 2+1+1$ dynamical Wilson--Clover twisted-mass fermions at four lattice spacings, allowing for a controlled continuum extrapolation at (near-)physical quark masses. The relevant matrix elements were computed including only quark-connected contributions.

We have determined all relevant form factors as functions of the momentum transfer $q^2$ over the full kinematic range. In the case of $\chi_{c1} \to J/\psi \ell^+ \ell^-$, this includes the multipole form factors $E_1(q^2)$, $M_2(q^2)$, and the longitudinal form factor $C_1(q^2)$, which does not contribute in the real-photon limit. Similarly, for $h_c \to \eta_c \ell^+ \ell^-$, we have computed both $F_1(q^2)$ and $\tilde{F}_2(q^2)$, the latter being accessible only for off-shell photons. The $q^2$-dependence of all form factors is smooth, and the continuum extrapolations are well described by an $a^2$ scaling behavior. The statistical precision after continuum-limit extrapolation is at the few-percent level, increasing slightly as $q^2$ approaches  $q^2_{\mathrm{max}}$.

Using the continuum-extrapolated form factors, we have computed the partial decay widths reported in Eq.~\eqref{eq:final_res_width_chi} for the $\chi_{c1}$ channel. These results are in very good agreement with the experimental determinations obtained from the PDG branching fractions in Eq.~\eqref{eq:PDG_branchings} or from the  BESIII branching fractions in Eq.~\eqref{eq:bes_branchings}.

In addition to the comparison of absolute decay widths, we have considered ratios of branching fractions, which provide a direct comparison with experiment without relying on external input for the total width of the $\chi_{c1}$. In particular, we have compared our lattice determination of the ratio between the Dalitz decay and the corresponding radiative decay,
\begin{align}
\frac{\mathrm{Br}(\chi_{c1} \to J/\psi e^+ e^-)}{\mathrm{Br}(\chi_{c1} \to J/\psi \gamma)}~,
\end{align}
given in Eq.~\eqref{eq:lqcd_D_1}, with the experimental result quoted in Eq.~\eqref{eq:bes_D_1}. We find a difference of about $2\sigma$. The lepton-flavor universality ratio reported in Eq.~\eqref{eq:lqcd_D_2} is, instead, in agreement with experiment.

We have also compared our predictions for the $q^2$-differential distribution in the $\chi_{c1} \to J/\psi e^+ e^-$ decay with the corresponding BESIII measurements. After applying the efficiency corrections, we find overall agreement between lattice QCD and experimental data. However, a few bins in the low-$q^2$ region exhibit tensions at the level of approximately $2\text{--}3\sigma$. More precise experimental measurements in this kinematic region would therefore be highly valuable to assess the significance of these differences.

Turning to the $h_c \to \eta_c e^+ e^-$ decay, our prediction for the total decay width, reported in Eq.~\eqref{eq:final_res_width_hc}, exceeds the experimental determination quoted in Eq.~\eqref{eq:rel_ph_hc} by approximately $2.5\text{--}3\sigma$. Given the current experimental uncertainties on both the branching fraction and the total width of the $h_c$, improved measurements of this channel would be particularly important to clarify whether this difference persists with increased precision.

The availability of form factors over the full kinematic range enables a variety of phenomenological analyses. In particular, our results provide the nonperturbative QCD input required for refined studies of possible contributions from light new mediators in charmonium Dalitz decays. In analogy with the recent analysis performed by the BESIII Collaboration in Ref.~\cite{BESIII:2025otp}, where bounds on dark-photon couplings were extracted from $\chi_{cJ} \to J/\psi e^+ e^-$ data, our lattice determination of the transition form factors allows for a fully first-principles treatment of the hadronic matrix elements entering such searches. This will make it possible to reassess and potentially sharpen constraints on dark-sector scenarios from charmonium decays. A dedicated phenomenological study along these lines will be presented elsewhere.

A natural extension of the present work is the study of the transition $\Psi(2S) \to \chi_{c1} \gamma$ together with its Dalitz counterpart in which the virtual photon converts into a lepton pair. The abundant production of $\Psi(2S)$ at BESIII would allow for precise experimental measurements of these processes. From the lattice QCD perspective, the $\Psi(2S)$ state presents additional challenges, as it is not the ground state in the corresponding spin and flavor channel. However, established techniques, such as the Generalized Eigenvalue method~\cite{Luscher:1990ck}, can be employed to face the problem. This extension will be investigated in future work.

\section{Acknowledgments}
\label{sec:akno}
We thank Gaofeng Fan for useful discussions and Zhang Jielei for kindly providing the efficiency correction factors applied to the BESIII data of Ref.~\cite{BESIII:2017ung} in our bin-by-bin comparison.  We thank the ETMC for the most enjoyable collaboration. V.L., G.G., R.F., and N.T. have been supported by the Italian Ministry
of University and Research (MUR) and the European
Union (EU) – Next Generation EU, Mission 4, Component 1, PRIN 2022, CUP F53D23001480006. 
F.S. is supported by ICSC – Centro Nazionale di Ricerca in High Performance Computing, Big Data and Quantum Computing, funded by European Union - Next Generation EU and by Italian  Ministry of University and Research (MUR) project FIS\_00001556. We acknowledge support from the LQCD123, ENP, and SPIF Scientific Initiatives of
the Italian Nuclear Physics Institute (INFN). 
This project has received support from the IN2P3 (CNRS) Master Project HighPTflavor.

The open-source packages tmLQCD~\cite{Jansen:2009xp,Abdel-Rehim:2013wba,Deuzeman:2013xaa,Kostrzewa:2022hsv}, LEMON~\cite{Deuzeman:2011wz}, DD-$\alpha$AMG~\cite{Frommer:2013fsa,Alexandrou:2016izb,Bacchio:2017pcp,Alexandrou:2018wiv}, QPhiX~\cite{joo2016optimizing,Schrock:2015gik} and QUDA~\cite{Clark:2009wm,Babich:2011np,Clark:2016rdz} have been used in the ensemble generation.

We gratefully acknowledge the ICSC - Centro Nazionale di Ricerca in High Performance Computing for providing computing time under the allocations RAC 1916318. We gratefully acknowledge CINECA for the provision of GPU time on Leonardo supercomputing facilities under the specific initiative INFN-LQCD123, and under project IscrB VITO-QCD, project IscrB SemBD, project IscrB ADSL2 and project IscrB Hcee. We gratefully acknowledge EuroHPC Joint Undertaking for awarding us access to MareNostrum5 through the project EHPC-EXT-2024E01-031. The authors gratefully acknowledge the Gauss Centre for Supercomputing e.V. (www.gauss-centre.eu) for funding this project by providing computing time on the GCS Supercomputers SuperMUC-NG at Leibniz Supercomputing Centre. The authors acknowledge the Texas Advanced Computing Center (TACC) at The University of Texas at Austin for providing HPC resources (Project ID PHY21001). We acknowledge EuroHPC Joint Undertaking for awarding the project ID EHPC-EXT-2023E02-052 access to MareNostrum5 hosted by at the Barcelona Supercomputing Center, Spain. We also thank the GENCI.fr for granting us access to the Jean Zay computers of the computing center IDRIS in Orsay.

\bibliographystyle{JHEP}
\bibliography{biblio}

\providecommand{\href}[2]{#2}\begingroup\raggedright\begin{thebibliography}{10}

\bibitem{Bodwin:1994jh}
G.T.~Bodwin, E.~Braaten and G.P.~Lepage, \emph{{Rigorous QCD analysis of inclusive annihilation and production of heavy quarkonium}}, \href{https://doi.org/10.1103/PhysRevD.55.5853}{\emph{Phys. Rev. D} {\bfseries 51} (1995) 1125} [\href{https://arxiv.org/abs/hep-ph/9407339}{{\ttfamily hep-ph/9407339}}].

\bibitem{Brambilla:1999xf}
N.~Brambilla, A.~Pineda, J.~Soto and A.~Vairo, \emph{{Potential NRQCD: An Effective theory for heavy quarkonium}}, \href{https://doi.org/10.1016/S0550-3213(99)00693-8}{\emph{Nucl. Phys. B} {\bfseries 566} (2000) 275} [\href{https://arxiv.org/abs/hep-ph/9907240}{{\ttfamily hep-ph/9907240}}].

\bibitem{Fleming:2000ib}
S.~Fleming, I.Z.~Rothstein and A.K.~Leibovich, \emph{{Power counting and effective field theory for charmonium}}, \href{https://doi.org/10.1103/PhysRevD.64.036002}{\emph{Phys. Rev. D} {\bfseries 64} (2001) 036002} [\href{https://arxiv.org/abs/hep-ph/0012062}{{\ttfamily hep-ph/0012062}}].

\bibitem{Brambilla:2004jw}
N.~Brambilla, A.~Pineda, J.~Soto and A.~Vairo, \emph{{Effective Field Theories for Heavy Quarkonium}}, \href{https://doi.org/10.1103/RevModPhys.77.1423}{\emph{Rev. Mod. Phys.} {\bfseries 77} (2005) 1423} [\href{https://arxiv.org/abs/hep-ph/0410047}{{\ttfamily hep-ph/0410047}}].

\bibitem{Dudek:2006ej}
J.J.~Dudek, R.G.~Edwards and D.G.~Richards, \emph{{Radiative transitions in charmonium from lattice QCD}}, \href{https://doi.org/10.1103/PhysRevD.73.074507}{\emph{Phys. Rev. D} {\bfseries 73} (2006) 074507} [\href{https://arxiv.org/abs/hep-ph/0601137}{{\ttfamily hep-ph/0601137}}].

\bibitem{Dudek:2009kk}
J.J.~Dudek, R.~Edwards and C.E.~Thomas, \emph{{Exotic and excited-state radiative transitions in charmonium from lattice QCD}}, \href{https://doi.org/10.1103/PhysRevD.79.094504}{\emph{Phys. Rev. D} {\bfseries 79} (2009) 094504} [\href{https://arxiv.org/abs/0902.2241}{{\ttfamily 0902.2241}}].

\bibitem{Becirevic:2012dc}
D.~Becirevic and F.~Sanfilippo, \emph{{Lattice QCD study of the radiative decays $J/\psi\to \eta_c\gamma$ and $h_c\to \eta_c\gamma$}}, \href{https://doi.org/10.1007/JHEP01(2013)028}{\emph{JHEP} {\bfseries 01} (2013) 028} [\href{https://arxiv.org/abs/1206.1445}{{\ttfamily 1206.1445}}].

\bibitem{Chen:2011kpa}
Y.~Chen et~al., \emph{{Radiative transitions in charmonium from $N_f=2$ twisted mass lattice QCD}}, \href{https://doi.org/10.1103/PhysRevD.84.034503}{\emph{Phys. Rev. D} {\bfseries 84} (2011) 034503} [\href{https://arxiv.org/abs/1104.2655}{{\ttfamily 1104.2655}}].

\bibitem{Li:2022cfy}
N.~Li, H.~Su and Y.-J.~Wu, \emph{{$\chi _{c1}\rightarrow {J/\psi \gamma }$ decay width on $N_f=2$ twisted mass lattice QCD}}, \href{https://doi.org/10.1140/epja/s10050-022-00768-w}{\emph{Eur. Phys. J. A} {\bfseries 58} (2022) 122}.

\bibitem{Becirevic:2025idm}
D.~Be{\v{c}}irevi{\'c}, R.~Di~Palma, R.~Frezzotti, G.~Gagliardi, V.~Lubicz, F.~Sanfilippo et~al., \emph{{Lattice QCD study of the {\ensuremath{\chi}}c1{\textrightarrow}J/{\ensuremath{\psi}}{\ensuremath{\gamma}} decay}}, \href{https://doi.org/10.1016/j.physletb.2025.139811}{\emph{Phys. Lett. B} {\bfseries 868} (2025) 139811} [\href{https://arxiv.org/abs/2506.17030}{{\ttfamily 2506.17030}}].

\bibitem{Becirevic:2025ocx}
D.~Be{\v{c}}irevi{\'c}, R.~Di~Palma, R.~Frezzotti, G.~Gagliardi, V.~Lubicz, F.~Sanfilippo et~al., \emph{{Lattice QCD determination of the radiative decay rates hc{\textrightarrow}{\ensuremath{\eta}}c{\ensuremath{\gamma}} and hb{\textrightarrow}{\ensuremath{\eta}}b{\ensuremath{\gamma}}}}, \href{https://doi.org/10.1103/ysxn-sxj7}{\emph{Phys. Rev. D} {\bfseries 112} (2025) 034505} [\href{https://arxiv.org/abs/2504.16807}{{\ttfamily 2504.16807}}].

\bibitem{Delaney:2023fsc}
{\scshape Hadron Spectrum} collaboration, \emph{{Radiative transitions in charmonium from lattice QCD}}, \href{https://doi.org/10.1007/JHEP05(2024)230}{\emph{JHEP} {\bfseries 05} (2024) 230} [\href{https://arxiv.org/abs/2301.08213}{{\ttfamily 2301.08213}}].

\bibitem{PhysRevD.86.094501}
{\scshape HPQCD} collaboration, \emph{Precision tests of the $j/\ensuremath{\psi}$ from full lattice qcd: Mass, leptonic width, and radiative decay rate to ${\ensuremath{\eta}}_{c}$}, \href{https://doi.org/10.1103/PhysRevD.86.094501}{\emph{Phys. Rev. D} {\bfseries 86} (2012) 094501}.

\bibitem{PhysRevD.108.014513}
{\scshape HPQCD} collaboration, \emph{Precise determination of decay rates for ${\ensuremath{\eta}}_{c}\ensuremath{\rightarrow}\ensuremath{\gamma}\ensuremath{\gamma}$, $j/\ensuremath{\psi}\ensuremath{\rightarrow}\ensuremath{\gamma}{\ensuremath{\eta}}_{c}$, and $j/\ensuremath{\psi}\ensuremath{\rightarrow}{\ensuremath{\eta}}_{c}{e}^{+}{e}^{\ensuremath{-}}$ from lattice qcd}, \href{https://doi.org/10.1103/PhysRevD.108.014513}{\emph{Phys. Rev. D} {\bfseries 108} (2023) 014513}.

\bibitem{BESIII:2017ung}
{\scshape BESIII} collaboration, \emph{{Observation of $\psi(3686) \rightarrow e^{+}e^{-}\chi_{cJ}$ and $\chi_{cJ} \rightarrow e^{+}e^{-}J/\psi$}}, \href{https://doi.org/10.1103/PhysRevLett.118.221802}{\emph{Phys. Rev. Lett.} {\bfseries 118} (2017) 221802} [\href{https://arxiv.org/abs/1701.05404}{{\ttfamily 1701.05404}}].

\bibitem{BESIII:2019yeu}
{\scshape BESIII} collaboration, \emph{{Study of electromagnetic Dalitz decays $\chi_{cJ} \rightarrow \mu^{+}\mu^{-}J/\psi$}}, \href{https://doi.org/10.1103/PhysRevD.99.051101}{\emph{Phys. Rev. D} {\bfseries 99} (2019) 051101} [\href{https://arxiv.org/abs/1901.06627}{{\ttfamily 1901.06627}}].

\bibitem{BESIII:2024kkf}
{\scshape BESIII} collaboration, \emph{{Observation of the electromagnetic Dalitz transition hc{\textrightarrow}e+e-{\ensuremath{\eta}}c}}, \href{https://doi.org/10.1103/PhysRevD.110.L111101}{\emph{Phys. Rev. D} {\bfseries 110} (2024) L111101} [\href{https://arxiv.org/abs/2407.00136}{{\ttfamily 2407.00136}}].

\bibitem{Colangelo:2025yud}
P.~Colangelo, F.~De~Fazio and R.~Pinto, \emph{{Two-lepton tales: Dalitz decays of heavy quarkonia}},  \href{https://arxiv.org/abs/2512.17672}{{\ttfamily 2512.17672}}.

\bibitem{PDG2024}
{\scshape Particle Data Group} collaboration, \emph{{Review of particle physics}}, \href{https://doi.org/10.1103/PhysRevD.110.030001}{\emph{Phys. Rev. D} {\bfseries 110} (2024) 030001}.

\bibitem{ExtendedTwistedMassCollaborationETMC:2024xdf}
{\scshape Extended Twisted Mass Collaboration (ETMC)} collaboration, \emph{{Strange and charm quark contributions to the muon anomalous magnetic moment in lattice QCD with twisted-mass fermions}},  \href{https://arxiv.org/abs/2411.08852}{{\ttfamily 2411.08852}}.

\bibitem{Frezzotti:2003ni}
R.~Frezzotti and G.C.~Rossi, \emph{{Chirally improving Wilson fermions. 1. O(a) improvement}}, \href{https://doi.org/10.1088/1126-6708/2004/08/007}{\emph{JHEP} {\bfseries 08} (2004) 007} [\href{https://arxiv.org/abs/hep-lat/0306014}{{\ttfamily hep-lat/0306014}}].

\bibitem{Frezzotti:2004wz}
R.~Frezzotti and G.C.~Rossi, \emph{{Chirally improving Wilson fermions. II. Four-quark operators}}, \href{https://doi.org/10.1088/1126-6708/2004/10/070}{\emph{JHEP} {\bfseries 10} (2004) 070} [\href{https://arxiv.org/abs/hep-lat/0407002}{{\ttfamily hep-lat/0407002}}].

\bibitem{Hatton:2020qhk}
{\scshape HPQCD} collaboration, \emph{{Charmonium properties from lattice $QCD$+QED : Hyperfine splitting, $J/\psi$ leptonic width, charm quark mass, and $a^c_\mu$}}, \href{https://doi.org/10.1103/PhysRevD.102.054511}{\emph{Phys. Rev. D} {\bfseries 102} (2020) 054511} [\href{https://arxiv.org/abs/2005.01845}{{\ttfamily 2005.01845}}].

\bibitem{deDivitiis:2004kq}
G.M.~de~Divitiis, R.~Petronzio and N.~Tantalo, \emph{{On the discretization of physical momenta in lattice QCD}}, \href{https://doi.org/10.1016/j.physletb.2004.06.035}{\emph{Phys. Lett. B} {\bfseries 595} (2004) 408} [\href{https://arxiv.org/abs/hep-lat/0405002}{{\ttfamily hep-lat/0405002}}].

\bibitem{Bedaque:2004kc}
P.F.~Bedaque, \emph{{Aharonov-Bohm effect and nucleon nucleon phase shifts on the lattice}}, \href{https://doi.org/10.1016/j.physletb.2004.04.045}{\emph{Phys. Lett. B} {\bfseries 593} (2004) 82} [\href{https://arxiv.org/abs/nucl-th/0402051}{{\ttfamily nucl-th/0402051}}].

\bibitem{Sachrajda:2004mi}
C.T.~Sachrajda and G.~Villadoro, \emph{{Twisted boundary conditions in lattice simulations}}, \href{https://doi.org/10.1016/j.physletb.2005.01.033}{\emph{Phys. Lett. B} {\bfseries 609} (2005) 73} [\href{https://arxiv.org/abs/hep-lat/0411033}{{\ttfamily hep-lat/0411033}}].

\bibitem{Sheikholeslami:1985ij}
B.~Sheikholeslami and R.~Wohlert, \emph{{Improved Continuum Limit Lattice Action for QCD with Wilson Fermions}}, \href{https://doi.org/10.1016/0550-3213(85)90002-1}{\emph{Nucl. Phys. B} {\bfseries 259} (1985) 572}.

\bibitem{Krasznahorkay:2015iga}
A.J.~Krasznahorkay et~al., \emph{{Observation of Anomalous Internal Pair Creation in Be8 : A Possible Indication of a Light, Neutral Boson}}, \href{https://doi.org/10.1103/PhysRevLett.116.042501}{\emph{Phys. Rev. Lett.} {\bfseries 116} (2016) 042501} [\href{https://arxiv.org/abs/1504.01527}{{\ttfamily 1504.01527}}].

\bibitem{PADME:2025dla}
{\scshape PADME} collaboration, \emph{{Search for a new 17 MeV resonance via e$^{+}$e$^{-}$ annihilation with the PADME experiment}}, \href{https://doi.org/10.1007/JHEP11(2025)007}{\emph{JHEP} {\bfseries 11} (2025) 007} [\href{https://arxiv.org/abs/2505.24797}{{\ttfamily 2505.24797}}].

\bibitem{BESIII:2025otp}
{\scshape BESIII} collaboration, \emph{{Search for a hypothetical gauge boson and dark photons in charmonium transitions}}, \href{https://doi.org/10.1103/76js-7s9f}{\emph{Phys. Rev. D} {\bfseries 113} (2026) 032009} [\href{https://arxiv.org/abs/2510.16531}{{\ttfamily 2510.16531}}].

\bibitem{Luscher:1990ck}
M.~Luscher and U.~Wolff, \emph{{How to Calculate the Elastic Scattering Matrix in Two-dimensional Quantum Field Theories by Numerical Simulation}}, \href{https://doi.org/10.1016/0550-3213(90)90540-T}{\emph{Nucl. Phys. B} {\bfseries 339} (1990) 222}.

\bibitem{Jansen:2009xp}
K.~Jansen and C.~Urbach, \emph{{tmLQCD: A Program suite to simulate Wilson Twisted mass Lattice QCD}}, \href{https://doi.org/10.1016/j.cpc.2009.05.016}{\emph{Comput. Phys. Commun.} {\bfseries 180} (2009) 2717} [\href{https://arxiv.org/abs/0905.3331}{{\ttfamily 0905.3331}}].

\bibitem{Abdel-Rehim:2013wba}
A.~Abdel-Rehim, F.~Burger, A.~Deuzeman, K.~Jansen, B.~Kostrzewa, L.~Scorzato et~al., \emph{{Recent developments in the tmLQCD software suite}}, \href{https://doi.org/10.22323/1.187.0414}{\emph{PoS} {\bfseries LATTICE2013} (2014) 414} [\href{https://arxiv.org/abs/1311.5495}{{\ttfamily 1311.5495}}].

\bibitem{Deuzeman:2013xaa}
A.~Deuzeman, K.~Jansen, B.~Kostrzewa and C.~Urbach, \emph{{Experiences with OpenMP in tmLQCD}}, \href{https://doi.org/10.22323/1.187.0416}{\emph{PoS} {\bfseries LATTICE2013} (2014) 416} [\href{https://arxiv.org/abs/1311.4521}{{\ttfamily 1311.4521}}].

\bibitem{Kostrzewa:2022hsv}
{\scshape ETM} collaboration, \emph{{Twisted mass ensemble generation on GPU machines}}, \href{https://doi.org/10.22323/1.430.0340}{\emph{PoS} {\bfseries LATTICE2022} (2023) 340} [\href{https://arxiv.org/abs/2212.06635}{{\ttfamily 2212.06635}}].

\bibitem{Deuzeman:2011wz}
{\scshape ETM} collaboration, \emph{{Lemon: an MPI parallel I/O library for data encapsulation using LIME}}, \href{https://doi.org/10.1016/j.cpc.2012.01.016}{\emph{Comput. Phys. Commun.} {\bfseries 183} (2012) 1321} [\href{https://arxiv.org/abs/1106.4177}{{\ttfamily 1106.4177}}].

\bibitem{Frommer:2013fsa}
A.~Frommer, K.~Kahl, S.~Krieg, B.~Leder and M.~Rottmann, \emph{{Adaptive Aggregation-Based Domain Decomposition Multigrid for the Lattice Wilson--Dirac Operator}}, \href{https://doi.org/10.1137/130919507}{\emph{SIAM J. Sci. Comput.} {\bfseries 36} (2014) A1581} [\href{https://arxiv.org/abs/1303.1377}{{\ttfamily 1303.1377}}].

\bibitem{Alexandrou:2016izb}
C.~Alexandrou, S.~Bacchio, J.~Finkenrath, A.~Frommer, K.~Kahl and M.~Rottmann, \emph{{Adaptive Aggregation-based Domain Decomposition Multigrid for Twisted Mass Fermions}}, \href{https://doi.org/10.1103/PhysRevD.94.114509}{\emph{Phys. Rev. D} {\bfseries 94} (2016) 114509} [\href{https://arxiv.org/abs/1610.02370}{{\ttfamily 1610.02370}}].

\bibitem{Bacchio:2017pcp}
S.~Bacchio, C.~Alexandrou and J.~Finkerath, \emph{{Multigrid accelerated simulations for Twisted Mass fermions}}, \href{https://doi.org/10.1051/epjconf/201817502002}{\emph{EPJ Web Conf.} {\bfseries 175} (2018) 02002} [\href{https://arxiv.org/abs/1710.06198}{{\ttfamily 1710.06198}}].

\bibitem{Alexandrou:2018wiv}
C.~Alexandrou, S.~Bacchio and J.~Finkenrath, \emph{{Multigrid approach in shifted linear systems for the non-degenerated twisted mass operator}}, \href{https://doi.org/10.1016/j.cpc.2018.10.013}{\emph{Comput. Phys. Commun.} {\bfseries 236} (2019) 51} [\href{https://arxiv.org/abs/1805.09584}{{\ttfamily 1805.09584}}].

\bibitem{joo2016optimizing}
B.~Jo{\'o}, D.D.~Kalamkar, T.~Kurth, K.~Vaidyanathan and A.~Walden, \emph{Optimizing wilson-dirac operator and linear solvers for intel{\textregistered} knl},  in \emph{High Performance Computing: ISC High Performance 2016 International Workshops, ExaComm, E-MuCoCoS, HPC-IODC, IXPUG, IWOPH, P\^{} 3MA, VHPC, WOPSSS, Frankfurt, Germany, June 19--23, 2016, Revised Selected Papers 31}, pp.~415--427, Springer, 2016.

\bibitem{Schrock:2015gik}
M.~Schr\"ock, S.~Simula and A.~Strelchenko, \emph{{Accelerating Twisted Mass LQCD with QPhiX}}, \href{https://doi.org/10.22323/1.251.0030}{\emph{PoS} {\bfseries LATTICE2015} (2016) 030} [\href{https://arxiv.org/abs/1510.08879}{{\ttfamily 1510.08879}}].

\bibitem{Clark:2009wm}
M.A.~Clark, R.~Babich, K.~Barros, R.C.~Brower and C.~Rebbi, \emph{{Solving Lattice QCD systems of equations using mixed precision solvers on GPUs}}, \href{https://doi.org/10.1016/j.cpc.2010.05.002}{\emph{Comput. Phys. Commun.} {\bfseries 181} (2010) 1517} [\href{https://arxiv.org/abs/0911.3191}{{\ttfamily 0911.3191}}].

\bibitem{Babich:2011np}
R.~Babich, M.A.~Clark, B.~Joo, G.~Shi, R.C.~Brower and S.~Gottlieb, \emph{{Scaling Lattice QCD beyond 100 GPUs}},  in \emph{{SC11 International Conference for High Performance Computing, Networking, Storage and Analysis Seattle, Washington, November 12-18, 2011}}, 2011, \href{https://doi.org/10.1145/2063384.2063478}{DOI} [\href{https://arxiv.org/abs/1109.2935}{{\ttfamily 1109.2935}}].

\bibitem{Clark:2016rdz}
M.A.~Clark, B.~Jo{\'o}, A.~Strelchenko, M.~Cheng, A.~Gambhir and R.C.~Brower, \emph{{Accelerating Lattice QCD Multigrid on GPUs Using Fine-Grained Parallelization}},  in \emph{SC '16: Proceedings of the International Conference for High Performance Computing, Networking, Storage and Analysis}, pp.~795--806, 2016, \href{https://doi.org/10.1109/SC.2016.67}{DOI} [\href{https://arxiv.org/abs/1612.07873}{{\ttfamily 1612.07873}}].

\end{thebibliography}\endgroup

\end{document}